\documentclass[11pt,a4paper]{article}
\pdfoutput=1
\usepackage{cite}
\usepackage{graphicx}
\usepackage{amsmath}
\usepackage{amssymb}
\usepackage{multirow}
\usepackage{color}

\newcommand{\nonu}{\nonumber}

\newcommand{\mrm}[1]{\mathrm{#1}}

\renewcommand{\b}{{\mathrm b}}
\renewcommand{\c}{{\mathrm c}}
\renewcommand{\d}{{\mathrm d}}
\newcommand{\e}{{\mathrm e}}

\newcommand{\g}{{\mathrm g}}

\newcommand{\p}{{\mathrm p}}
\newcommand{\q}{{\mathrm q}}
\newcommand{\s}{{\mathrm s}}

\renewcommand{\u}{{\mathrm u}}
\newcommand{\A}{{\mathrm A}}

\newcommand{\Pb}{{\mathrm{Pb}}}

\newcommand{\pbar}{\overline{\mathrm p}}
\newcommand{\qbar}{\overline{\mathrm q}}

\newcommand{\pp}{\p\p}
\newcommand{\ppbar}{\p\pbar}

\newcommand{\Pom}{\mathbb{P}}

\newcommand{\TeV}{\mathrm{TeV}}
\newcommand{\GeV}{\mathrm{GeV}}
\newcommand{\MeV}{\mathrm{MeV}}

\newcommand{\aem}{\alpha_{\mathrm{em}}}

\newcommand{\pT}{p_{\perp}}

\newcommand{\pTo}{p_{\perp 0}}

{\end{list}}
\newcounter{enumct}

\newlength{\abstwidth}
\begin{document}
\sloppy
 
\pagestyle{empty}

\begin{flushright}
LU TP 19-06\\
MCnet-19-01\\
January 2019
\end{flushright}

\vspace{\fill}

\begin{center}
{\Huge\bf Hard diffraction in photoproduction with \textsc{Pythia}~8}\\[4mm]
{\Large Ilkka Helenius\footnote{Institute for Theoretical Physics,
T\"ubingen University, Auf der Morgenstelle 14, 72076 T\"ubingen, 
Germany}$^,$\footnote{University of Jyvaskyla, Department of Physics, 
P.O. Box 35, FI-40014 University of Jyvaskyla, Finland}}, 
{\Large Christine O. Rasmussen\footnote{\label{LU}{Department of 
Astronomy and Theoretical Physics, Lund University, S\"olvegatan 
14A, SE-223 62 Lund, Sweden}} }

\end{center}

\begin{center}
\begin{minipage}{\abstwidth}
{\bf Abstract}
We present a new framework for modeling hard diffractive events in
photoproduction, implemented in the general purpose event
generator \textsc{Pythia}~8. The model is an 
extension of the model for hard diffraction with dynamical gap survival 
in $\p\p$ and $\p\pbar$ collisions proposed in 2015, now also allowing 
for other beam types. It thus relies on several existing ideas: the 
Ingelman-Schlein approach, the framework for multiparton interactions 
and the recently developed framework for photoproduction in 
$\gamma\p$, $\gamma\gamma$, $\e\p$ and $\e^+\e^-$ collisions. The model 
proposes an explanation for the observed factorization breaking in 
photoproduced diffractive dijet events at HERA, showing an overall good 
agreement with data. The model is also applicable to ultraperipheral 
collisions with $\p\p$ and $\p\Pb$ beams, and predictions are made for 
such events at the LHC. 

\end{minipage}
\end{center}

\vspace{\fill}

\phantom{dummy}

\clearpage

\pagestyle{plain}
\setcounter{page}{1}

\section{\label{Sec:1}Introduction}

Diffractive excitations represent large fractions of the total cross
section in a wide range of collisions. A part of these has been seen to
have a hard scale, as in e.g.\ the case of diffractive dijet production.
These \textit{hard diffractive} events allow for a perturbative
calculation of the scattering subprocess, but still require some 
phenomenological modeling. This includes modeling 
of the Pomeron, expected to be responsible for the 
color-neutral momentum transfer between the 
beam and the diffractive system $X$. In the framework of collinear 
factorization, a diffractive parton distribution 
function (dPDF) may be defined. This can further 
be factorized into a Pomeron flux and a PDF, describing the flux of 
Pomerons from the beam and the parton density within the Pomeron, 
respectively.

\begin{figure}[!ht]
\begin{minipage}[c]{0.475\linewidth}
\centering
\includegraphics[width=\linewidth]{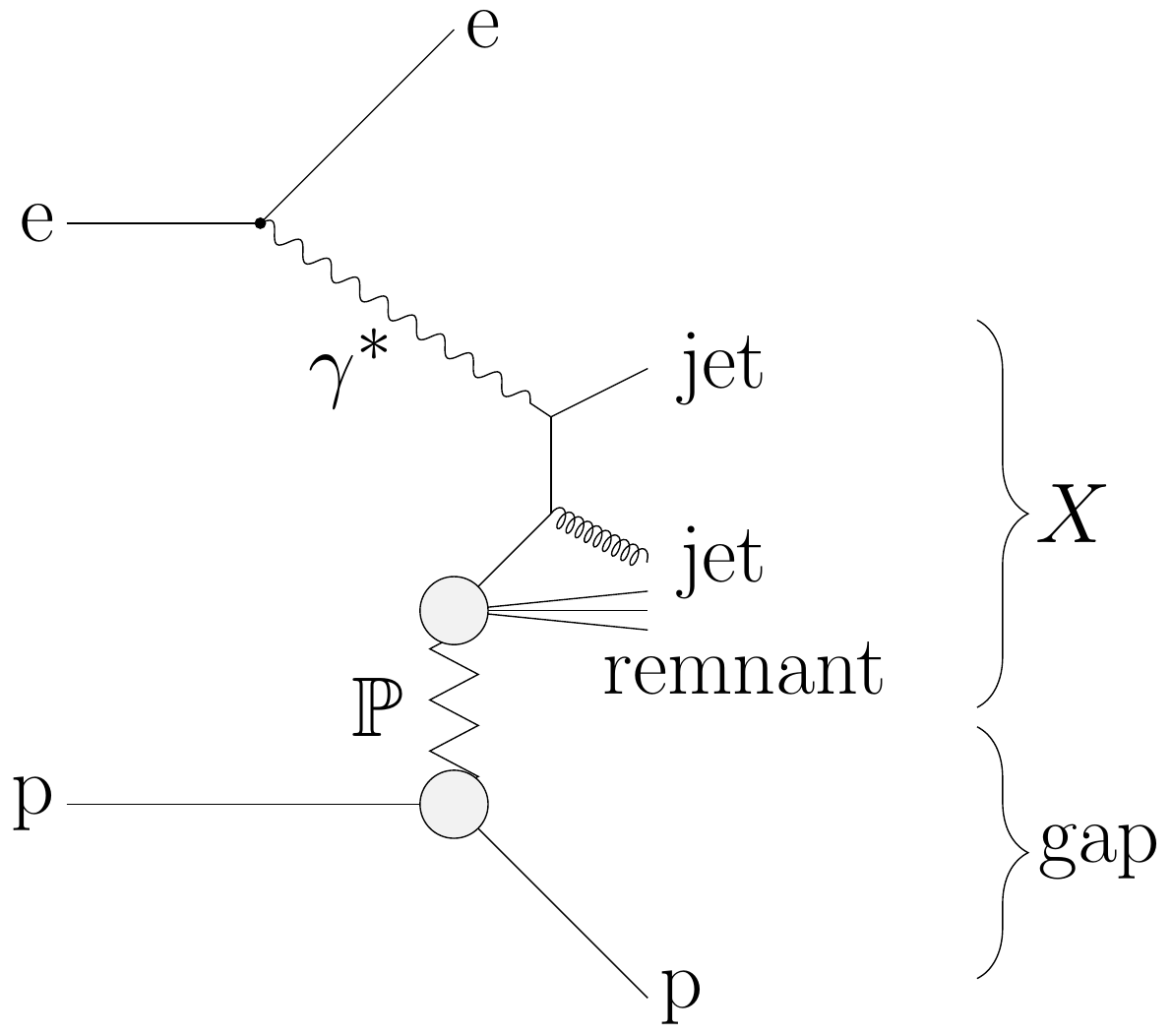}\\
(a)
\end{minipage}
\hfill
\begin{minipage}[c]{0.475\linewidth}
\centering
\includegraphics[width=\linewidth]{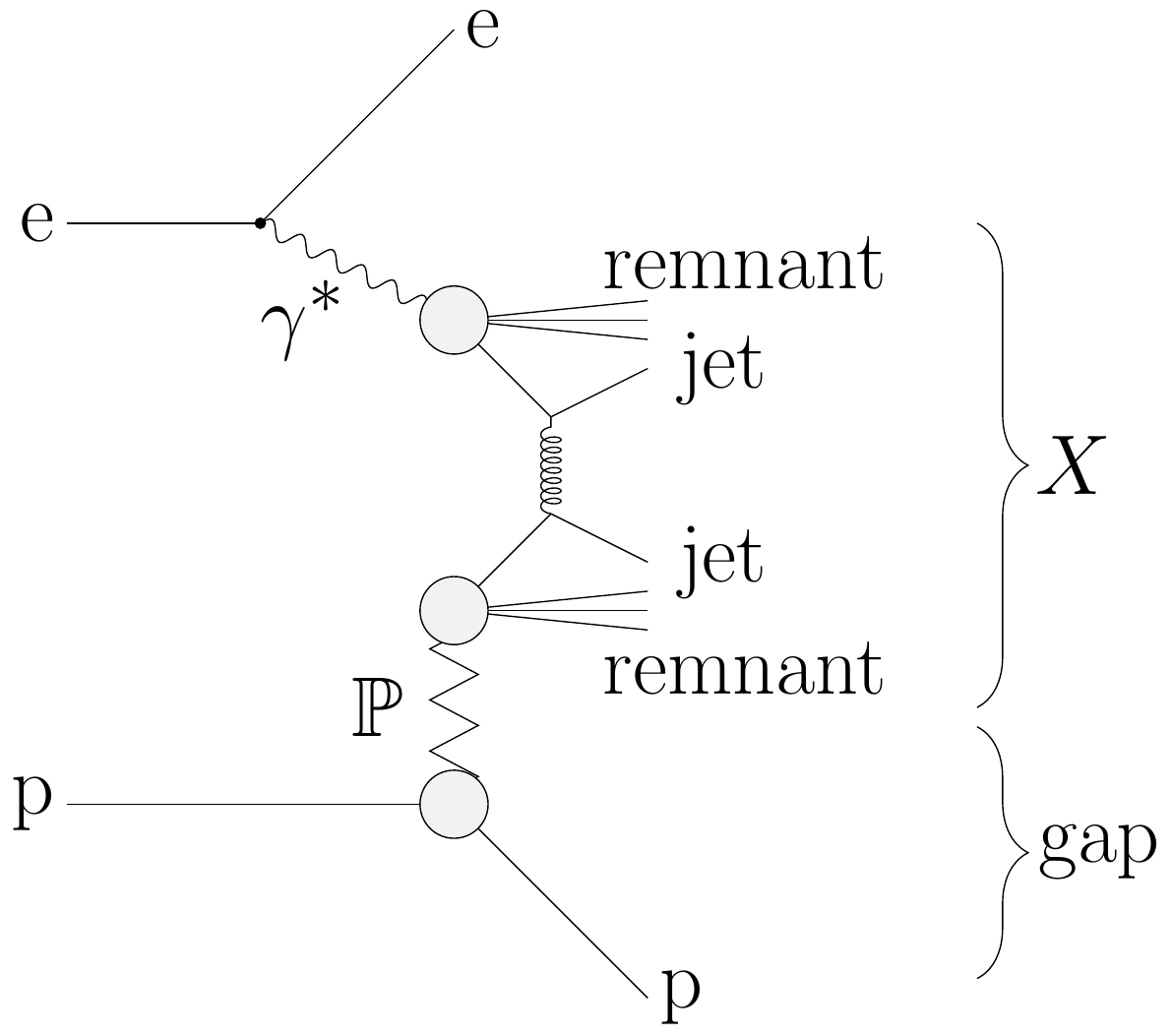}\\
(b)
\end{minipage}
\label{Fig:diff_systems}
\caption{Leading order Feynman diagrams for 
diffractive dijet production with photons in $\e\p$ collisions. 
Either the photon participates directly in the 
hard scattering matrix element (a) or a parton from the resolved 
photon participates (b). 
}
\end{figure}

Here we focus mainly on photoproduced diffractive dijets in $\e\p$
collisions. This scattering process can be separated into different 
subsystems, visualized in Fig.~\ref{Fig:diff_systems}. The initial state consists of an 
electron and a proton, with the former radiating off a (virtual) 
photon. If the photon is highly virtual, we are in the range of deep 
inelastic scattering (DIS), while a photon with low enough virtuality 
can be considered (quasi-)real. This is the photoproduction regime. 
No clear distinction between the two regimes exists, however, and 
photons of intermediate virtuality require careful consideration to 
avoid double-counting. A special feature in the photoproduction 
regime is that there is a non-negligible probability for the photon 
to fluctuate into a hadronic state. These resolved photons open up 
for all possible hadron-hadron processes, including diffractive ones.

The next subsystem shown in Fig.~\ref{Fig:diff_systems} is the photon-proton scattering 
system. Here, diffraction could in principle occur on both sides if 
the photon is resolved. In direct photoproduction (and in DIS) the 
diffractive system can only be present on the photon side, as no 
Pomeron flux can be defined for point-like photons. In this article 
the emphasis will be on Pomeron emission from the proton.

The final subsystem is the hard scattering generated
inside the diffractive system $X$. For direct photoproduction
(and DIS) this includes the photon as an incoming parton, see Fig.~\ref{Fig:diff_systems}
(a). In the resolved case, Fig.~\ref{Fig:diff_systems} (b), a parton is extracted from the
hadronic photon, which then proceeds to initiate the hard scattering along
with a parton extracted from the Pomeron. In both cases a beam remnant
is left behind from the Pomeron, while resolved photoproduction also gives 
rise to a beam remnant from the hadronic photon. Multiple scatterings or
multiparton interactions (MPIs) are expected between the remnants, but
also in the larger photon-proton system. 
The particles produced by the latter type of MPIs may destroy the
diffractive signature, the \textit{rapidity gap} between the diffractive
system and the elastically scattered proton (or meson, depending on the
side of the diffractive system).\\

The model for photoproduced diffractive dijets presented here is 
based on the general-purpose event ge\-nerator 
\textsc{Pythia}~8 \cite{Sjostrand:2014zea}. It combines the
existing frameworks for photoproduction and hard diffraction, the latter 
originally introduced for purely hadronic collisions. The new model thus allows 
for event generation of photon-induced hard diffraction with different 
beam configurations. The model is highly dependent on the components of 
\textsc{Pythia}~8. The relevant ones -- the model for MPIs, 
photoproduction and hard diffraction -- are described in the following 
sections.\\

The first measurements of diffractive dijets was done by the UA8 
experiment at the S$\ppbar$S collider at CERN \cite{Bonino:1988ae}. 
Later on, similar events have been observed in $\e\p$ collisions at HERA
\cite{Derrick:1994ze}, in $\ppbar$ collisions at the Tevatron 
\cite{Affolder:2000vb}, and nowadays also in $\p\p$ collisions at 
the LHC \cite{Aad:2015xis}. Similarly, diffractively produced
$W^{\pm}$ and $Z^0$ bosons have been observed at the Tevatron 
\cite{Aaltonen:2010qe}. All of these processes are expected to be 
calculable within a perturbative framework, such as the 
Ingelman-Schlein picture \cite{Ingelman:1984ns}. 
A model for such \textit{hard diffractive} events was 
included in \textsc{Pythia}~8 \cite{Rasmussen:2015qgr}, based on
the Ingelman-Schlein approach and the rapidity gap survival idea
of Bjorken \cite{Bjorken:1992er}.
The model proposed an explanation of the observed \textit{factorization 
breaking} in hard diffractive $\p\pbar$ collisions -- the observation 
that with the Pomeron PDFs and fluxes derived from HERA DIS data, the 
factorization-based calculation was an order of magnitude above the 
measurement. The suppression factor required on top of the dPDF-based
calculation, was dynamically generated by requiring no additional MPIs 
in the $\p\p$ (or $\p\pbar$) system. The model predicted production rates in agreement with 
$\p\p$ and $\p\pbar$ measurements, albeit some differential distributions 
did show room for improvement when comparing to Tevatron data. The 
latest preliminary analysis on diffractive dijets by CMS 
\cite{CMS:2018udy} finds a very good agreement between the model 
and data in all differential distributions. \\

First evidence of factorization breaking for diffractive dijets 
in $\e\p$ collisions was observed by an H1 measurement 
\cite{Adloff:1998gg}, where a suppression factor of 0.6 was required to 
describe the dijet data in the photoproduction region, whereas the 
analysis for the DIS region was, by construction, well described by the 
factorization-based model without a corresponding suppression factor. 
Advances in the formulation of the dPDFs improved the description of 
data in the DIS regime, but the discrepancies remained in the 
photoproduction limit. Several analyses have been performed by H1 and 
ZEUS for diffractive dijet production \cite{Aktas:2007hn, 
Chekanov:2007rh, Chekanov:2009aa, Aaron:2010su, Andreev:2015cwa}, all 
requiring a suppression factor between $0.5-0.9$ in order for the 
factorization-based calculations to describe data.

The extension of the hard diffraction model in this article, to 
collisions with (intermediate) photons, makes it possible to explain the 
factorization-breaking in the photoproduction regime. 
The model is also applicable to the DIS regime, 
but here no further suppression is added since the highly virtual 
photons do not have any partonic structure that would give rise to 
the MPIs. Furthermore, the framework can also be applied to 
diff\-ractive photoproduction in purely hadronic collisions, 
usually referred to as ultra-peripheral collisions (UPCs) 
\cite{Baltz:2007kq}. The model predicts a substantial suppression 
for diffractive dijets in UPCs at the LHC.\\ 

The article is structured as follows: After the introduction in 
sec.~\ref{Sec:1}, we briefly describe in sec.~\ref{Sec:2} the 
event generation procedure in \textsc{Pythia}~8. We then proceed 
in sec.~\ref{Sec:3} to the photoproduction framework available 
in \textsc{Pythia}~8 and continue to a short description of the 
hard diffraction model in sec.~\ref{Sec:4}. We present results with our 
model compared to data from HERA on diffractive dijets in 
photoproduction in sec.~\ref{Sec:5}, and show some predictions for 
photoproduction in UPCs at the LHC in sec.~\ref{Sec:6}. We end with 
sec.~\ref{Sec:7} where we summarize our work and provide an outlook for 
further studies.

\section{\label{Sec:2}Event generation with \textsc{Pythia}~8}
Recently, \textsc{Pythia}~8 has undergone a drastic expansion. Where the 
earlier version, \textsc{Pythia}~6 \cite{Sjostrand:2006za}, was designed 
to accommodate several types of collisions (lepton-lepton, hadron-hadron 
and lepton-hadron, excluding nuclei), the rewrite to C++ focused mainly on 
the hadronic physics at the Tevatron and the LHC. While the LHC will run 
for years to come, there are several future collider projects under 
consideration. A common feature between the projected colliders is that 
they will be using lepton beams either primarily (linear $\e^+\e^-$ 
colliders: CLIC and ILC \cite{Linssen:2012hp, Baer:2013cma} or 
Electron-Ion Collider (EIC) \cite{Accardi:2012qut}), or as a first 
phase towards a hadronic collider (FCC 
\cite{Gomez-Ceballos:2013zzn, Mangano:2016jyj}). To enable studies 
related to these future colliders, \textsc{Pythia}~8 has been extended 
to handle many processes involving lepton beams.
Another major facility has been the extension from $\p\p$ to $\p\A$ and 
$\A\A$ collisions with the inclusion of the \textsc{Angantyr} model 
for heavy ion collisions \cite{Bierlich:2018xfw}. 
Combining the heavy-ion machinery with the recent 
developments related to lepton beams will also allow simulations of 
$\e\A$ collisions and ultra-peripheral $\A\A$ collisions. Work in 
this direction has been started within the \textsc{Pythia} collaboration. 

The \textsc{Pythia}~6 description of lepton-lepton and lepton-hadron 
collisions included a sophisticated model for merging of the DIS regime 
(high-virtuality photons) and the photoproduction regime 
(low-virtuality photons) \cite{Friberg:2000ra}. 
This, however, created upwards of 25 different event classes, each of 
which had to be set up differently. The model for the transition from 
photoproduction to DIS turned out not to agree so well with data, and 
the division of the different event classes was somewhat artificial.
The aim for the \textsc{Pythia}~8 implementation of these processes has 
been to reduce the number of hard-coded event classes and increase 
robustness. The present framework, however, does not yet include a 
smooth merging of the high- and low-virtuality events and therefore the 
events with intermediate virtualities are not addressed. Work towards  
such a combined framework is currently ongoing. In addition, there is 
progress towards improving the parton showers for DIS events 
(see e.g.\ \cite{Cabouat:2017rzi} and \cite{Hoche:2015sya}). 
In this paper we focus on the photoproduction regime, which is mature 
and well tested for hard-process events with virtuality 
$\lesssim 1~\GeV$ against LEP and HERA data \cite{Helenius:2017aqz, 
Helenius:2018bai, IHinProgress}.\\

The generation of \textit{non-diffractive} (ND) $\p\p$ or $\p\pbar$ 
events proceeds with the following steps.
First, the incoming beams are set up with (possible) PDFs at a given
(user-defined) energy. Then the hard scattering of interest is 
generated based on the matrix element (ME) of the process
and the PDFs. The generated partonic system is then 
evolved with a parton shower (PS), in \textsc{Pythia}~8 using the 
interleaved evolution of both initial and final state showers (ISR, FSR) 
\cite{Sjostrand:2004ef}~ and MPIs \cite{Sjostrand:2004pf}. The splitting probabilities for 
the FSR and ISR are obtained from the standard collinear DGLAP evolution 
equations. The ISR probabilities also depend on the PDFs of the
incoming beams, as the evolution is backwards from a high scale, set by 
the hard process, to a lower scale. Similarly, the MPI probabilities 
depend on the PDFs of the incoming beams, and these have to be adjusted
whenever an MPI has removed a parton from the beam. Colour reconnection 
(CR) is allowed after the evolution to mimic the finite-color effects 
that are not taken into account in the infinite-color PS.
After the partonic evolution, a minimal number of partons
are added as beam remnants in order to conserve color, flavor 
and the total momentum of the event. Lastly, the generated partons are 
hadronized using the Lund string model \cite{Andersson:1983ia} along 
with decays of unstable particles.

In $\e\p$ events, \textsc{Pythia}~8 operates with two regimes: the DIS
regime, where the electron emits a highly virtual photon ($Q^2\gg 1$
GeV$^2$), and the photoproduction regime, where the photon is
(quasi-)real ($Q^2\lesssim1$ GeV$^2$). Currently no description is
available for inter\-mediate-virtuality photons. In DIS events, the hard 
scattering occurs between the incoming lepton and a parton from the 
hadron beam by an exchange of a virtual photon (or another EW boson).
The photon can thus be considered devoid of any internal structure.
In the photoproduction regime, the photon flux can be factorized from 
the hard scattering, such that the intermediate photon can be regarded 
as a particle initiating the hard scattering. In this regime, both
point-like and hadron-like states of the photon occur. This
significantly increases the complexity of the event generation, thus the
photoproduction regime is thoroughly described in the next section.

\section{\label{Sec:3}The photoproduction framework}

The (quasi-)real photon contains a point-like, direct part without 
substructure as well as a hadron-like part with internal structure.
The latter part, the resolved photon, dominates the total cross 
section of the physical photon. The total
cross section is expected to contain all types of hadronic collisions,
including elastic (el), single- and double diffractive (SD, DD) and
inelastic ND collisions. The ND collisions 
contain both hard and soft events, where the former can be calculated
perturbatively, while the latter are modeled using the MPI framework 
in \textsc{Pythia}~8 \cite{Sjostrand:1987su}. Elastic and diffractive 
collisions require a phenomenological model for the hadronic photon.

The ND processes were first introduced in \\\textsc{Pythia}~8.215 
\cite{IHinProgress}, with a cross section given as a fraction of the 
total cross section, $\sigma_{\mrm{ND}}=f\sigma_{\mrm{tot}}$, 
$f<1$. The framework for photoproduction has since been expanded to 
include all soft QCD processes using the Schuler-Sj\"ostrand model 
\cite{Schuler:1993wr} in \textsc{Pythia}~8.235, and with this the cross 
sections for each of the event classes is calculated separately. The 
full description of these event classes is postponed to a forthcoming 
paper \cite{IHinProgress}, as we here concentrate on
diffractive processes with 
a hard scale. Between the two versions, the $\gamma\p$ and 
$\gamma\gamma$ frameworks were extended to $\e\p$ and $\e^+\e^-$ by 
the introduction of a photon flux within a lepton, now giving a 
complete description of all photoproduction events in $\gamma\p$, 
$\gamma\gamma$, $\e\p$ and $\e^+\e^-$ collisions 
in the latest release, 8.240. Furthermore, an 
option to provide an external photon flux has been included, allowing 
the user to study photoproduction also in 
UPCs, where the virtuality of the intermediate photon 
is always small and thus the photoproduction framework directly 
applicable. An internal setup for these cases is under way.\\

The resolved
photon is usually split into two: one describing a fluctuation of the 
photon into a low-mass meson and the other describing a fluctuation 
into a $\q\qbar$ pair of higher virtuality. The former is usually 
treated according to a vector-meson dominance (VMD) model 
\cite{Stodolsky:1966am, Joos:1967ony}, where the photon is a 
superposition of the lightest vector mesons (usually $\rho$, $\omega$ 
and $\phi$), whereas the latter, the anomalous part of the photon, is 
treated as ``the remainder'', $\sigma_{\mrm{anom}}=\sigma_{\mrm{tot}} - 
\sigma_{\mrm{direct}} - \sigma_{\mrm{VMD}}$. A generalization of the 
VMD exists (the GVMD model) which takes into account also higher-mass 
mesons with the same quantum numbers as photons \cite{Sakurai:1972wk}. 
Note, however, that if the resonances are broad and closely spaced, 
they would look like a smooth continuum.\\

The event generation for the direct photons begins by sampling the 
hard scattering between the incoming photon 
and a parton (or another direct photon in case of $\gamma\gamma$),
e.g.\ $\q\gamma\rightarrow\q \g$. The subsequent parton-shower 
generation always include FSR and in $\gamma\p$ case also ISR
for the hadronic beam. The whole photon momentum goes into the hard
process, $x_{\gamma}\sim 1$, as direct photons do not have any 
internal structure. Hence there is no energy left for MPIs and no 
photon remnant is left behind. The hadronization is then performed with 
the Lund string model as usual.

For resolved photons, a model for the partonic
content of the hadronic photon, the photon PDF, needs to be taken into
account. This PDF includes both the 
VMD and the anomalous contributions, the latter being 
calculable within perturbative QCD, the former requiring a 
non-perturbative input. As in the case of protons, the non-perturbative 
input is fixed in a global QCD analysis using experimental data. There 
are several PDF analyses available for photons \cite{Gluck:1991jc, 
Aurenche:1994in, Schuler:1995fk, Cornet:2002iy} using mainly data from 
LEP, but some also exploiting HERA data to constrain the gluonic part 
of the PDF \cite{Slominski:2005bw}. 
Ideally one would have a PDF for each of the VMD states, 
in practice one uses the same parametrization for all -- or 
approximates these with pion PDFs.

After the setup of the photon PDFs, the hard collision kinematics
has to be chosen. Here, a parton from the photon PDF
initiates the hard process, carrying a fraction of the photon momentum, 
$x_{i}<1$, with parton $i$ being extracted from the photon. Thus energy is 
still available in the fluctuation after the initial hard process, 
opening up for additional MPIs along with ISR and FSR in the subsequent 
evolution. As with other hadronic processes, a remnant is left 
behind, with its structure being derived from the flavor content of 
the original meson or $\q\qbar$ state and the kicked-out partons.\\

As in pp collisions, the PS splitting probabilites
with resolved photons are based on the DGLAP equations.
The DGLAP equation governing the scale evolution of resolved photon
PDFs can be written as \cite{DeWitt:1978wn}
\begin{align}
\frac{\partial f_{i/\gamma}(x_{i}, Q^2)}{\partial\log(Q^2)} =&
  \frac{\aem(Q^2)}{2\pi}e_i^2P_{i\gamma}(x_{i}) \nonu\\
  + \frac{\alpha_{\mrm{s}}(Q^2)}{2\pi}&\sum_j\int_{x_i}^1\frac{\d z}{z}
  P_{ij}(z)f_{j/\gamma}(\frac{x_{i}}{z}, Q^2)~,
\end{align}
where $f_{i(j)/\gamma}$ corresponds to the PDF of the photon,
$x_{i}$ the fractional momenta of the photon 
carried by the parton $i$, $\aem$, 
$\alpha_{\mrm{s}}$ the electromagnetic and strong couplings, 
$e_i$ the charge of parton $i$ and $P_{ij}$, $P_{i\gamma}$ the DGLAP and 
$\gamma\rightarrow\q\qbar$ splitting kernels, respectively. 
The term proportional to $P_{i\gamma}$ gives rise to the 
anomalous part of the photon PDF.
In \textsc{Pythia}~8 the separation into VMD and anomalous contributions 
is not explicitly performed. By the backwards evolution of ISR, however, 
a resolved parton can be traced back to the original photon by a 
$\gamma\rightarrow\q\qbar$ branching at some scale $Q^2$. Post facto, 
an event where this happens for $Q^2>Q_0^2$ can then be associated with 
an anomalous photon state, and where not with a VMD state. The dividing
scale $Q_0$ is arbitrary to some extent, but would be of the order of 
the $\rho^0$-meson mass. In the interleaved evolution of the parton showers
and MPIs, additional MPIs and ISR splittings on the photon side 
become impossible below the scale where the photon became unresolved.
This reduces the average number of MPIs for resolved
photons compared to hadrons, and therefore has an impact also for the
hard diffraction model as discussed in sec.~\ref{Sec:MPIphoton}.

\subsection{MPIs with photons}
\label{Sec:MPIphoton}

When the photon becomes resolved it is possible to have several 
partonic interactions in the same event. MPIs in \textsc{Pythia}~8 
are generated according to the leading-order (LO) QCD cross sections, 
albeit being regularized by introducing a screening parameter $\pTo$ 
\cite{Sjostrand:2004pf}, 
\begin{align}
\frac{\d\sigma}{\d\pT^2}\sim \frac{\alpha_{\mrm{s}}^2(\pT^2)}{\pT^4} 
  \rightarrow \frac{\alpha_{\mrm{s}}^2(\pTo^2+\pT^2)}{(\pTo^2+\pT^2)^2}~.
\end{align}
Note here that $\pTo$ can be related to the size $d$ of the colliding
objects, $\pTo\sim1/d$, thus a different value of the screening
parameter could be motivated if the photon has a different size than the
proton. Further, one could imagine working with different matter 
profiles for both the proton and the photon, and possibly also for each 
of the components of the photon. For now the shape is kept common for 
all systems, but possibly with different scale factors, i.e.\ average radii.

The screening parameter is allowed to vary with 
center-of-mass energy $\sqrt{s}$, 
\begin{align}
\pTo(\sqrt{s})=\pTo^{\mrm{ref}}
\left(\frac{\sqrt{s}}{\sqrt{s_{\mrm{ref}}}}\right)^p~,
\end{align}
with $\pTo^{\mrm{ref}}$, $p$ tunable parameters and 
$\sqrt{s_{\mrm{ref}}}$ a reference scale. Thus both the parameters 
from the matter profile and the parameters related to $\pTo$ require 
input from data. These parameters can be fixed by a global tune, with 
the \textsc{Monash} tune \cite{Skands:2014pea} being the current 
default. The MPI parameters in this tune, however, are derived using 
only data from $\pp$ and $\ppbar$ collisions. As the partonic structure 
and matter profile of resolved photons can be very different from that 
of protons, the values for the MPI parameters should be revisited for 
$\gamma\gamma$ and $\gamma\p$ collisions. The limitation is that 
there are only a few data sets sensitive to the MPIs available for 
these processes, and therefore it is not possible to perform a global 
retune for all the relevant parameters. Thus we have chosen to use the 
same form of the impact-parameter profile as for protons and study only 
the $\pTo$ parameters (which allow for different scale factors). 

For $\gamma\gamma$ collisions, LEP data is available for charged-hadron 
$\pT$ spectra in different $W_{\gamma\gamma}$ bins, allowing studies of 
the energy dependence of $\pTo$ as shown in \cite{Helenius:2017aqz}. In 
the $\gamma\p$ case the HERA data for charged-hadron production is 
averaged over a rather narrow $W_{\gamma\p}$ bin. Hence a similar study 
of the energy dependence is not possible for $\gamma\p$, and it becomes
necessary to assume the same energy dependence for $\pTo$ in $\gamma\p$ 
as for $\p\p$ collisions. The value of the $\pTo$-parameter, however, 
can be retuned with the available data. As discussed in 
\cite{Helenius:2018bai} a good description of the H1 data from HERA 
can be obtained with a slightly larger $\pTo^{\mrm{ref}}$ in $\gamma\p$
than what is used in the $\p\p$ tune, $\pTo^{\mrm{ref}}(\gamma\p) = 
3.00~\GeV$ versus $\pTo^{\mrm{ref}}(\p\p)=2.28~\GeV$. Thus the
photon-tune is consistent with a smaller size of the photon, i.e.\ that
the photon does not quite reach a typical hadron size during its 
fluctuation.

The rule of thumb is that a larger screening parameter gives less MPI 
activity in an event, thus a smaller probability for MPIs with resolved 
photons is expected compared to proton-proton collisions. As the model 
for hard diffraction is highly dependent on the MPI framework, we 
expect that the increased screening parameter gives less gap-suppression 
in photoproduction than what was found in the proton-proton study. This 
is simply because there is a larger probability for the event to have 
no additional MPIs when the $\pTo^{\mrm{ref}}$-value is larger. 
Furthermore, since the ISR splittings may collapse the resolved photon 
into an unresolved state and, by construction, the direct-photon induced 
processes do not give rise to additional interactions, the role of 
MPIs is suppressed for photoproduction compared to purely hadronic
collisions. Also, the invariant mass of the photon-proton system in
the photoproduction data from HERA is typically an order of magnitude 
smaller than that in previously considered (anti-)proton-proton data, 
which further reduces the probability for MPIs. Anticipating results 
to be shown below, this is in accordance with what is seen in 
diffractive dijet production at HERA, where the suppression 
factor is much smaller than that at the Tevatron.

\subsection{Photon flux in different beam configurations}

In the photoproduction regime one can factorize the flux of photons 
from the hard-process cross section. For lepton beams a 
virtuality-dependent flux is used,
\begin{equation}
f_{\gamma/\e}(x,Q^2) = \frac{\aem}{2 \pi} 
  \frac{1 + (1-x)^2}{x} \frac{1}{Q^2}~,
\label{eq:photonFluxEdiff}
\end{equation}
where $x$ is the momentum fraction of the photon w.r.t.\ the lepton.
Integration from the kinematically allowed minimum virtuality up to 
the maximum $Q^2_{\mrm{max}}$ allowed by the photoproduction framework, 
yields the well-known Weizs\"acker--Williams flux 
\cite{vonWeizsacker:1934nji, Williams:1934ad}
\begin{equation}
f_{\gamma/\e}(x) = \frac{\aem}{2 \pi} 
  \frac{1 + (1-x)^2}{x} \log 
  \left[ \frac{Q^2_{\mathrm{max}} (1-x)}{m_{\e}^2 x^2} \right]~,
\label{eq:photonFluxEint}
\end{equation}
where $m_{\e}$ is the mass of the lepton.

In $\p\p$ collisions the electric form factor arising from the finite 
size of the proton, or equivalently that the proton should not break up
by the photon emission recoil, needs to be taken into account. A good 
approximation of a $Q^2$-differential flux is given by
\begin{equation}
f_{\gamma/\p}(x,Q^2) = \frac{\aem}{2 \pi} 
  \frac{1 + (1-x)^2}{x} 
  \frac{1}{Q^2} \frac{1}{(1+Q^2/Q^2_0)^4}~,
\label{eq:photonFluxPdiff}
\end{equation}
where $Q^2_0 = 0.71~\GeV^2$. Integration over the virtuality provides 
the flux derived by Drees and Zeppenfeld \cite{Drees:1988pp}, 
\begin{align}
f_{\gamma/\p}(x) =& \frac{\aem}{2 \pi} 
  \frac{1 + (1-x)^2}{x} \nonu\\
  \times& \left[ \log(A) - \frac{11}{6} + \frac{3}{A} - \frac{3}{2A^2} + \frac{1}{3A^3} \right]~,
\end{align}
where $A = 1 + Q_0^2/Q_{\mrm{min}}^2$ and
$Q_{\mrm{min}}^2$ is the minimum scale limited by the kinematics of 
a photon emission.
Due to the form factor the photon flux drops
rapidly with increasing virtuality and becomes negligible already at
$Q^2\sim 2~\GeV^2$. This ensures that the photons from protons are well
within the photoproduction regime and there is no need to introduce any
cut on maximal photon virtuality.\\

\begin{figure}[ht!]
\centering
\includegraphics[width=0.7\linewidth]{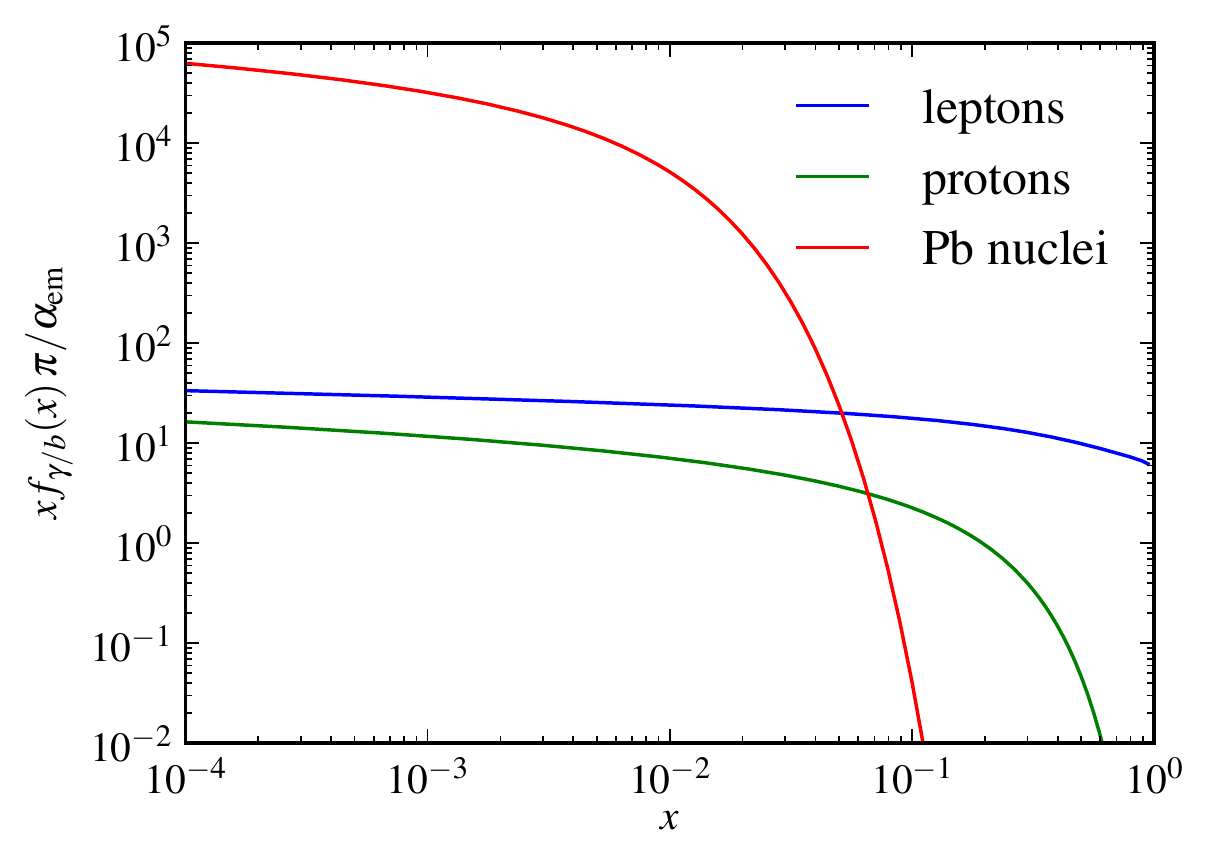}\\
\caption{\label{Fig:PhotonFlux}
The photon fluxes used for different beam types. Here $f_{\gamma/b}$
is the photon flux obtained from the beam $b$.
}
\end{figure}

In case of heavy ions it is more convenient to work in impact-parameter 
space. The size of a heavy nucleus is a better defined quantity than it
is for protons, so the impact parameter $b$ of the collision can be used 
to reject the events where additional hadronic interactions would 
overwhelm the electromagnetic interaction. Simply rejecting the events 
for which the minimal impact parameter, $b_{\mathrm{min}}$, is smaller 
than the sum of the radii of the colliding nuclei (or colliding 
hadron and nucleus for $\p\A$) provides a $b$-integrated flux,
\begin{align}
f_{\gamma/\A}(x) =& \frac{\aem Z^2}{\pi x} \times\nonu\\
  &\left[ 2 \xi K_1(\xi) K_0(\xi)  - \xi^2(K^2_1(\xi) - 
  K^2_0(\xi))\right]~,
\label{eq:photonFluxAint}
\end{align}
where $Z$ is the charge of the emitting nucleus, $\:K_i$ are the modified 
Bessel functions of the second kind and 
$\xi = b_{\mathrm{min}} \,x\, m_N$, where $x$ is a per-nucleon energy
fraction and $m_N$ a per-nucleon mass. The downside of working in 
the impact-parameter space is that the virtuality cannot be sampled 
according to the flux, as virtuality and impact parameter are conjugate 
variables. For heavy-ions, however, the maximal virtuality is very 
small (of the order of $60~\MeV$ \cite{Baltz:2007kq}), and can be safely 
neglected for the considered applications. The different photon 
fluxes are shown in Fig.~\ref{Fig:PhotonFlux}.\\

When extending the photoproduction regime from pure photon-induced
processes to 
collisions where the photon is emitted by a beam particle, some
additions are needed. In direct photoproduction, the partonic processes 
can be generated by using the photon flux directly in the factorized 
cross-section formula, similar to what is done with the PDFs in a 
usual hadronic collision. In resolved photoproduction, a PDF for the 
partons from the photons emitted from the beam particle is needed. This 
can be found by convoluting the photon flux from the beam particle $b$, 
$f_{\gamma/b}(x)$, with the photon PDFs, $f_{i/\gamma}(x_{\gamma},Q^2)$,
where $Q^2$ refers to the scale at which the resolved photon is probed. 
This scale can be linked to the scale of the hard(est) process, e.g.\ 
the $\pT$ of the leading jet in jet-production processes. The 
convolution yields
\begin{equation}
x_{i}f_{i/b}(x_{i},Q^2) = \int_{x_{i}}^1 
  \frac{\d x}{x} x f_{\gamma/b}(x) 
  \frac{x_i}{x}f_{i/\gamma}(\frac{x_i}{x},Q^2)~,
\label{eq:photonFluxConvol}
\end{equation}
with $x_{i}$ the energy fraction of beam particle momentum carried by 
parton $i$ and $x$ the energy fraction of the photon w.r.t.\ the beam. 
In practice the intermediate photon kinematics is sampled according to 
the appropriate flux during the event generation, thus taking care of 
the convolution on the fly. 

\section{\label{Sec:4}Hard diffraction in \textsc{Pythia}~8}

The \textsc{Pythia} model for hard diffractive events in $\p\p$ 
collisions was introduced as an explanation for the factorization 
breaking between diffractive DIS at HERA and the Tevatron 
\cite{Rasmussen:2015qgr}. The model can be applied to any process 
with sufficiently hard scales, including production of dijets, 
$Z^0,W^{\pm},H$ etc. It begins with the Ingelman-Schlein picture, where 
the diffractive cross section factorizes into a Pomeron-particle cross 
section and a Pomeron flux. Based on this ansatz a tentative 
probability for diffraction is defined as the ratio of diffractive PDF 
(dPDF) to inclusive PDF, as it is assumed that the proton PDF can be 
split into a diffractive and a non-diffractive part,
\begin{align}\label{Eq:PDFsel}
f_{i/\p}(x_i, Q^2) =& f_{i/\p}^{\mrm{ND}}(x_i, Q^2) + 
  f_{i/\p}^{\mrm{D}}(x_i, Q^2)\nonu~,\\
f_{i/\p}^{\mrm{D}}(x_i, Q^2) =& \int_{x_i}^1 
  \frac{\d x_{\Pom}}{x_{\Pom}}f_{\Pom/\p}(x_{\Pom})
  f_{i/\Pom}(\frac{x_i}{x_{\Pom}}, Q^2)\nonu~,\\
P^{\mrm{D}}_{A} =& 
  \frac{f_{i/B}^{\mrm{D}}(x_i, Q^2)}{f_{i/B}(x_i,Q^2)}\nonu~,\\
P^{\mrm{D}}_{B} =& 
  \frac{f_{i/A}^{\mrm{D}}(x_i, Q^2)}{f_{i/A}(x_i,Q^2)}~,
\end{align}
with $f_{i/\p}$ describing the PDF of the proton, $f_{i/\p}^{\mrm{D}}$ 
being the diffractive part of the proton PDF defined as a convolution of 
the Pomeron flux in a proton ($f_{\Pom/\p}$) and the Pomeron PDFs ($f_{i/\Pom}$). 
The probabilities for side $A,B$ to be the diffractive system are given 
as $P^{\mrm{D}}_{A,B}$ and each relies on the variables of the opposite 
side. 

This tentative probability is then used to classify an event as
preliminary diffractive or non-diffractive. If non-diffractive, 
the events are handled as usual non-diffractive ones. If diffractive,
the interleaved evolution of ISR, FSR and MPIs is applied, but only 
events surviving without additional MPIs are considered as fully 
diffractive events. The reasoning behind this is that additional MPIs 
in the $\p\p$ system would destroy the rapidity gap between the 
diffractive system and the elastically scattered proton. The gap
survives if no further MPIs occur, and the event can be experimentally 
quantified as being diffractive, with e.g.\ the large rapidity gap 
method. This no-MPI requirement suppresses the probability for 
diffraction with respect to the tentative dPDF-based probability, and 
can thus be seen as a gap-survival factor. Unlike other methods of 
gap survival (e.g.\ \cite{Bjorken:1992er, Khoze:2000wk, Gotsman:2005rt, 
Jones:2013pga}) this method is performed on an event-by-event basis, 
thus inherently is a dynamical effect. Furthermore, it does not 
include any new parameters, but relies solely on the existing and well 
tested (for $\p\p/\p\pbar$) MPI framework. Once the system is 
classified as diffractive, the full interleaved evolution is performed 
in the $\Pom\p$ subsystem. Here the model does not restrict the number 
of MPIs, as these will not destroy the rapidity gap between the 
scattered proton and the Pomeron remnant.\\

\subsection{Hard diffraction with photons}

In this article we extend the hard diffraction model to collisions 
involving one or two (intermediate) photons. The extension is 
straightforward. Changing the proton PDF in 
eqs.~(\ref{Eq:PDFsel}) to 
a photon PDF on one side, it is possible to describe hard diffraction 
in $\gamma\p$ interactions. Changing on both sides, the model is 
extended to $\gamma\gamma$ collisions. Thus 
eq.~(\ref{Eq:PDFsel}) is valid in events with (intermediate) photons 
with the change $\p\rightarrow\gamma$. Connecting the event generation with an 
appropriate photon flux allows to study hard diffraction in both 
$\e\p$ and $\e^+\e^-$ collisions as well as in ultra-peripheral 
collisions of protons and nuclei. The differential cross section of the
\textit{hard scattering} ($X_{\mrm{h}}$) in a diffractive system $X$, 
e.g.\ the dijet system within the diffractive system, for direct (dir) 
and resolved (res) photoproduction can then schematically be written as,
\begin{align}\label{Eq:XSconv}
\d\sigma^{AB\rightarrow X_\mrm{h}B}_{\mrm{dir}}=&f_{\gamma/A}(x)\otimes 
  f_{\Pom/p}(x_{\Pom},t)\otimes f_{j/\Pom}(x_j,Q^2) \nonu\\
  \otimes&\, \d\sigma^{\gamma j\rightarrow X_{\mrm{h}}} , \nonu\\
\d\sigma^{AB\rightarrow X_\mrm{h}B}_{\mrm{res}}=&f_{\gamma/A}(x)\otimes 
  f_{i/\gamma}(x_{\gamma}, Q^2)\otimes 
  f_{\Pom/B}(x_{\Pom},t)\nonu\\\otimes&\, f_{j/\Pom}(x_j,Q^2)\otimes  
  \d\sigma^{i j\rightarrow X_{\mrm{h}}},
\end{align}
with beam $A$ emitting a photon, beam $B$ emitting a Pomeron, 
and $AB\rightarrow X_{\mrm{h}}B$ denoting that the diffractive system is present
on side $A$. Changing $A\rightarrow B$ in eqs.~(\ref{Eq:XSconv}) thus
results in a diffractive system on side $B$. In the
above, $f_{\gamma/A}$ denotes the photon flux from beam $A$, 
$f_{i/\gamma}$ the photon PDF, while $f_{\Pom/B}$ and $f_{j/\Pom}$ are
the Pomeron flux and PDF, respectively. $\d\sigma^{\gamma(i) j\rightarrow
X_{\mrm{h}}}$ are the partonic cross sections calculated from the hard 
scattering MEs. The full diffractive system $X$ also
contains partons from MPIs and beam remnants that also have 
to be taken into account, thus eqs.~(\ref{Eq:XSconv}) only represent the 
hard subprocess part of the diffractive system. Presently, neither the 
double diffractive process $AB\rightarrow X_{\mrm{h}}^AX_{\mrm{h}}^B$ nor the central 
diffractive process $AB\rightarrow AX_{\mrm{h}}B$ are modelled, and the Pomeron 
can only be extracted from protons and resolved photons.
As the model is based on dPDFs and the dynamical gap 
survival derived from the MPI framework inside \textsc{Pythia}~8, the
extension does not require any further modelling or parameters.\\

The dynamical gap survival is present only in the cases where the 
photon fluctuates into a hadronic state. Hence the tentative 
probability, eqs.~(\ref{Eq:PDFsel}), equates the 
final probability for diffraction in direct photoproduction and in 
the DIS regime, where no MPIs occur. In resolved photoproduction, 
the dynamical gap survival suppresses the tentative probability for 
diffraction, offering an explanation for the discrepancies between 
next-to-leading order (NLO) predictions for dijets in 
photoproduction compared to measured quantities at HERA, 
see e.g.\ \cite{Adloff:1998gg, Chekanov:2007rh, Andreev:2015cwa}. The 
observed factorization breaking is not as striking as in $\pp$ 
collisions, but the factorization-based calculation still overshoots 
the latest H1 analysis by roughly a factor of two 
\cite{Andreev:2015cwa}.\\ 

\begin{figure}[!ht]
\begin{minipage}[c]{0.475\linewidth}
\centering
\includegraphics[width=0.9\linewidth]{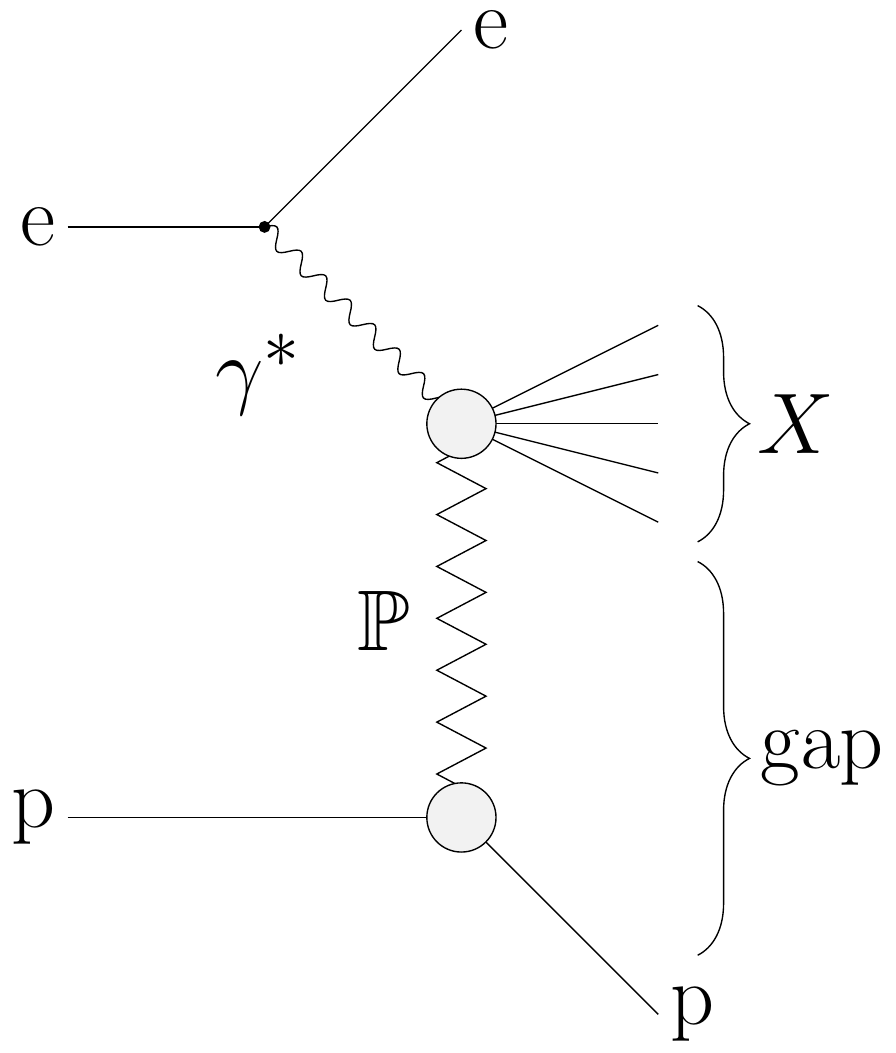}\\
(a)
\end{minipage}
\hfill
\begin{minipage}[c]{0.475\linewidth}
\centering
\includegraphics[width=0.9\linewidth]{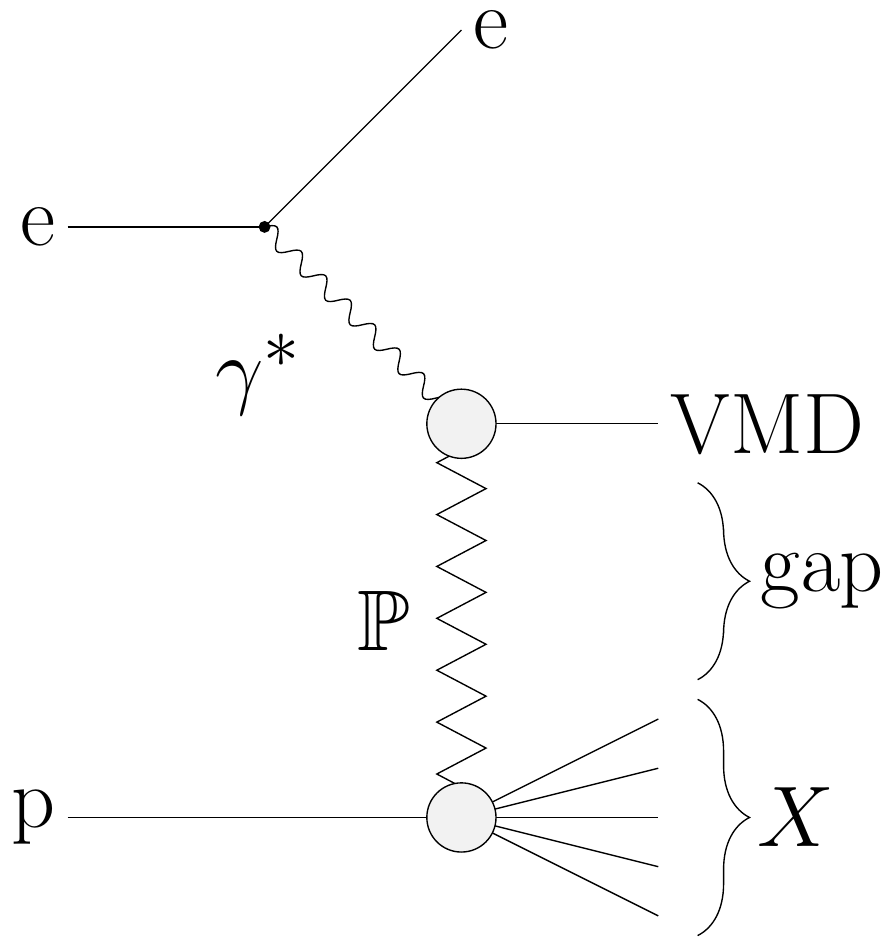}\\
(b)
\end{minipage}
\caption{\label{Fig:diff_sides} The two diffractive systems available
for resolved photoproduction: either the proton is elastically scattered
and the photon side contains the diffractive system (a), or the vector
meson is elastically scattered and the proton side contains the
diffractive system (b).}
\end{figure}

It should be noted that this extension allows for diffraction on both 
sides, i.e.\ the Pomeron can be extracted from the hadronic photon and/or 
the proton, see Fig.~\ref{Fig:diff_sides}. Typically, the experiments only 
considered diffractive events where the diffractive system consists of a 
photon and a Pomeron, with a rapidity gap on the proton side (and a 
surviving proton, whether observed or not). The option to generate 
diffractive events on only one of the sides exist in \textsc{Pythia}~8, 
such as to avoid needless event generation.

\subsection{Recent improvements in dPDFs}

Since the publication of the hard diffraction model for $\p\p/\p\pbar$, 
several improvements have been made for dPDFs. Work has been put into 
the inclusion of NLO corrections to the splitting kernels describing 
the evolution of the partons inside the Pomeron. Other work includes 
more recent fits to combined HERA data, or includes additional data 
samples into experiment-specific fits, so as to constrain some of the 
distributions in the dPDFs. A subset of these new dPDFs have been added 
to \textsc{Py\-thia}~8 recently and are briefly introduced below.

Specifically two new sets of dPDFs have been introduced, along with the 
Pomeron fluxes used in these fits. The first set, the GKG18 dPDFs by 
Goharipour et. al. \cite{Goharipour:2018yov}, consists of two LO and two 
NLO dPDFs fitted to two different combined HERA data sets available, 
using the \textsc{xFitter} tool \cite{Alekhin:2014irh} recently extended 
to dPDFs. 
In addition, we consider an analysis released by the ZEUS
collaboration offering three 
NLO dPDFs fitted to a larger sample of data. One of these, denoted
ZEUS SJ, includes also diffractive DIS dijets from 
\cite{Chekanov:2007aa} in order to have better constraints for the gluon 
dPDF \cite{Chekanov:2009aa}. Using PDFs derived at NLO is not perfectly 
consistent with the LO matrix elements available in \textsc{Pythia}~8, 
but since the ZEUS SJ dPDF analysis is the only of the considered 
dPDF analyses including dijet data\footnote{H1 has also performed a dPDF 
analysis with DIS dijets at NLO \cite{Aktas:2007bv} with very similar 
results as ZEUS SJ.}, it is interesting to compare the results to other 
dPDFs. 

\begin{figure*}[!ht]
\begin{minipage}[c]{0.475\linewidth}
\centering
\includegraphics[width=0.9\linewidth]{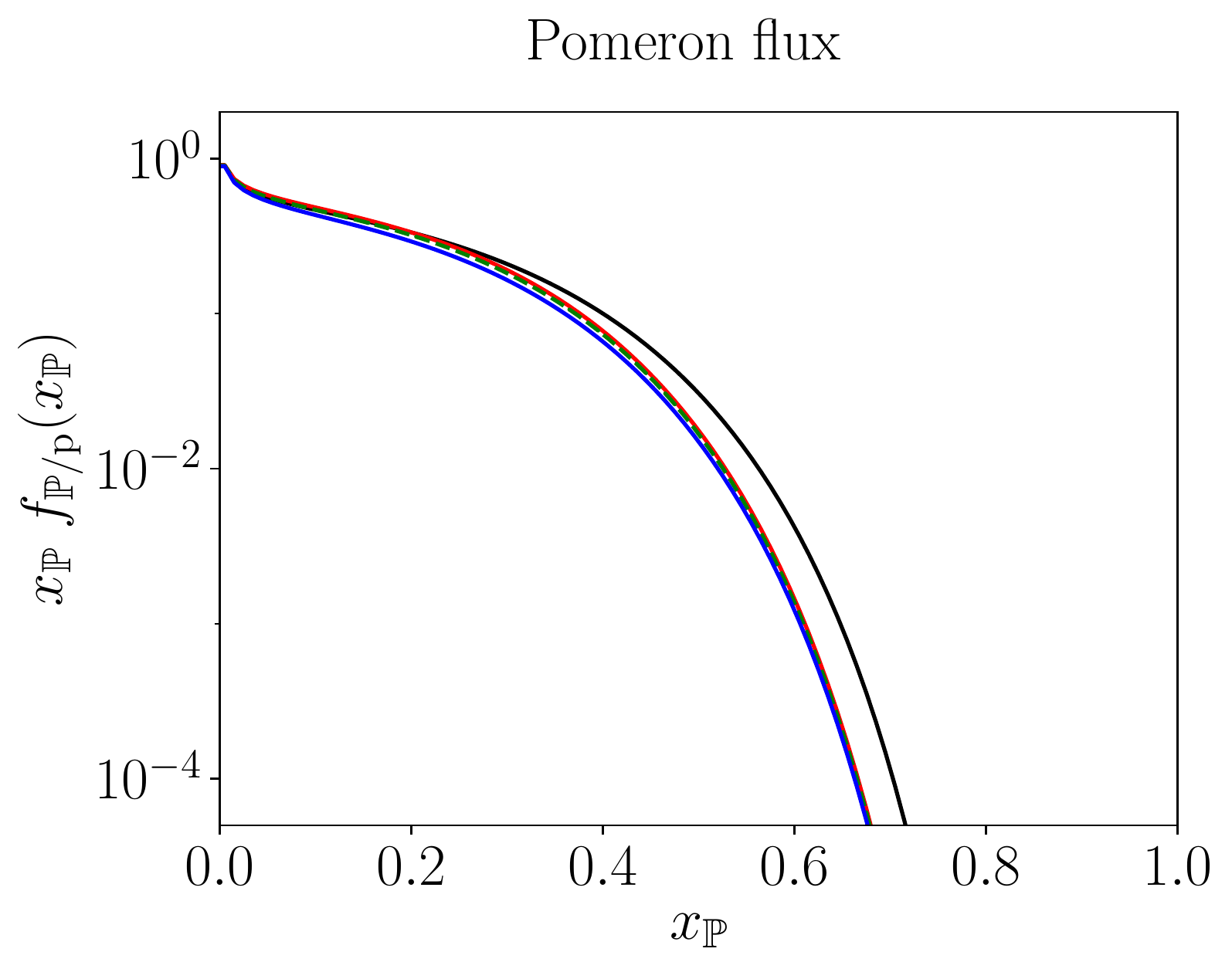}\\
(a)
\end{minipage}
\hfill
\begin{minipage}[c]{0.475\linewidth}
\centering
\includegraphics[width=0.9\linewidth]{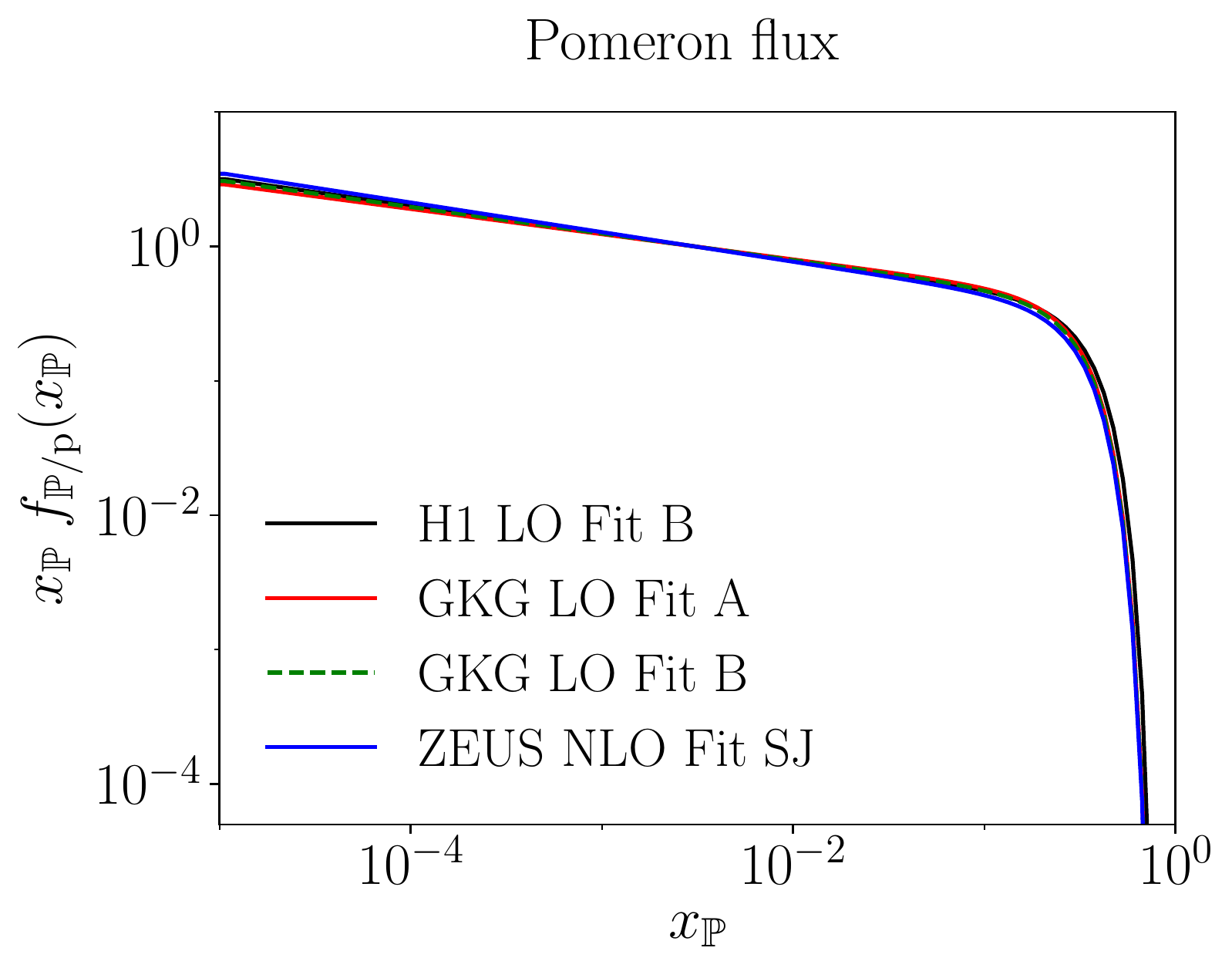}\\
(b)
\end{minipage}
\caption{\label{Fig:Fluxes}The GKG18 LO Fit A, B and ZEUS SJ fluxes on a linear 
(a) and logarithmic (b) scale in $x_{\Pom}$. Note that $t$ has been
integrated over its kinematical range, 
$f(x_{\Pom})=\int \d t f(x_{\Pom},t)$.}
\end{figure*}

\begin{figure*}[!ht]
\begin{minipage}[c]{0.475\linewidth}
\centering
\includegraphics[width=0.9\linewidth]{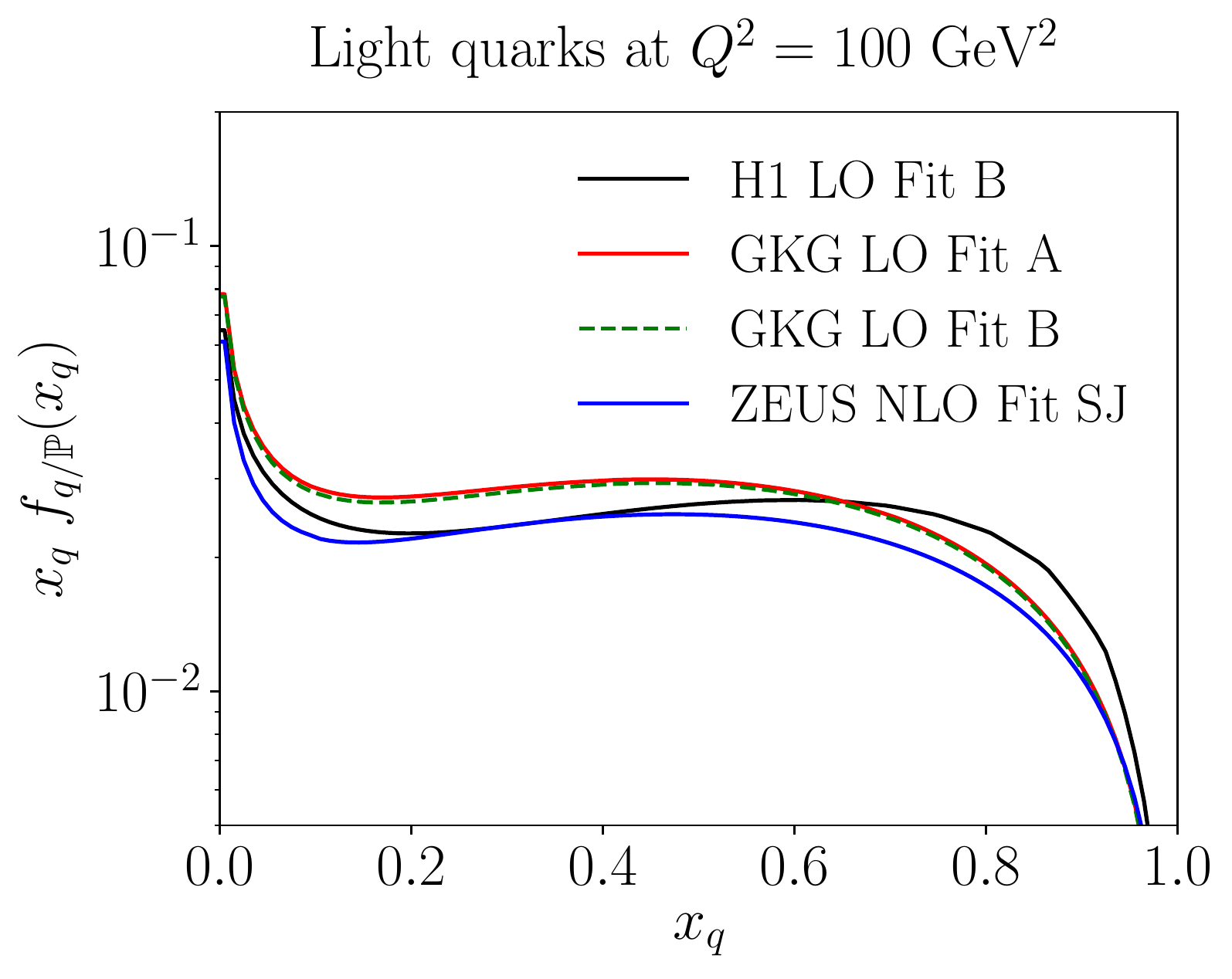}\\
(a)
\end{minipage}
\hfill
\begin{minipage}[c]{0.475\linewidth}
\centering
\includegraphics[width=0.9\linewidth]{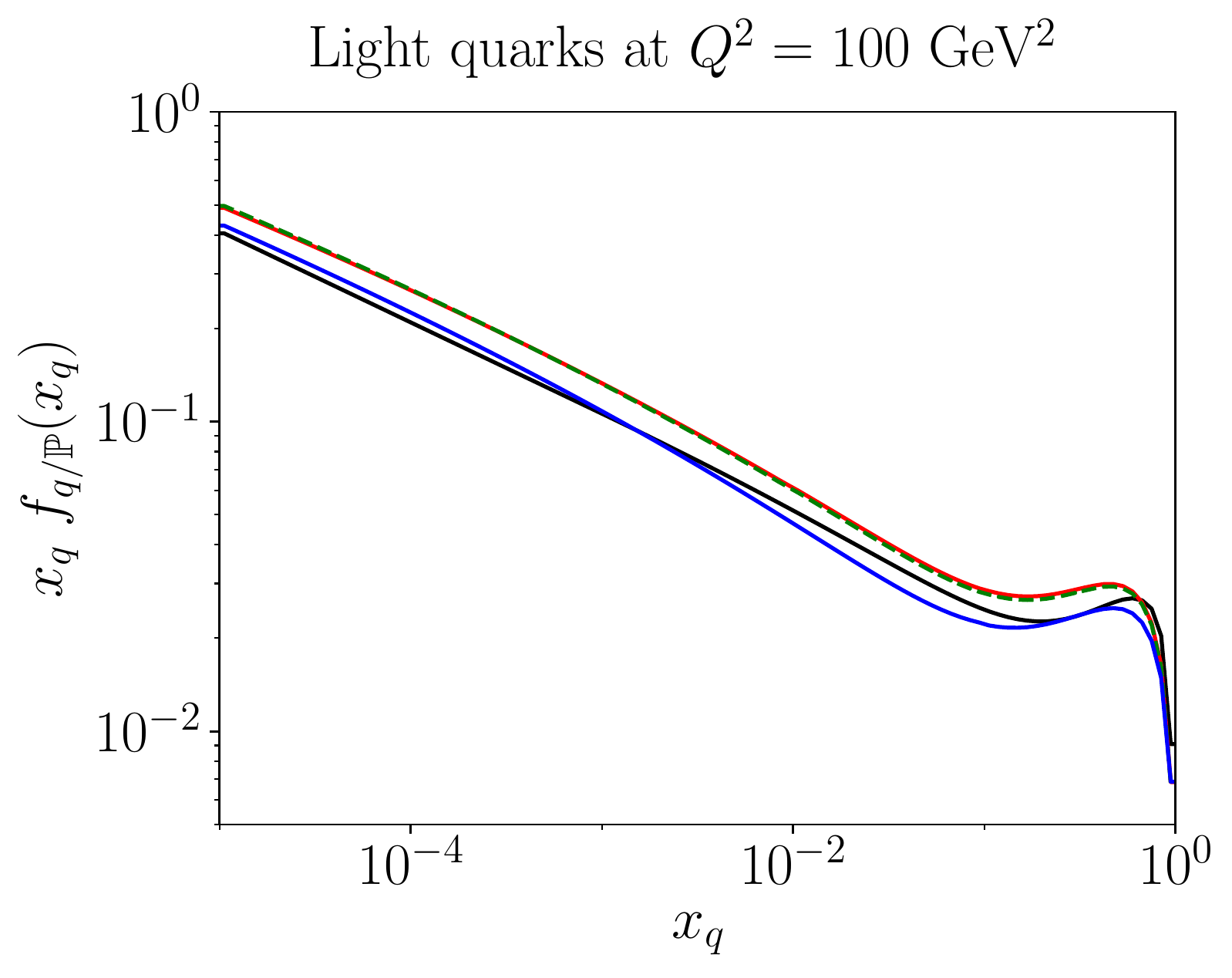}\\
(b)
\end{minipage}
\begin{minipage}[c]{0.475\linewidth}
\centering
\includegraphics[width=0.9\linewidth]{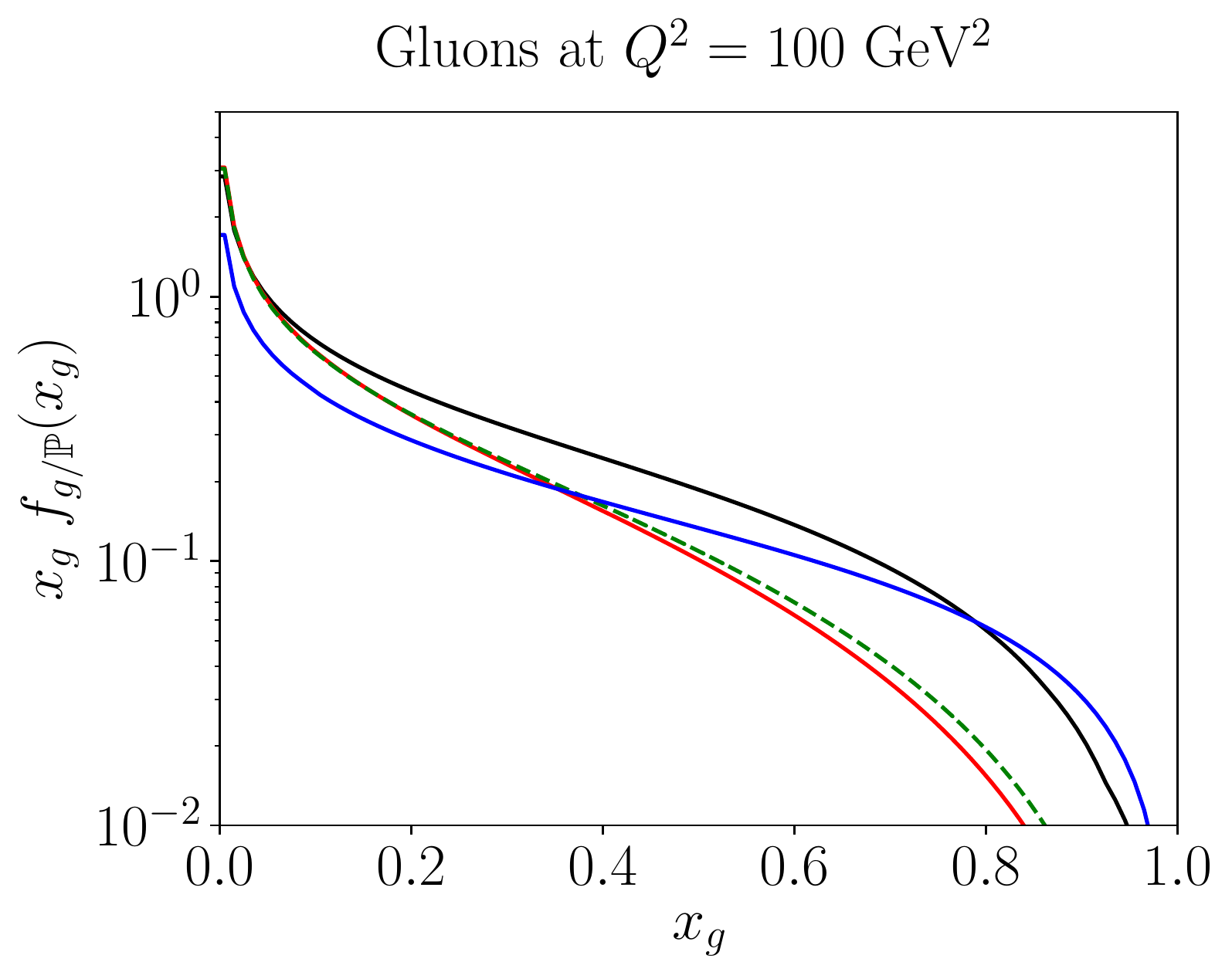}\\
(c)
\end{minipage}
\hfill
\begin{minipage}[c]{0.475\linewidth}
\centering
\includegraphics[width=0.9\linewidth]{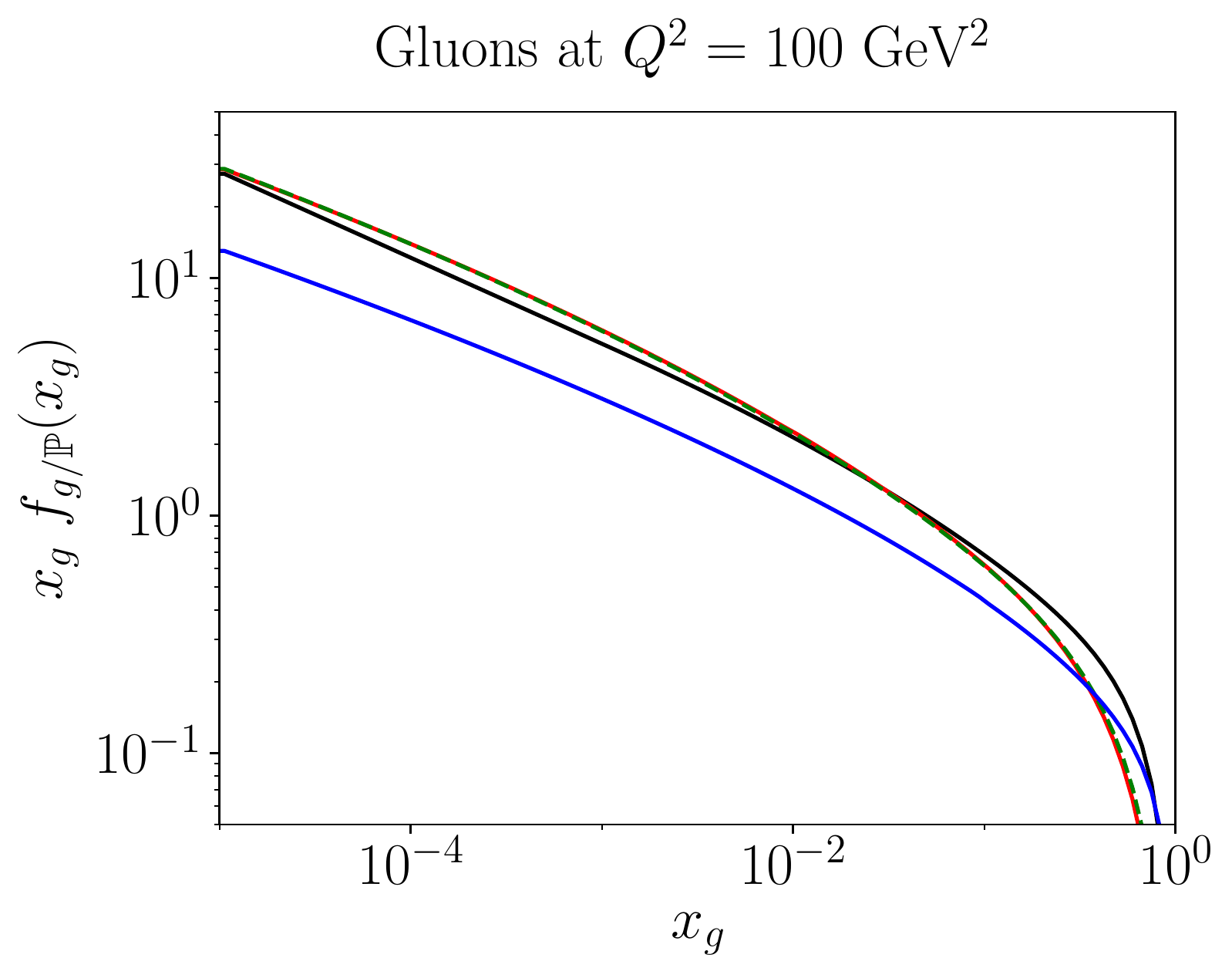}\\
(d)
\end{minipage}
\caption{\label{Fig:dPDFs}
The GKG18 LO Fit A, B and ZEUS SJ dPDFs on a linear (a,c) and 
logarithmic (b,d) scale. The upper figures shows the light quark 
content, the lower the gluonic content. 
}
\end{figure*}

Both the GKG18 and the ZEUS SJ fits uses the following 
parametrization for the Pomeron flux,
\begin{align}
f_{\Pom}(x_{\Pom}, t) =& A_{\Pom}\frac{\exp(B_{\Pom}
t)}{x_{\Pom}^{2\alpha_{\Pom}-1}}~,
\end{align} 
with $\alpha_{\Pom}=\alpha_{\Pom}(0) + \alpha_{\Pom}'t$
and $A,B$ being parameters to be included in the fits. The dPDFs are
typically parametrized as 
\begin{align}
zf_{i}(z, Q_0^2) =& A_iz^{B_i}(1-z)^{C_i}~,
\end{align}
again with $A_i,B_i,C_i$ being parameters to be determined in the fits. 
The dPDFs are then evolved using standard DGLAP evolution 
\cite{Lipatov:1974qm, Gribov:1972ri, Altarelli:1977zs, 
Dokshitzer:1977sg} to higher $Q^2$. Different schemes for the inclusion 
of heavy quarks were invoked in the two fits; see the original papers 
for details. In both dPDFs the light quarks ($\u$, $\d$, $\s$) have 
been assumed equal at the starting scale, while heavy quarks 
($\c$, $\b$) are generated 
dynamically above their mass thresholds. We show the new Pomeron PDFs and 
fluxes in figs. \ref{Fig:Fluxes} and \ref{Fig:dPDFs}, along with the H1 
Fit B LO PDF \cite{Aktas:2006hy} used as a default in \textsc{Pythia}~8. 
The GKG18 dPDFs are available with \textsc{Pythia}~8.240, while 
the ZEUS SJ set is expected in a forthcoming release.

\section{\label{Sec:5}Diffractive dijets in the photoproduction range}

The production of dijets in a diffractive system is particularly
interesting, as it provides valuable information on the validity of
factorization theorems widely used in particle physics. These
factorization theorems are not expected to hold in the case of
diffractive dijets arising from resolved photoproduction, as this
process essentially is a hadron-hadron collision, where the hard
scattering factorization fails.
 
Both H1 and ZEUS have measured the production of diffractive dijets 
in both the photoproduction and DIS range. We here limit ourselves to 
showing results from two analyses, the H1 2007 and ZEUS 2008 analyses 
on diffractive dijets \cite{Aktas:2007hn, Chekanov:2007rh}. Other 
analyses have been presented, including several ones examining only the 
DIS regime, but as the analysis codes or even the data itself have not 
always been preserved, we limit ourselves to reconstructing only a 
subset of these analyses. We aim to validate and provide the analyses 
used in this article within the Rivet framework \cite{Buckley:2010ar}. 

Both experiments have data on $\e\p$ collisions at $\sqrt{s}=318$ GeV
using $27.5$ GeV electrons and 920 GeV protons, with the proton moving 
in the $+z$ direction. Both use the large rapidity gap method for 
selecting diffractive systems. The experimental cuts in the two 
analyses are shown in table \ref{Tab:HERAcuts}. In the H1 analysis 
we concentrate on the differential cross sections as a function of 
four variables: invariant mass of the photon-proton system ($W$), 
transverse energy of the leading jet 
($E_{\perp}^{*\,\mathrm{jet}\,1}$) and momentum fractions 
$z_{\Pom}^{\mathrm{obs}}$ and $x_{\gamma}^{\mathrm{obs}}$, both 
constructed from the measured jets as
\begin{align}
x_{\gamma}^{\mathrm{obs}}=&\frac{\sum_{i=1}^2 
  (E^{\mathrm{jet},i} - p^{\mathrm{jet},i}_{z})}{2 y E_e} ~,
  \nonu\\
z_{\Pom}^{\mathrm{obs}}=&\frac{\sum_{i=1}^2 
  (E^{\mathrm{jet},i} + p^{\mathrm{jet},i}_{z})}{2 x_{\Pom} E_p} ~,
\label{eq:xGmzPom}
\end{align}
where $E_e$ ($E_p$) is the energy of the beam electron (proton) and the
summation includes the two leading jets, i.e.\ the two with highest 
$E_{\perp}$. The inelasticity $y$ and Pomeron momentum fraction w.r.t.\ 
the proton $x_{\Pom}$ are determined from the hadronic final state. In 
the ZEUS analysis the momentum fractions $z_{\Pom}^{\mathrm{obs}}$ and 
$x_{\gamma}^{\mathrm{obs}}$ are defined in terms of transverse energy 
and pseudorapidity of the jets, 
\begin{align}
x_{\gamma}^{\mathrm{obs}}=&\frac{\sum_{i=1}^2 
  E_{\perp}^{\mrm{jet},\,i}\exp(-\eta^{\mathrm{jet},i})}{2 y E_e}
  ~,\nonu\\
z_{\Pom}^{\mathrm{obs}}=&\frac{\sum_{i=1}^2 
  E_{\perp}^{\mrm{jet},\,i}\exp(\eta^{\mathrm{jet},i})}{2 x_{\Pom} E_p}~,
\label{eq:xGmzPomJets}
\end{align}
equivalent to the definitions in eq.~(\ref{eq:xGmzPom}), if the jets 
are massless. In a LO parton-level calculation these definitions would 
exactly correspond to the momentum fraction of partons inside a photon
($x_{\gamma}$) and Pomeron ($z_{\Pom}$). Due to the underlying event, 
parton-shower emissions and hadronization effects, however, the 
connection between the measured $z_{\Pom}^{\mathrm{obs}}$ and 
$x_{\gamma}^{\mathrm{obs}}$ and the actual $x_{\gamma}$ and $z_{\Pom}$ 
is slightly smeared, but still eqs.~(\ref{eq:xGmzPom}) and
(\ref{eq:xGmzPomJets}) serve as decent hadron-level estimates 
for the quantities. In place of $W$ the ZEUS analysis provides the 
differential cross section in terms of invariant mass of the 
photon-Pomeron system, $M_X$.\\

\begin{table}
\begin{center}
\caption{
Kinematical cuts used in the experimental analyses by H1 
\cite{Aktas:2007hn} and ZEUS \cite{Chekanov:2007rh}. An asterisk ($^*$)
indicates that the observable is evaluated in the photon-proton rest
frame. $x_{\Pom}, M_Y, t$ are found in the rest frame of the hadronic
system $X$, while the remaining are found in the laboratory frame.}
\label{Tab:HERAcuts}
\begin{tabular}{ll}
\hline\noalign{\smallskip}
H1 2007 & ZEUS 2008 \\
\noalign{\smallskip}\hline\noalign{\smallskip}
$Q^2< 0.01~\GeV^2$                 & $Q^2< 1~\GeV^2$                      \\
-                                  & $0.2 < y < 0.85$                     \\
$165~\GeV< W < 242~\GeV$           & -                                    \\
$N_{\mrm{jet}}\geq2$               & $N_{\mrm{jet}}\geq2$                 \\
$E_{\perp}^{*\,\mrm{jet\,1}} > 5.0~\GeV$   & $E_{\perp}^{\mrm{jet\,1}} > 7.5~\GeV$     \\
$E_{\perp}^{*\,\mrm{jet\,2}} > 4.0~\GeV$   & $E_{\perp}^{\mrm{jet\,2}} > 6.5~\GeV$     \\
$-1 < \eta^{\mrm{jet\,1,2}} < 2.0$ & $-1.5 < \eta^{\mrm{jet\,1,2}} < 1.5$ \\
$x_{\Pom} < 0.03$                  & $x_{\Pom} < 0.025$                   \\
$M_Y \leq 1.6~\GeV$                & -                                    \\
$|t| < 1.0~\GeV^2$                 & -                                    \\
\noalign{\smallskip}\hline
\end{tabular}
\end{center}
\end{table}

There are several theoretical uncertainties affecting the distributions 
of the diffractive events. Here we focus on the most important ones: 
\begin{itemize}
\item Renormalization- and factorization-scale variations, estimating 
      the uncertainties of the LO descriptions in \textsc{Pythia}~8.
\item dPDF variations affecting especially the 
      $z^{\mathrm{obs}}_{\Pom}$ distribution 
      and indirectly the number of events through the cuts on the
      squared momentum transfer, $t$, the momentum fraction of the beam
      carried by the Po\-meron, $x_{\Pom}$ and the mass of the scattered 
      (and possibly excited) proton, $M_Y$. 
\item $\pTo^{\mrm{ref}}$-variations, affecting the gap survival factor.
\end{itemize}

Other relevant parameters and distributions have also been varied, 
showing little or no effect on the end distributions. Remarkably, 
one of these was the choice of photon PDF. \textsc{Pythia}~8 uses 
the CJKL parametrization \cite{Cornet:2002iy} as a default both in 
the hard process and in the shower and remnant description. 
As the MPI and ISR generation in the current photoproduction framework 
require some further approximations for the PDFs, that are not 
universal and thus cannot be determined for an arbitrary PDF set, only 
the hard-process generation is affected by a change of photon PDF.
Thus the effect of a different photon PDF on the various
observables is not fully addressed with the present framework. The 
hard-process generation should, however, provide the leading photon 
PDF dependence. We find only a minimal change to the final 
distributions when changing to either the SaS \cite{Schuler:1995fk}, 
GRV \cite{Gluck:1991jc} or GS-G \cite{Gordon:1996pm} provided with 
LHAPDF5 \cite{Whalley:2005nh}. There are two reasons for the weak 
dependence on photon PDFs. Firstly, 
the cuts applied by the experimental analyses presented here forces 
$x_{\gamma}$ to be rather large, where the photon PDFs are relatively 
well constrained by the LEP data. Secondly, the no-MPI requirement 
rejects mainly events from the low-$x_{\gamma}$ region, where the 
differences between the mentioned photon PDFs are more pronounced. \\

Two other analyses from HERA \cite{Adloff:1998gg, Andreev:2015cwa} 
have also been used to check the 
current framework, giving results similar to the analyses presented 
here. For our baseline setup we show comparisons to both the H1 and ZEUS
analyses, while for the more detailed variations we focus on 
comparisons to ZEUS.

\subsection{Baseline results}

\begin{figure*}[!ht]
\begin{minipage}[c]{0.475\linewidth}
\centering
\includegraphics[width=0.9\linewidth]{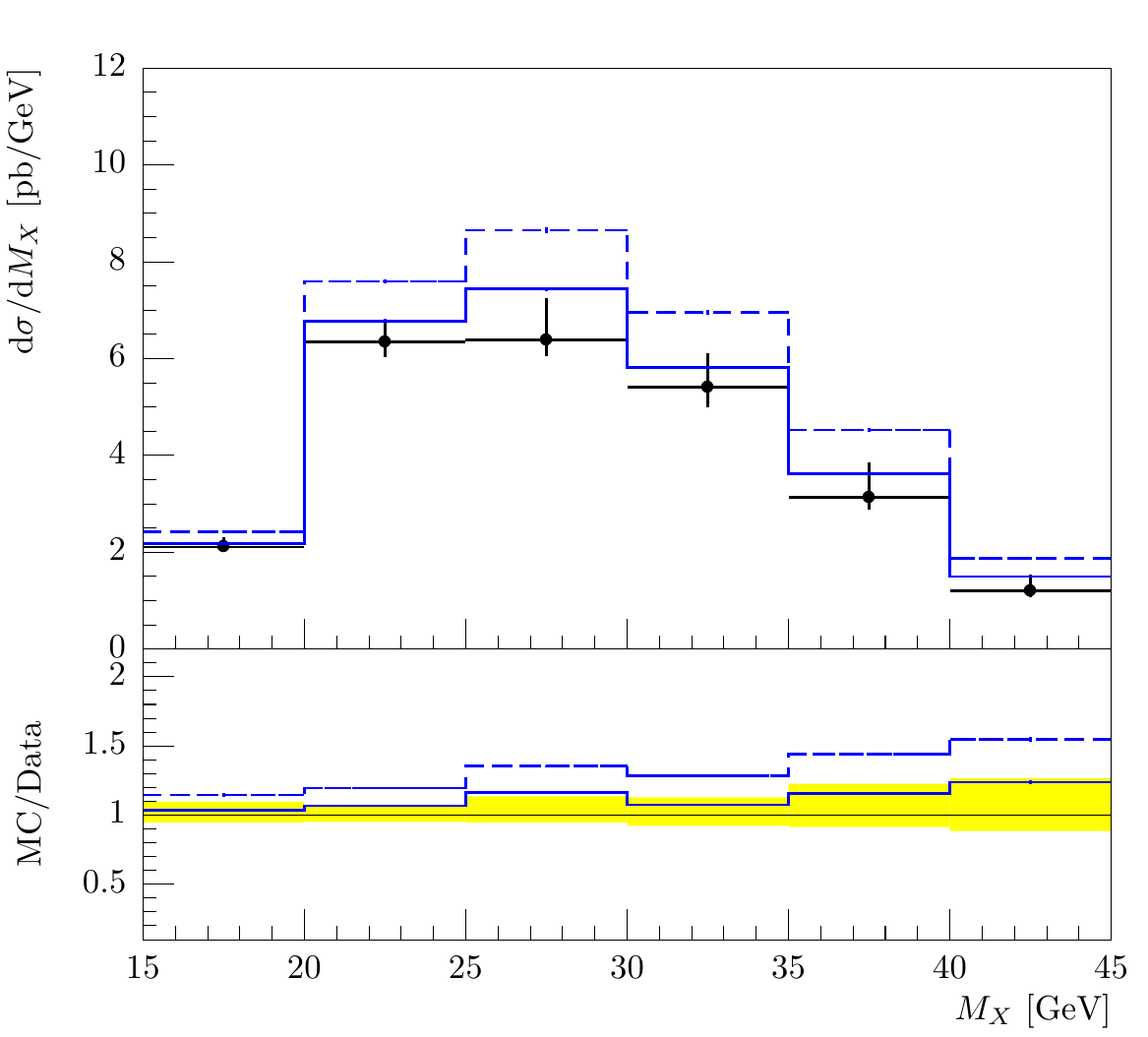}\\
(a)
\end{minipage}
\hfill
\begin{minipage}[c]{0.475\linewidth}
\centering
\includegraphics[width=0.9\linewidth]{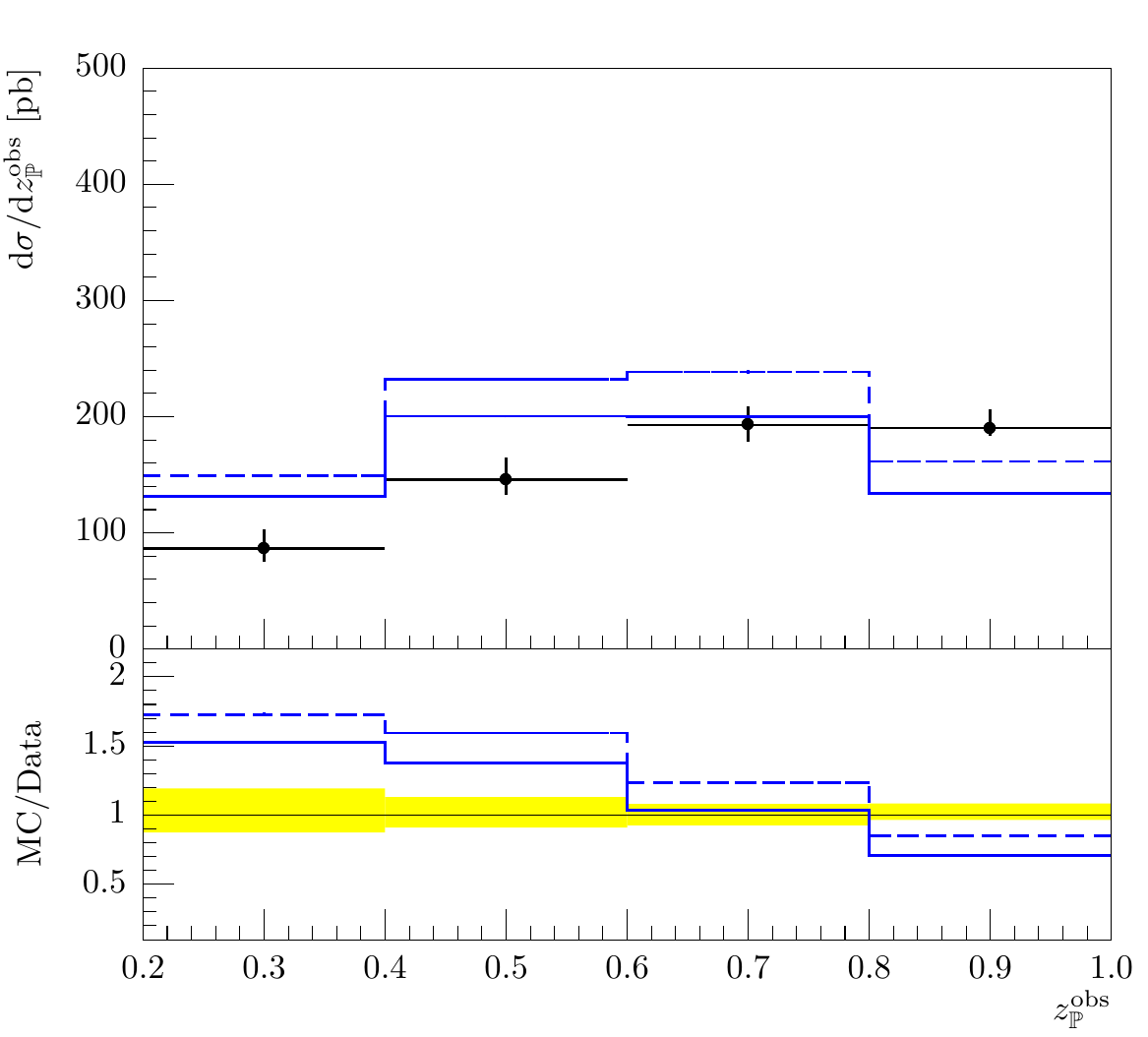}\\
(b)
\end{minipage}
\begin{minipage}[c]{0.475\linewidth}
\centering
\includegraphics[width=0.9\linewidth]{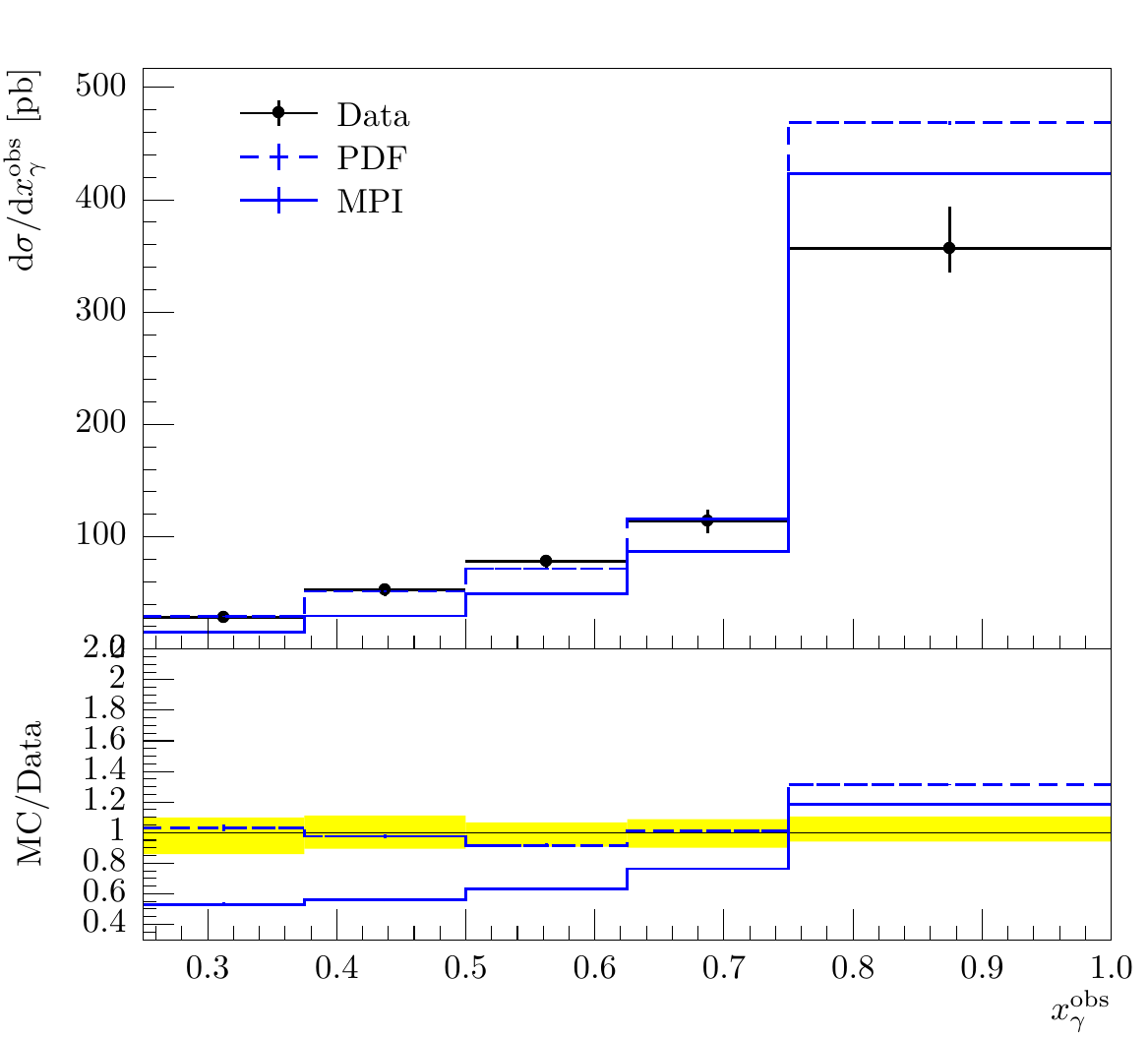}\\
(c)
\end{minipage}
\hfill
\begin{minipage}[c]{0.475\linewidth}
\centering
\includegraphics[width=0.9\linewidth]{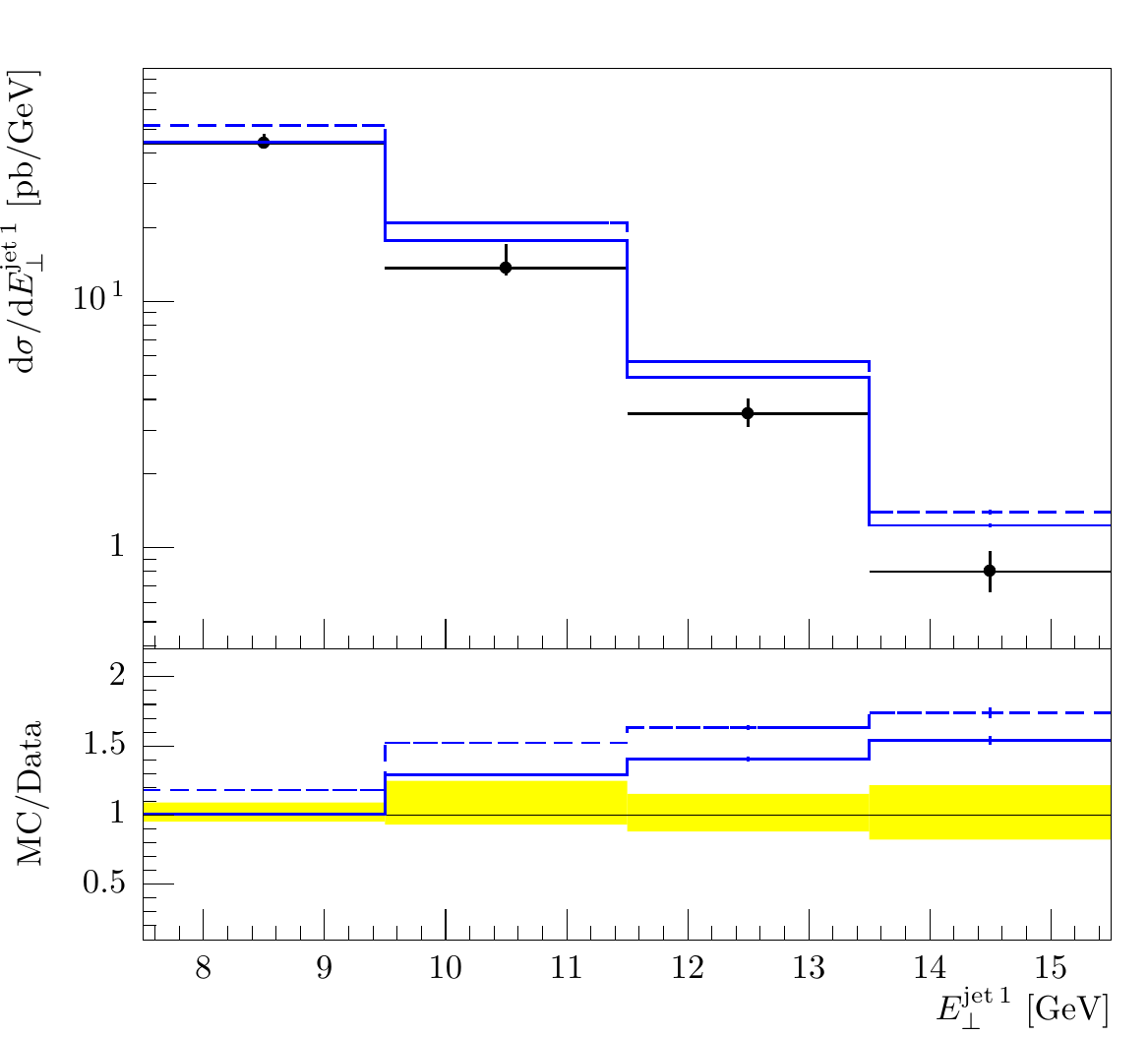}\\
(d)
\end{minipage}
\caption{\label{Fig:PDFandMPI_ZEUS}
The model with (solid lines) and without (dashed lines) gap survival 
compared to ZEUS data on $M_X$ (a), $z^{\mathrm{obs}}_{\Pom}$
(b), $x^{\mathrm{obs}}_{\gamma}$ (c) and 
$E_{\perp}^{\mrm{jet\,1}}$ (d).
}
\end{figure*}

\begin{figure*}[!ht]
\begin{minipage}[c]{0.475\linewidth}
\centering
\includegraphics[width=0.9\linewidth]{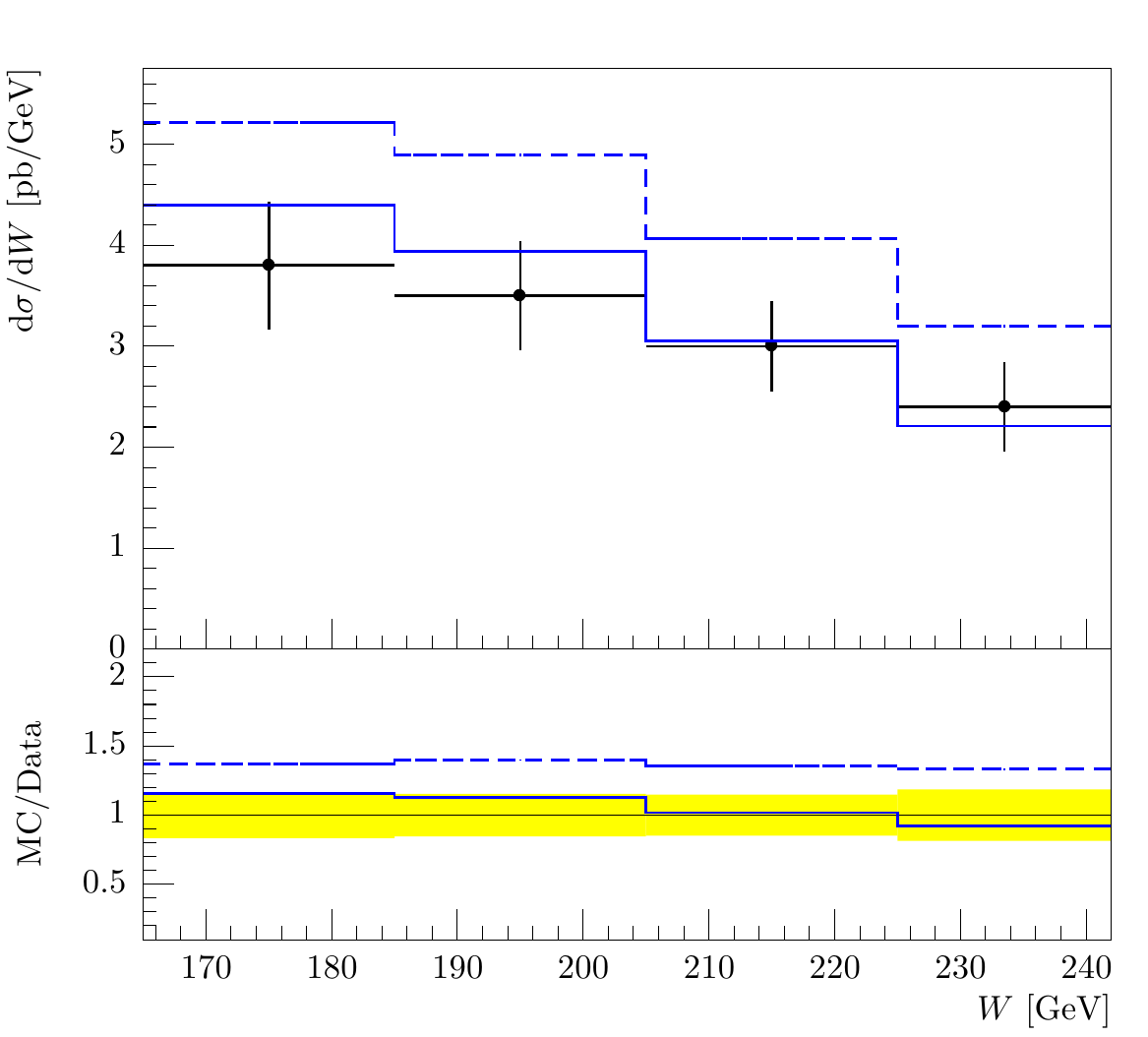}\\
(a)
\end{minipage}
\hfill
\begin{minipage}[c]{0.475\linewidth}
\centering
\includegraphics[width=0.9\linewidth]{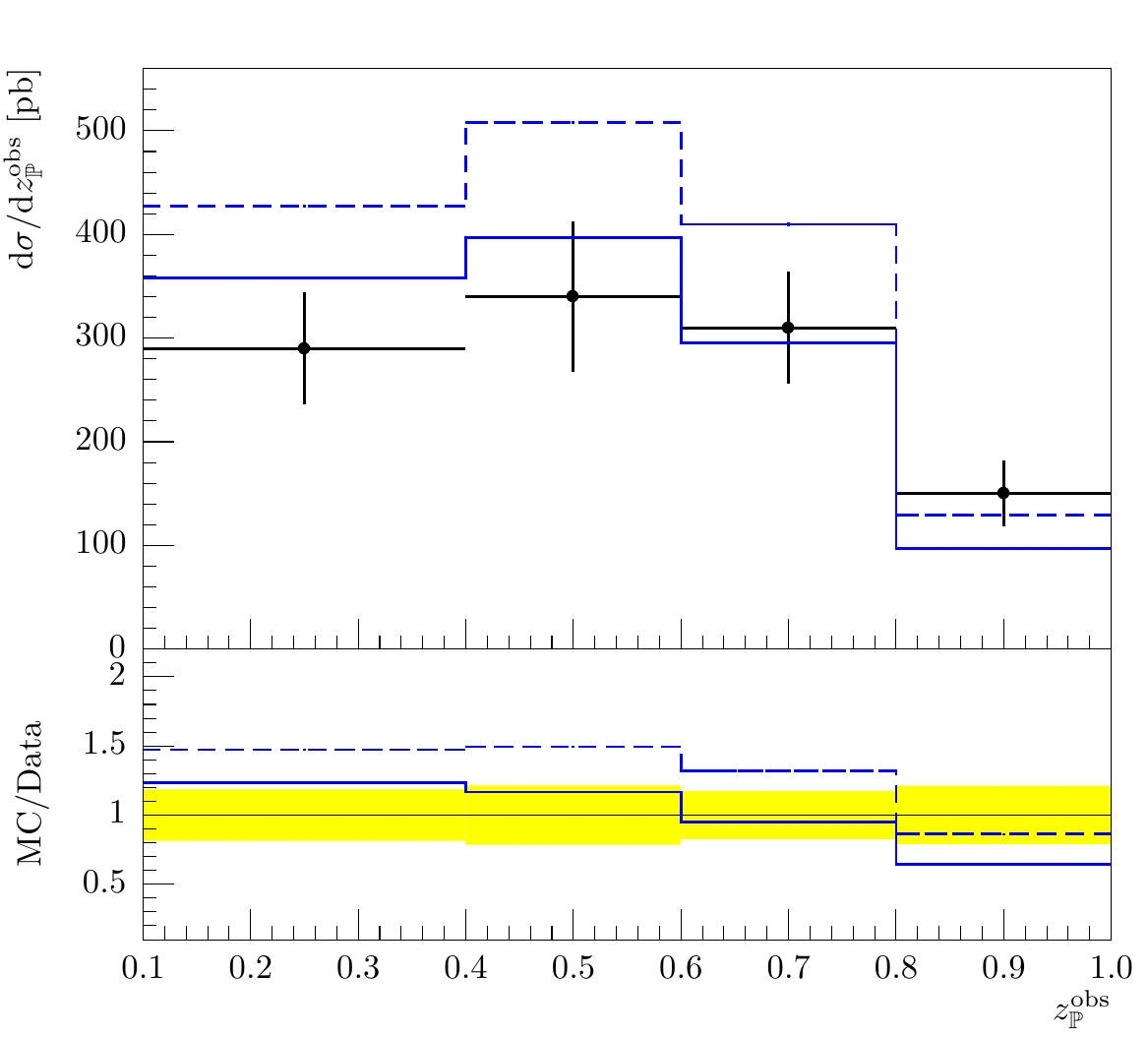}\\
(b)
\end{minipage}
\begin{minipage}[c]{0.475\linewidth}
\centering
\includegraphics[width=0.9\linewidth]{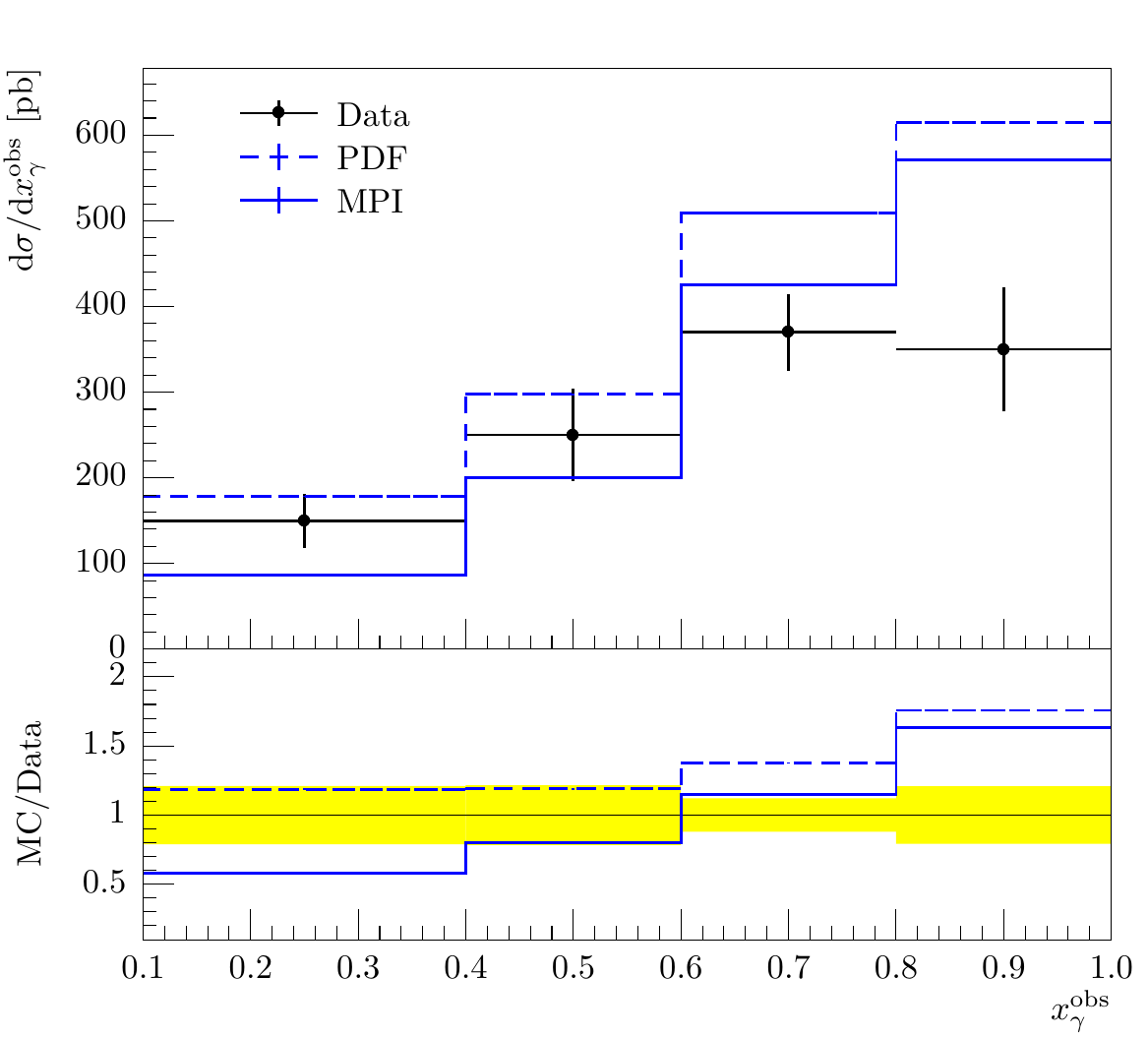}\\
(c)
\end{minipage}
\hfill
\begin{minipage}[c]{0.475\linewidth}
\centering
\includegraphics[width=0.9\linewidth]{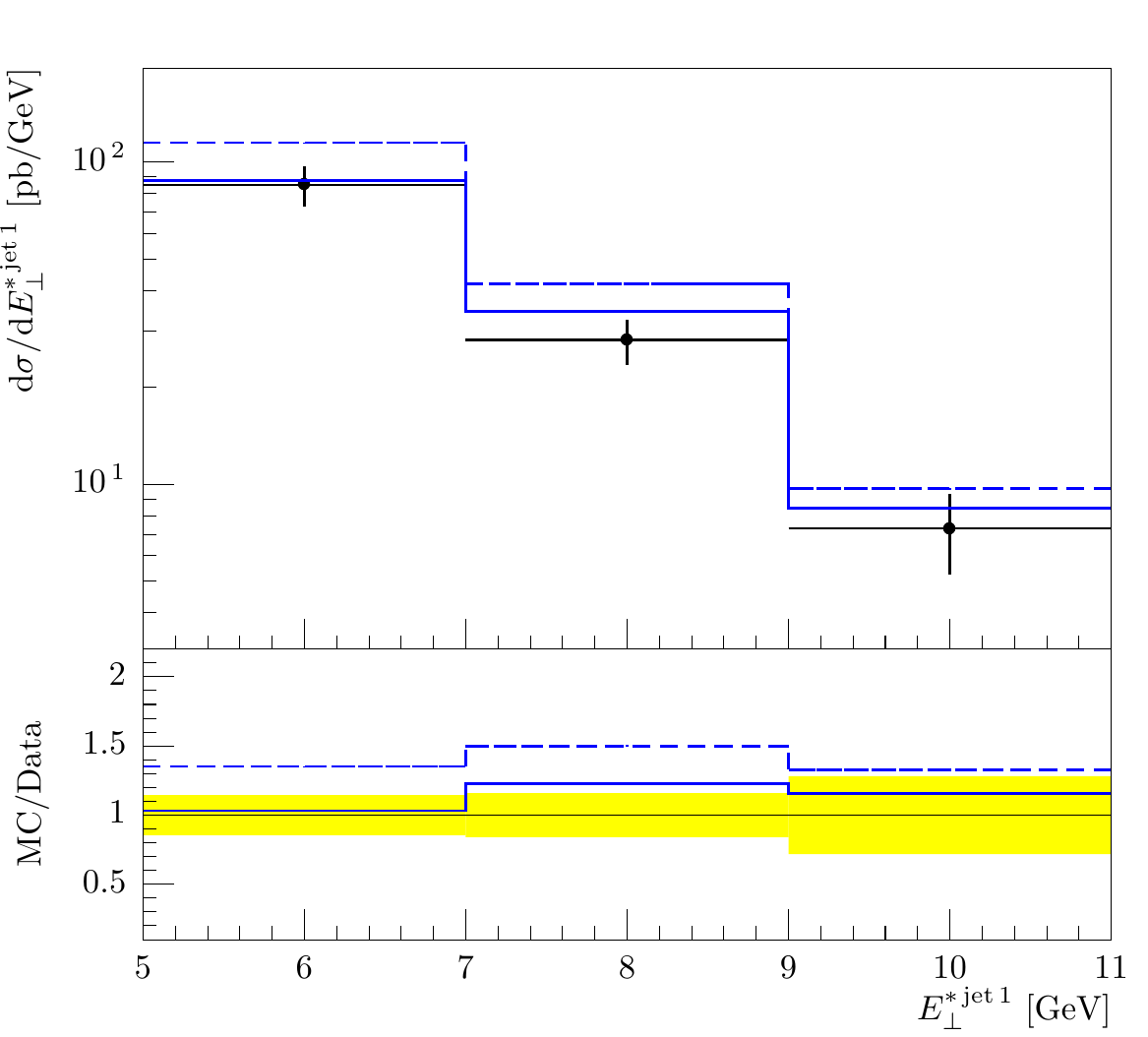}\\
(d)
\end{minipage}
\caption{\label{Fig:PDFandMPI_H1}
The model with (solid lines) and without (dashed lines) gap survival 
compared to H1 data on $W$ (a), $z^{\mathrm{obs}}_{\Pom}$
(b), $x^{\mathrm{obs}}_{\gamma}$ (c) and 
$E_{\perp}^{*\,\mrm{jet\,1}}$ (d).
}
\end{figure*}

In figs. \ref{Fig:PDFandMPI_ZEUS} and \ref{Fig:PDFandMPI_H1} we show 
the results obtained with \textsc{Pythia}~8 along with the experimental 
measurements. We show two simulated samples, one based on dPDFs solely
without the dynamic gap survival (the ``PDF'' sample, dashed lines), 
and one including the dynamic gap survival (the ``MPI'' sample, solid 
lines). The results show that the ``PDF'' sample 
is too large compared to data in all distributions
except for $x_{\gamma}$, thus showing evidence of 
factorization breaking. The ``MPI'' sample, however, seems to give a 
reasonably good description of data as
the ratio of MC/data is smaller for the ``MPI'' sample than the ``PDF''
sample, thus hinting that it is the additional 
probability for multiparton interactions between the photon remnant 
and the proton that causes the factorization breaking.

A $\chi^2$-test have been performed in order to quantify
which of the models do better. Here, we have performed three 
different tests; using only either of the H1 or ZEUS datasets, or using
both, table \ref{Tab:Chi2}. It is evident that the ``MPI'' model including the gap 
survival effect does a better job than the ``PDF'' model without it,
within our baseline setup. The calulcation of the 
$\chi^2$ values include all differential cross sections provided by the 
experimental analyses, excluding the additional 
$x^{\mathrm{obs}}_{\gamma}$-binned distributions in ZEUS analysis to avoid 
counting the same data twice. Error correlations are not provided and so
not considered. 

\begin{table}
\begin{center}
\caption{$\chi^2$ tests using three different datasets.}
\label{Tab:Chi2}
\begin{tabular}{llll}
\hline\noalign{\smallskip}
$\chi^2/n_{\mrm{DOF}}$ & H1 & ZEUS & Combined \\
\noalign{\smallskip}\hline\noalign{\smallskip}
PDF & 5.20 & 9.64 & 7.6\\
MPI & 1.42 & 5.10 & 3.44 \\
\noalign{\smallskip}\hline
\end{tabular}
\end{center}
\end{table}

In general, most distributions are well described by the model including
dynamical gap survival. The invariant mass distributions for the 
photon-Pomeron system ($M_X$) and for the photon-proton system ($W$) in 
figs.~\ref{Fig:PDFandMPI_ZEUS} and \ref{Fig:PDFandMPI_H1} (a) are both 
sensitive to the form of the photon flux from leptons. Both data sets 
are well compatible with the MPI samples, indicating that the standard 
Weizs\"{a}cker-Williams formula provide a good description of the flux.

It is, however, evident that in some observables the shape of the
data is poorly described. Examples are 
$z^{\mathrm{obs}}_{\Pom}$ and $x^{\mathrm{obs}}_{\gamma}$, 
figs.~\ref{Fig:PDFandMPI_ZEUS}, \ref{Fig:PDFandMPI_H1} (b, c). The
former is sensitive to the dPDFs used in the event generation. The 
baseline samples use the LO H1 Fit B flux and dPDF, fitted to data 
that is mainly sensitive to quarks. As the Pomeron is assumed to be 
primarily of gluonic content, it is expected that the vast majority of 
the dijets arise from gluon-induced processes. Thus a 
poorly-constrained gluon dPDF is expected to give discrepancies with 
distributions sensitive to this parameter, such as $z_{\Pom}$. In both
the H1 and ZEUS analyses $z^{\mathrm{obs}}_{\Pom}$ 
is overestimated in the low end, while being underestimated in the 
high-$z^{\mathrm{obs}}_{\Pom}$ end. If the measured 
jets are dominantly gluon-induced, then it is expected that changing 
from the H1 LO Fit B dPDF to the ZEUS SJ fit should improve on the
$z^{\mathrm{obs}}_{\Pom}$-distribution, as the 
low-$z^{\mathrm{obs}}_{\Pom}$ gluons are suppressed in
this dPDF.\\ 

\begin{figure*}[!ht]
\begin{minipage}[c]{0.475\linewidth}
\centering
\includegraphics[width=0.9\linewidth]{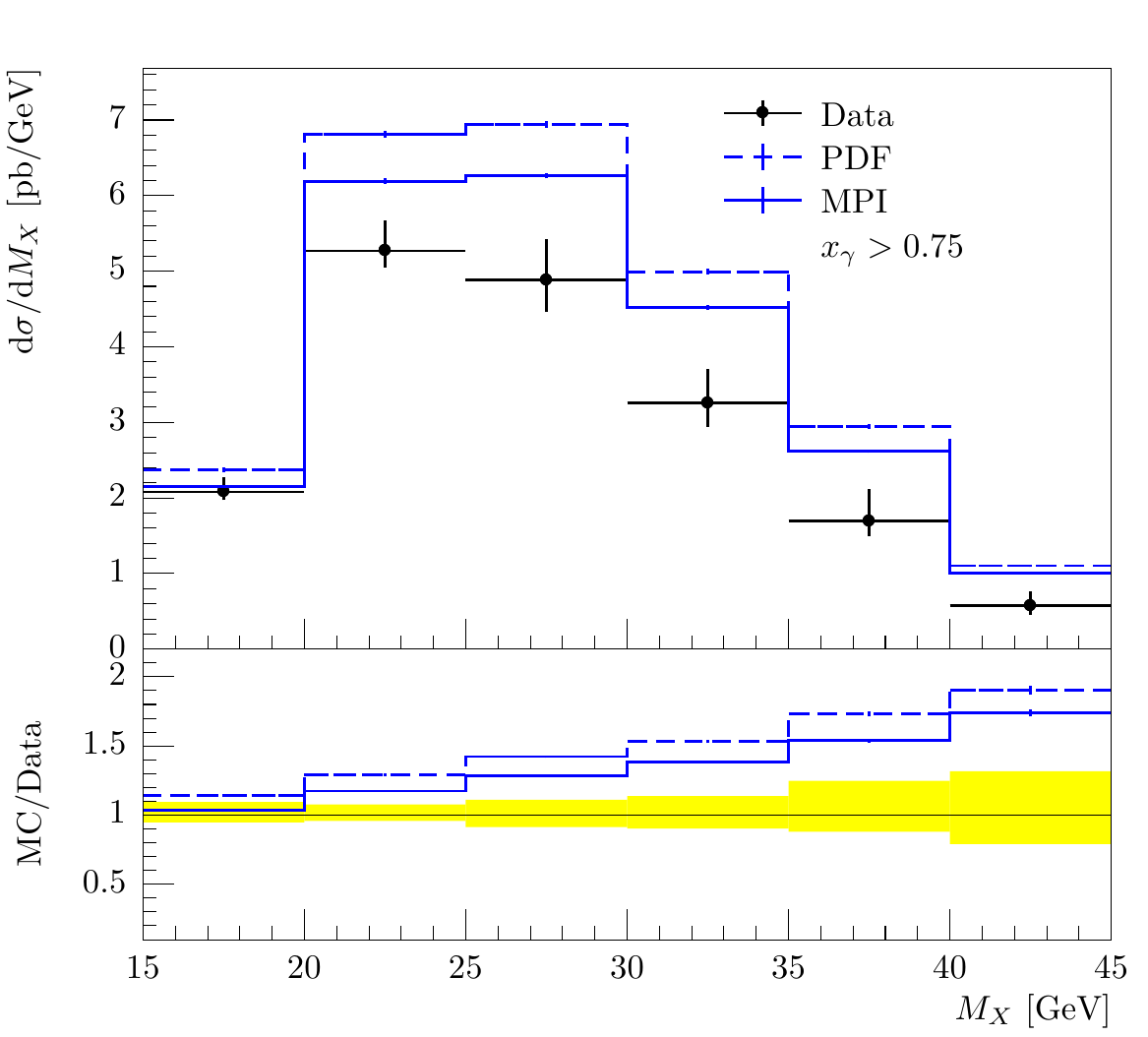}\\
(a)
\end{minipage}
\hfill
\begin{minipage}[c]{0.475\linewidth}
\centering
\includegraphics[width=0.9\linewidth]{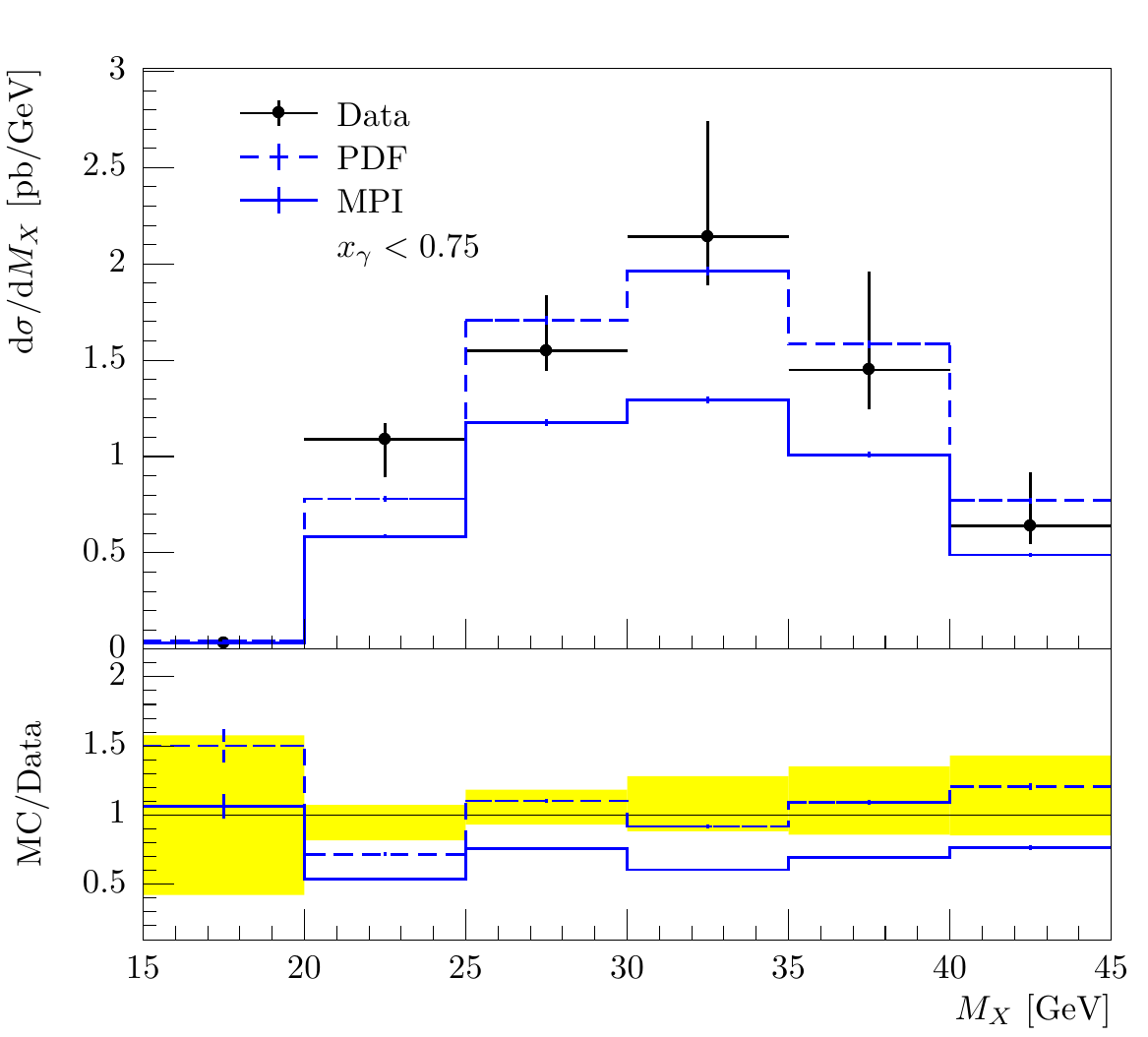}\\
(b)
\end{minipage}
\caption{\label{Fig:ZEUS_xgamma_regions}
The model with (solid lines) and without (dashed lines) gap survival 
compared to ZEUS data on $M_X$ in the direct-enhanced (a) and
resolved-enhanced (b) regions. 
}
\end{figure*}

The latter observable, $x^{\mathrm{obs}}_{\gamma}$,
is similarly underestimated in the 
low end and overestimated in the high end. The tight cut on $x_{\Pom}$ 
together with the requirement of high-$E_{\perp}$ jets reduces the 
contribution from lower values of $x^{\mathrm{obs}}_{\gamma}$. This 
suppresses the resolved contribution and therefore increases the 
relative contribution from direct processes, which typically are 
close to $x^{\mathrm{obs}}_{\gamma}=1$. The additional no-MPI 
requirement further suppresses the already low resolved contribution, 
and we end up with not being able to describe the shape of 
$x^{\mathrm{obs}}_{\gamma}$. As already discussed, the 
discrepancy cannot be explained with the uncertainties in the photon 
PDFs, as the sensitivity to different PDF analyses was found to be very 
low. The issue seems to be a problem with the 
relative normalizations of the direct and resolved contributions. This 
is evident from Fig.~\ref{Fig:ZEUS_xgamma_regions}, where the ZEUS 
analysis conveniently splits the data into two regions, a direct- and a 
resolved-enhanced region with the division at 
$x^{\mathrm{obs}}_{\gamma}=0.75$. Here, 
the model underestimates the resolved-enriched part of the 
cross section and overestimates the direct-enriched part, confirming 
what we already observed in figs.~\ref{Fig:PDFandMPI_ZEUS}, 
\ref{Fig:PDFandMPI_H1} (c). 
\\ 

Future measurements could shed more light on this issue, especially 
experimental setups in which the events passing the kinematical cuts would not be
dominated by the direct contribution. In the experimental analyses considered 
here, a similar observation was made when comparing to a NLO 
calculation: the shape of $x^{\mathrm{obs}}_{\gamma}$ 
was well described by the NLO 
calculation (corresponding to our PDF selection) in the direct-enhanced
region, but applying a constant suppression factor for the resolved 
contribution undershot the data at 
$x^{\mathrm{obs}}_{\gamma} < 0.75$, similar to what
we observe. It is worth pointing out that both poorly-described
distributions, $x^{\mathrm{obs}}_{\gamma}$ and 
$z^{\mathrm{obs}}_{\Pom}$, are constructed from the jet kinematics.
Therefore further studies on jet reconstruction and their $\eta$ 
distributions could offer some insights for the observed 
discrepancies.\\

The jet variable $E_{\perp}$ can be used to check if the amount of activity
within the diffractive system is properly described. As this system
contains a Pomeron, it might very well be that the MPI parameters here
could be different from the MPI parameters in the $\gamma\p$-system. It
seems that using the same parameters for the $\gamma\Pom$ system as for
$\gamma\p$ slightly overestimates the high-$E_{\perp}$ tail. This indicates
that there might be too much MPI activity in the events, thus requiring
a slightly larger $\pTo^{\mrm{ref}}$ value in the diffractive system
than in the $\gamma\p$ system. The argument for a different
$\pTo^{\mrm{ref}}$-value for $\gamma\p$ as compared to $\p\p$ can also
be applied here: if the Pomeron has a smaller size than the proton, then
the $\pTo^{\mrm{ref}}$-value can be increased. Having too much MPI 
activity in the $\gamma\Pom$-system may also push the 
$x^{\mathrm{obs}}_{\gamma}$ 
distribution towards higher values, as the $E_{\perp}$ of the 
jets may increase due to the underlying event. A full discussion of 
the MPI parameters in the diffractive system in $\p\p$ collisions has 
been provided in \cite{Rasmussen:2015qgr}, but have not been pursued 
further here. 

\subsection{Scale variations}

\begin{figure*}[!ht]
\begin{minipage}[c]{0.475\linewidth}
\centering
\includegraphics[width=0.9\linewidth]{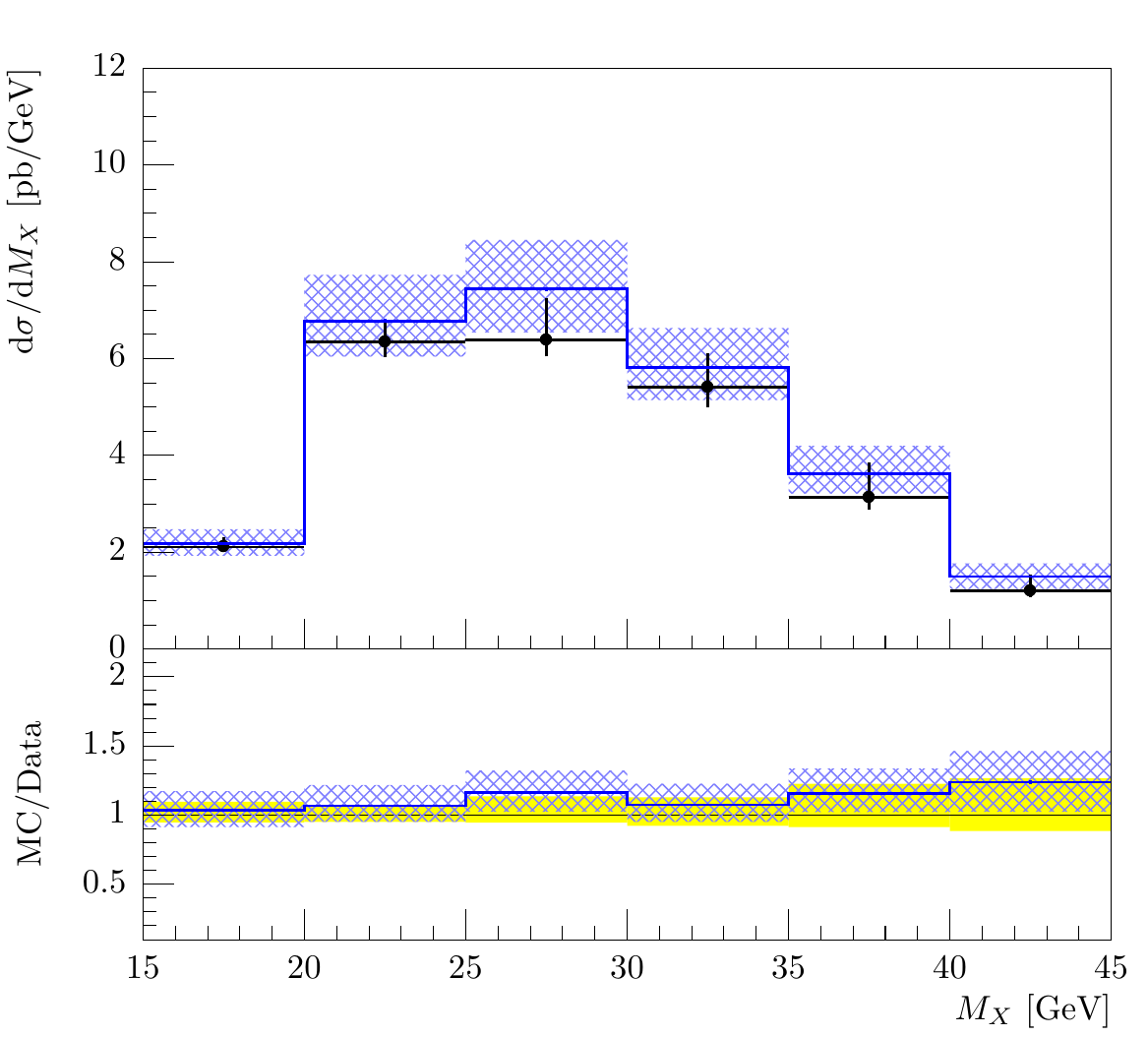}\\
(a)
\end{minipage}
\hfill
\begin{minipage}[c]{0.475\linewidth}
\centering
\includegraphics[width=0.9\linewidth]{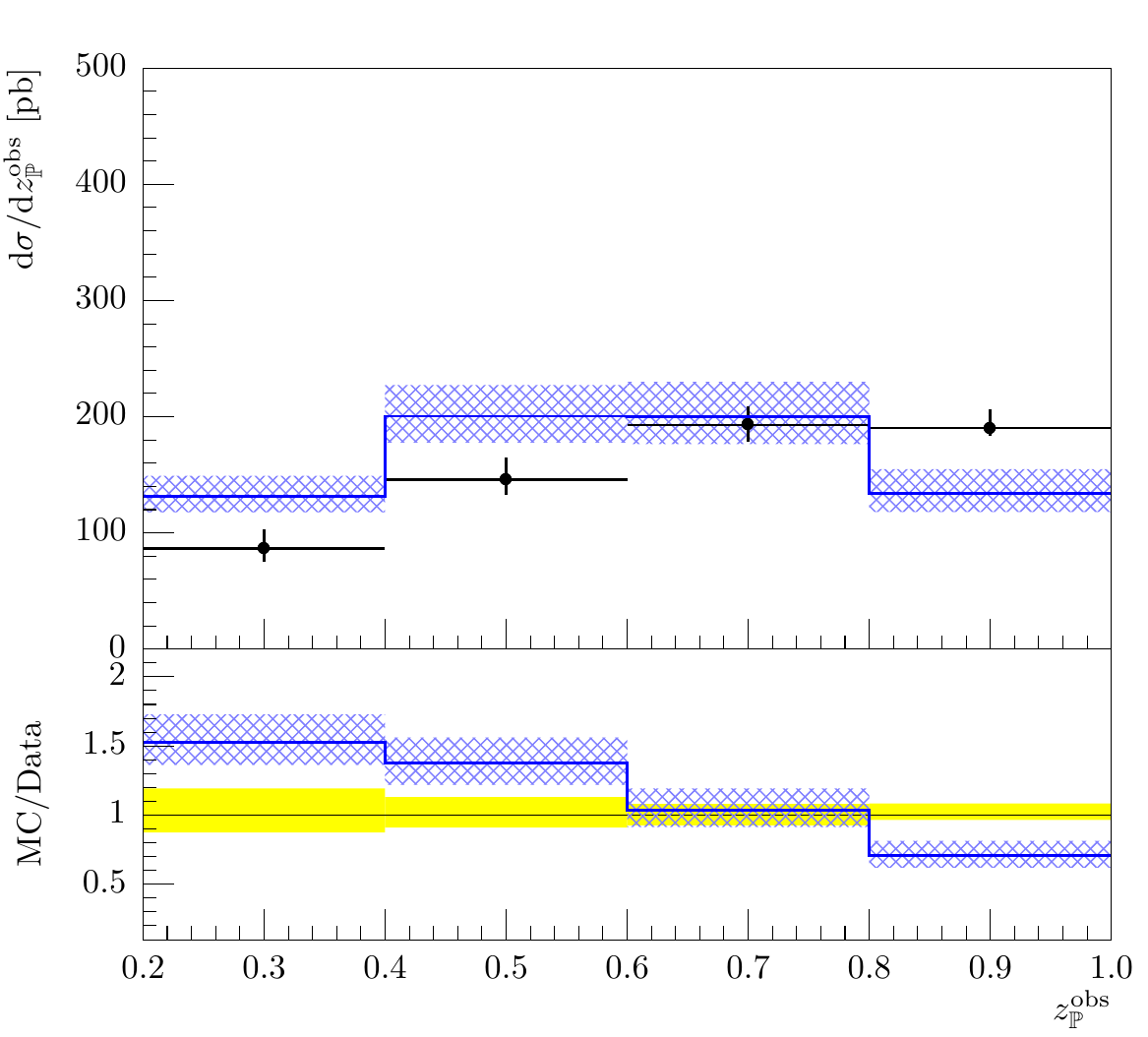}\\
(b)
\end{minipage}
\begin{minipage}[c]{0.475\linewidth}
\centering
\includegraphics[width=0.9\linewidth]{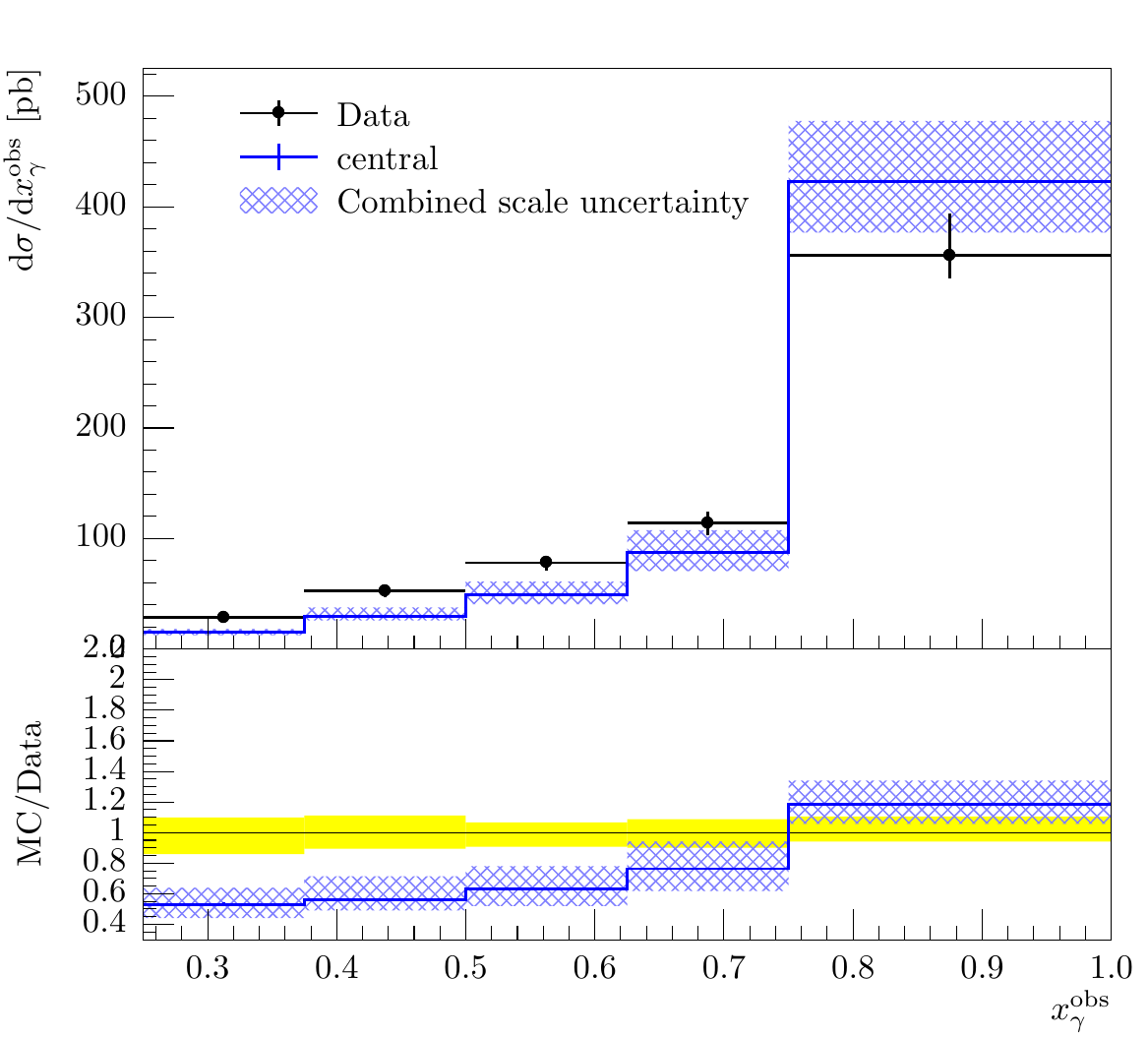}\\
(c)
\end{minipage}
\hfill
\begin{minipage}[c]{0.475\linewidth}
\centering
\includegraphics[width=0.9\linewidth]{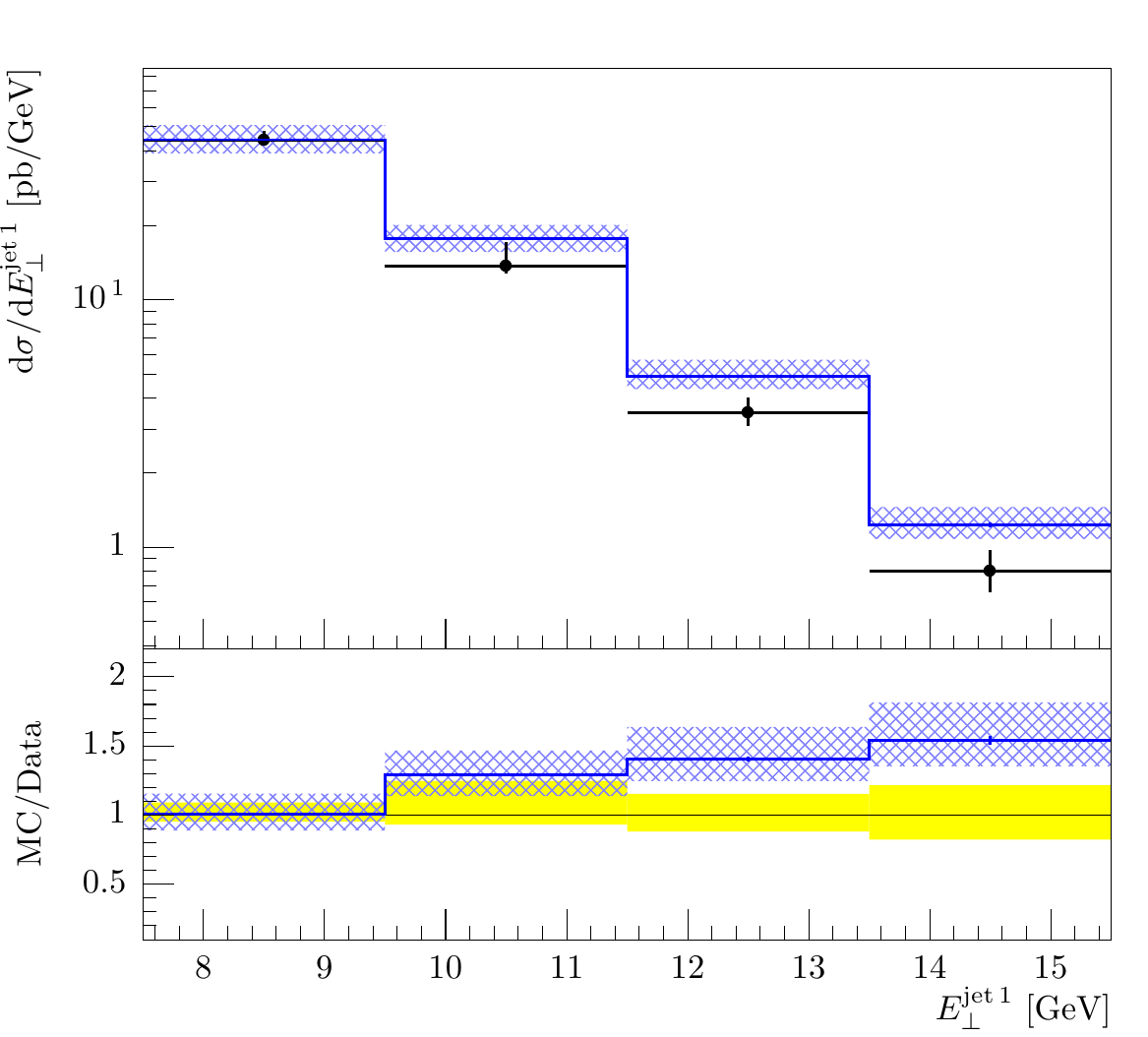}\\
(d)
\end{minipage}
\caption{\label{Fig:PDFandMPI_scales}
The model along with the uncertainty bands arising from varying the
renormalization- and factorization scales compared to ZEUS data on 
$M_X$ (a), $z^{\mathrm{obs}}_{\Pom}$ (b), 
$x^{\mathrm{obs}}_{\gamma}$ (c) and 
$E_{\perp}^{\mrm{jet\,1}}$ (d).
}
\end{figure*}

To probe the uncertainties in the choice of renormalization and
factorization scales, $\mu_R$ and $\mu_F$, we employ the usual method of 
varying the scales up and down with a factor of two. Each is probed 
individually, such that one scale is kept fixed while the other is 
varied. Only the scales at matrix-element level are varied; thus 
the shower and MPI scales have been excluded from these variations. Each
variation gives rise to an uncertainty band, and in 
Fig.~\ref{Fig:PDFandMPI_scales} we show the envelope using the maximal
value obtained from either of the two uncertainty bands. The envelope is 
dominated by the renormalization scale, giving the largest uncertainty
in most of the figures shown -- not unusual in a LO calculation. 
Note, however, that the scale uncertainty in the 
high-$x^{\mathrm{obs}}_{\gamma}$ bin
actually reaches the upper error of the data point,
essentially hinting that the model is able to describe the
direct-enhanced region within theoretical uncertainties. The resolved
region, however, cannot be fully accounted for within these theoretical
uncertainties. 

\subsection{Variations of the dPDFs}

\begin{figure*}[!ht]
\begin{minipage}[c]{0.475\linewidth}
\centering
\includegraphics[width=0.9\linewidth]{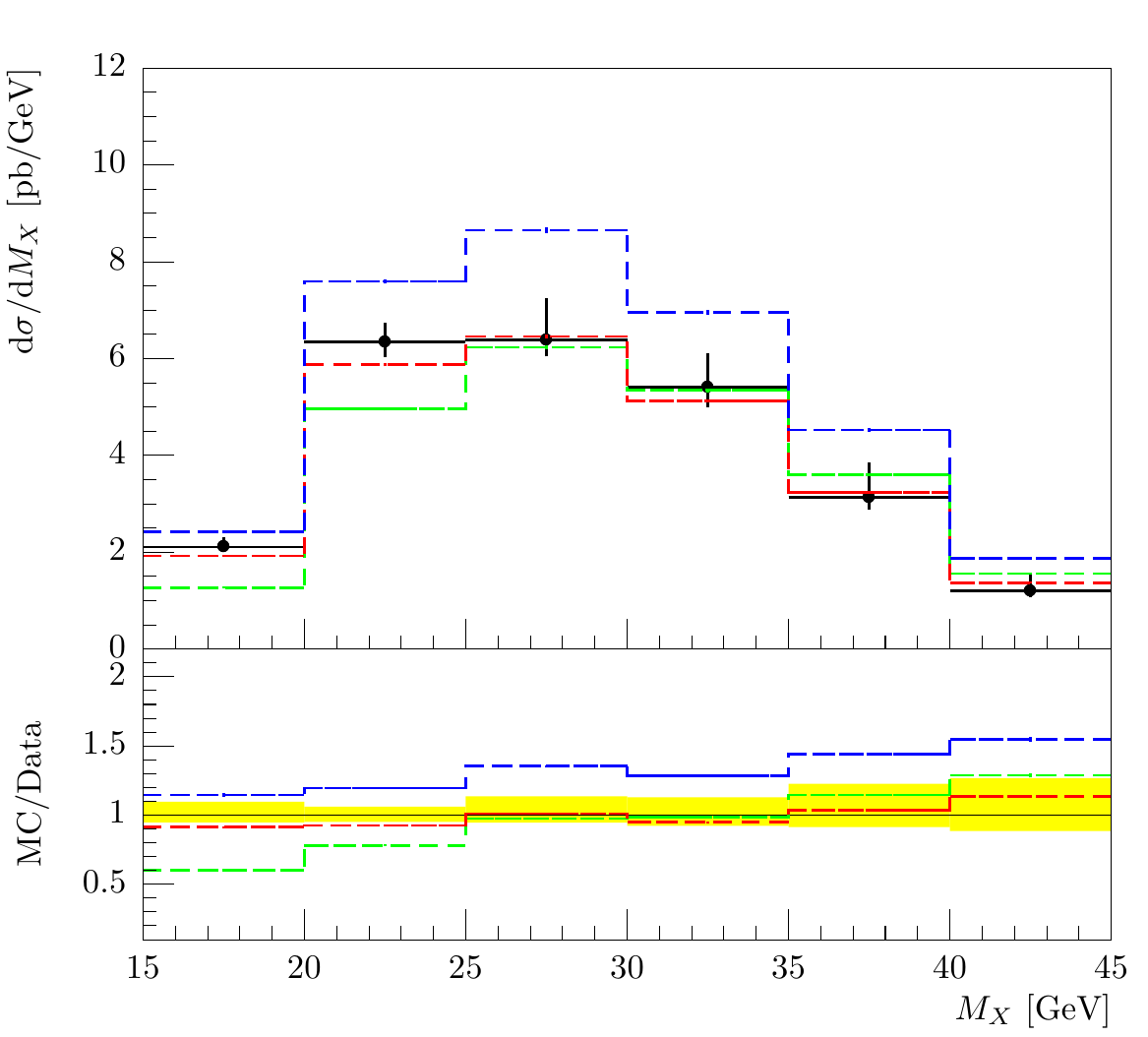}\\
(a)
\end{minipage}
\hfill
\begin{minipage}[c]{0.475\linewidth}
\centering
\includegraphics[width=0.9\linewidth]{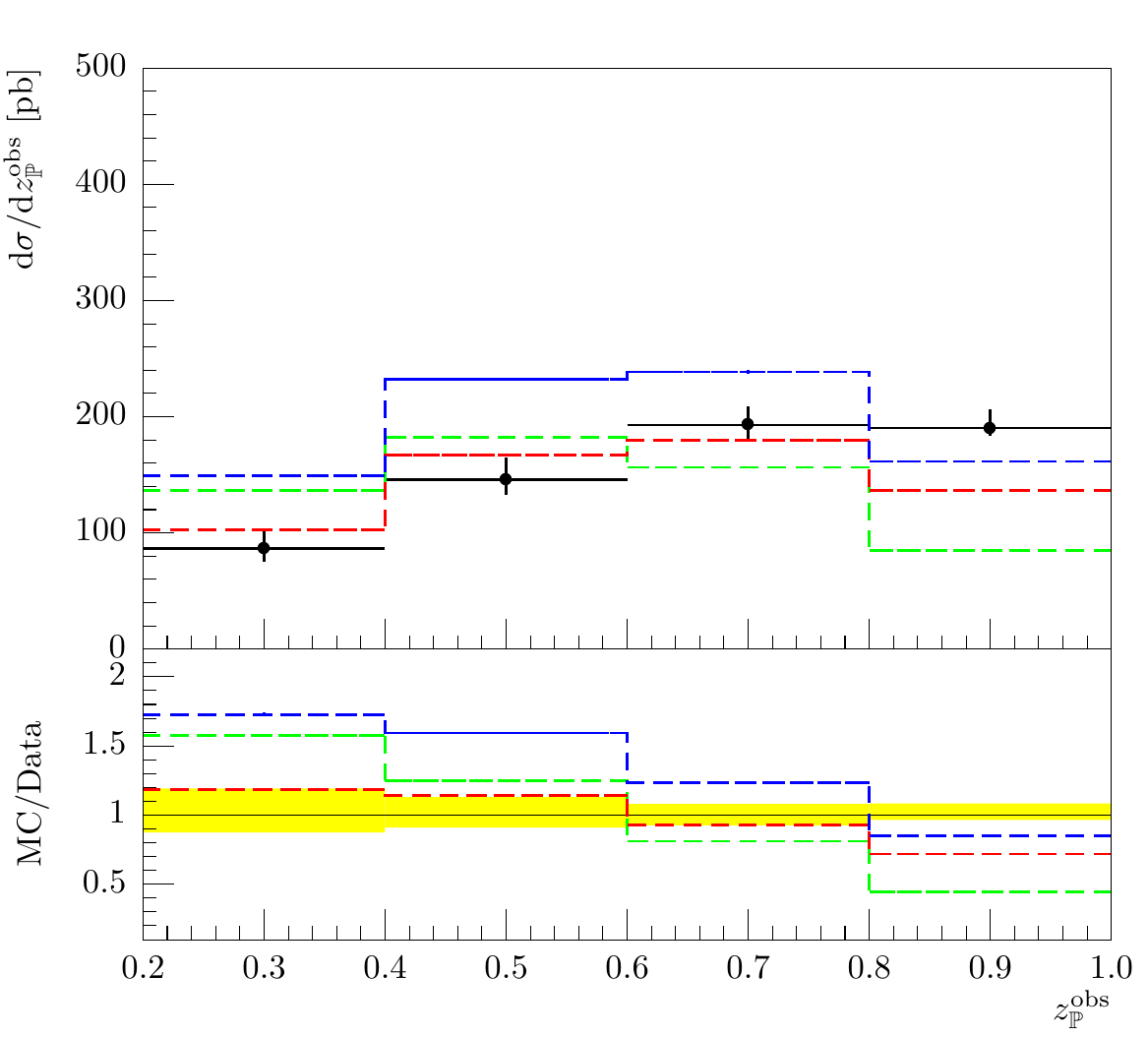}\\
(b)
\end{minipage}
\begin{minipage}[c]{0.475\linewidth}
\centering
\includegraphics[width=0.9\linewidth]{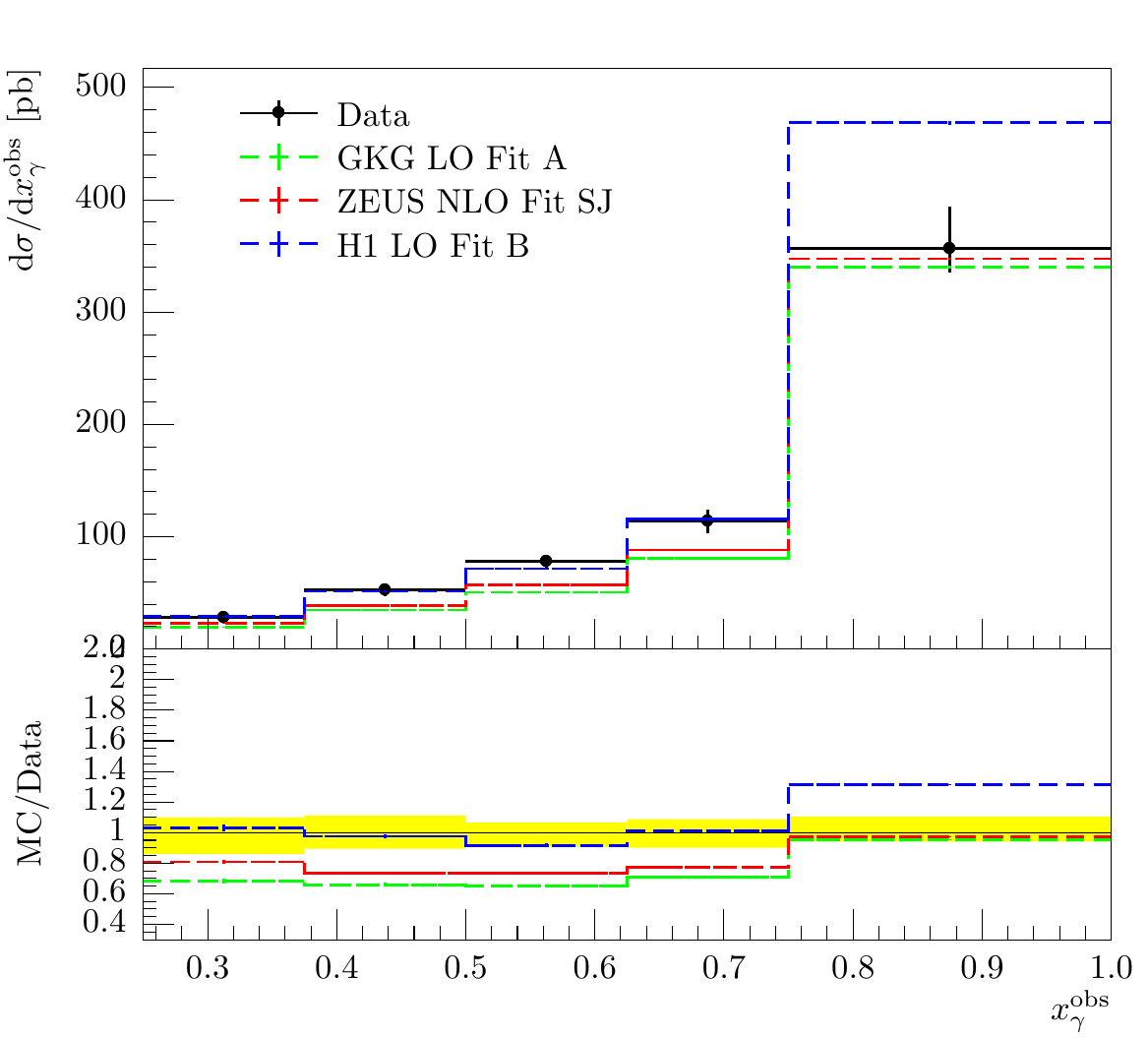}\\
(c)
\end{minipage}
\hfill
\begin{minipage}[c]{0.475\linewidth}
\centering
\includegraphics[width=0.9\linewidth]{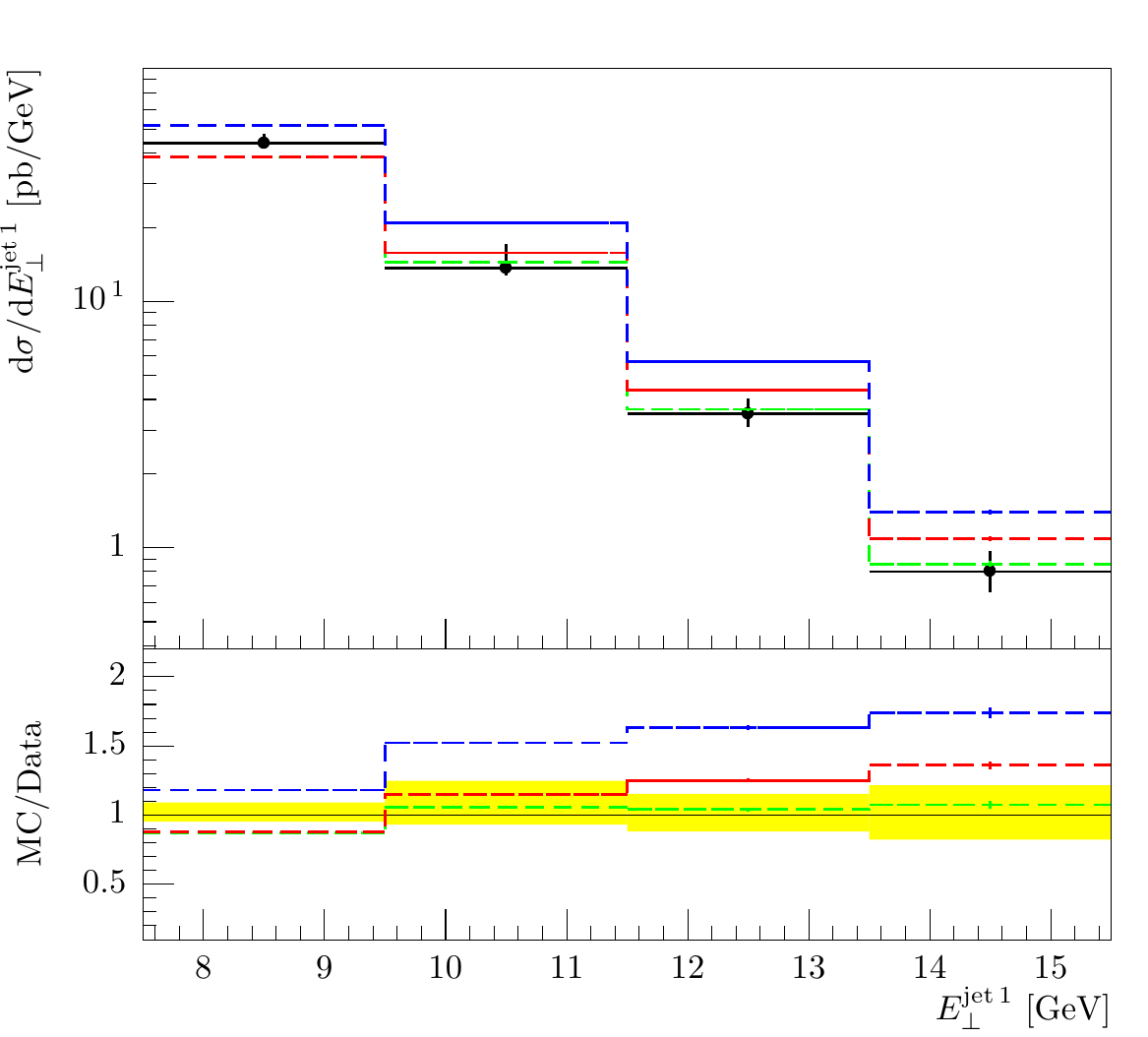}\\
(d)
\end{minipage}
\caption{\label{Fig:PDFandMPI_GKG}
The model without gap suppression using three different dPDFs: H1 LO 
Fit B (blue lines), GKG LO Fit A (green lines) and ZEUS NLO SJ 
(red lines) compared to ZEUS data on $M_X$ (a), 
$z^{\mathrm{obs}}_{\Pom}$ (b), 
$x^{\mathrm{obs}}_{\gamma}$ (c) and 
$E_{\perp}^{\mrm{jet\,1}}$ (d).
}
\end{figure*}

\begin{figure*}[!ht]
\begin{minipage}[c]{0.475\linewidth}
\centering
\includegraphics[width=0.9\linewidth]{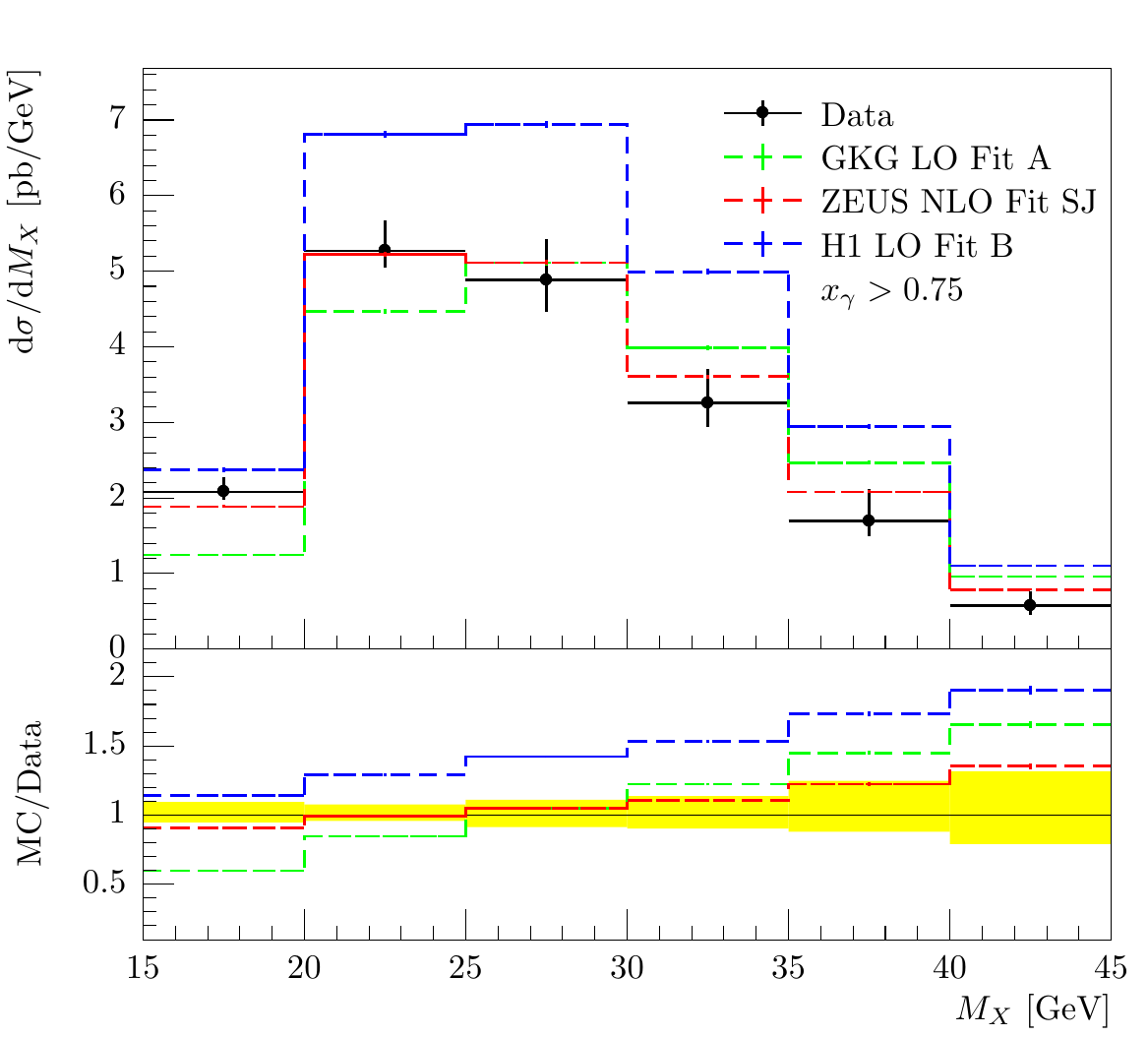}\\
(a)
\end{minipage}
\hfill
\begin{minipage}[c]{0.475\linewidth}
\centering
\includegraphics[width=0.9\linewidth]{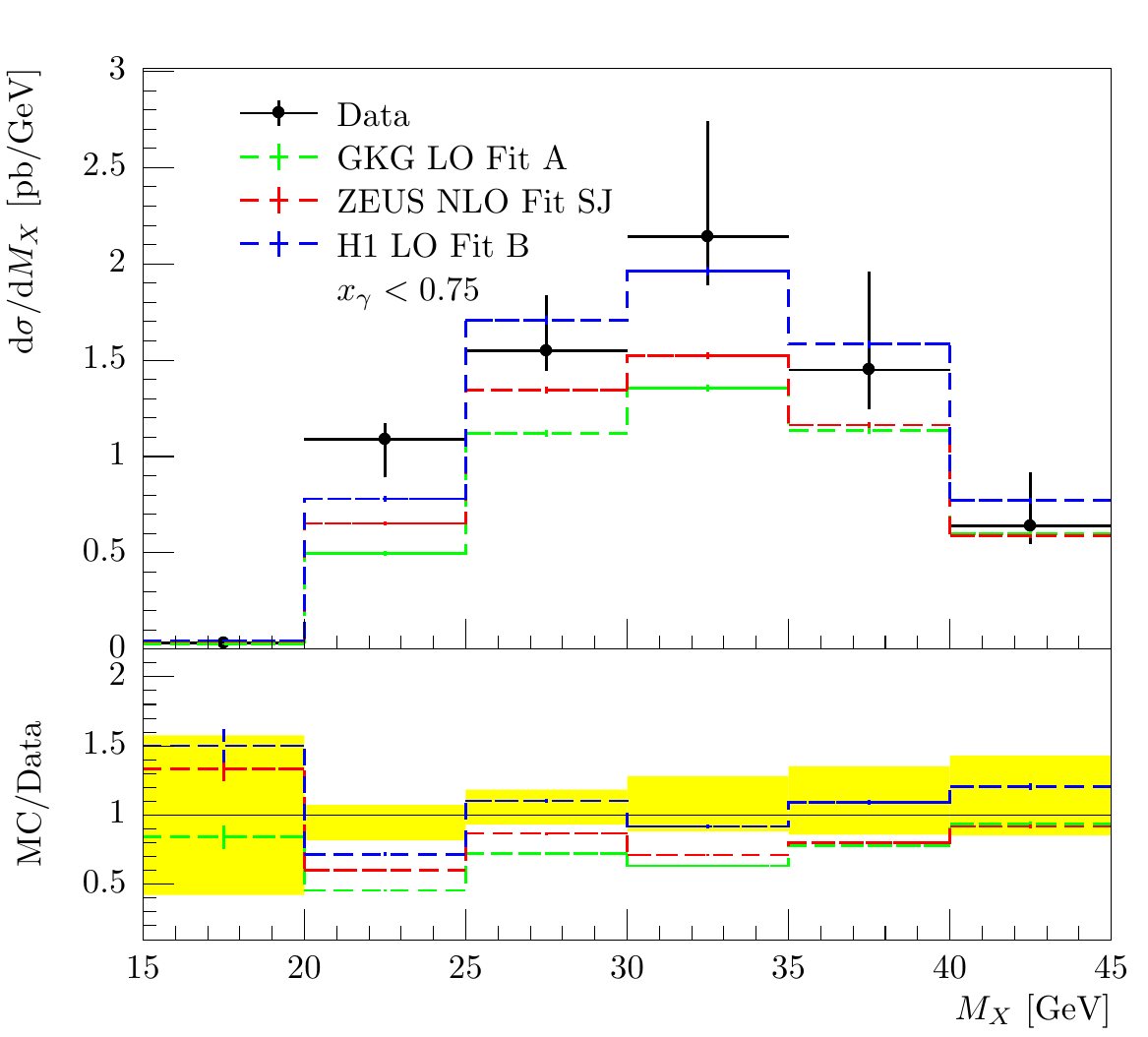}\\
(b)
\end{minipage}
\caption{\label{Fig:GKG_xgamma_regions}
The model without gap suppression using the three dPDFs: H1 Fit B LO 
(blue lines), GKG LO Fit A (green lines) and ZEUS SJ (red lines) 
compared to ZEUS data on $M_X$ in the direct-enhanced (a) and
resolved-enhanced (b) regions. 
}
\end{figure*}

As explained above, the considered observables are sensitive to the 
dPDFs, especially the fractional momentum carried by the parton from 
the Pomeron, $z^{\mathrm{obs}}_{\Pom}$. We here 
investigate if the increased amount of diffractive DIS data in the 
GKG LO dPDFs will provide a better description of the data than the 
less constrained H1 LO Fit B dPDF. We also show results obtained when 
using the NLO dPDF and flux from
ZEUS SJ, as this dPDF includes data on diffractive dijets that is 
directly sensitive to the gluon distributions. Note, however, that a
combination of NLO PDFs and LO matrix elements is still only accurate 
to LO and mixing different orders may result in different
results compared to a situation where the matrix elements and PDF 
determination are consistently at the same perturbative order.

In Fig.~\ref{Fig:PDFandMPI_GKG} we show results using two of the new 
dPDFs, ZEUS NLO SJ \cite{Chekanov:2009aa} and GKG LO Fit A 
\cite{Goharipour:2018yov} \textit{without} the gap suppression factor.
At first glance, the new dPDFs improve the overall description of data
without a further need for suppression. Overall the new dPDFs seem to 
suppress the distributions as compared to H1 Fit B LO dPDF, with the
ZEUS SJ dPDF performing slightly better than GKG LO Fit A as seen e.g.\ in
the $z^{\mathrm{obs}}_{\Pom}$ distribution. Here, the 
ZEUS SJ dPDF flattens out at high $z^{\mathrm{obs}}_{\Pom}$
as compared to the GKG and H1 dPDFs, having a slightly larger 
$x_g$-distribution in this regime.

The distributions that the baseline study did not fully describe, also 
the new dPDFs fail to describe. Especially the 
$x^{\mathrm{obs}}_{\gamma}$ distribution is still underestimated at 
$x^{\mathrm{obs}}_{\gamma}<0.75$, which underlines the discrepancies 
with the relative normalization between the direct and resolved 
contributions. The $E_{\perp}^{\mrm{jet\,1}}$ distribution is now 
well described with the GKG set. With the ZEUS SJ set the 
normalization is improved compared to the H1 Fit B but the shape of the 
distribution is similarly off.

\begin{figure*}[!ht]
\begin{minipage}[c]{0.475\linewidth}
\centering
\includegraphics[width=0.9\linewidth]{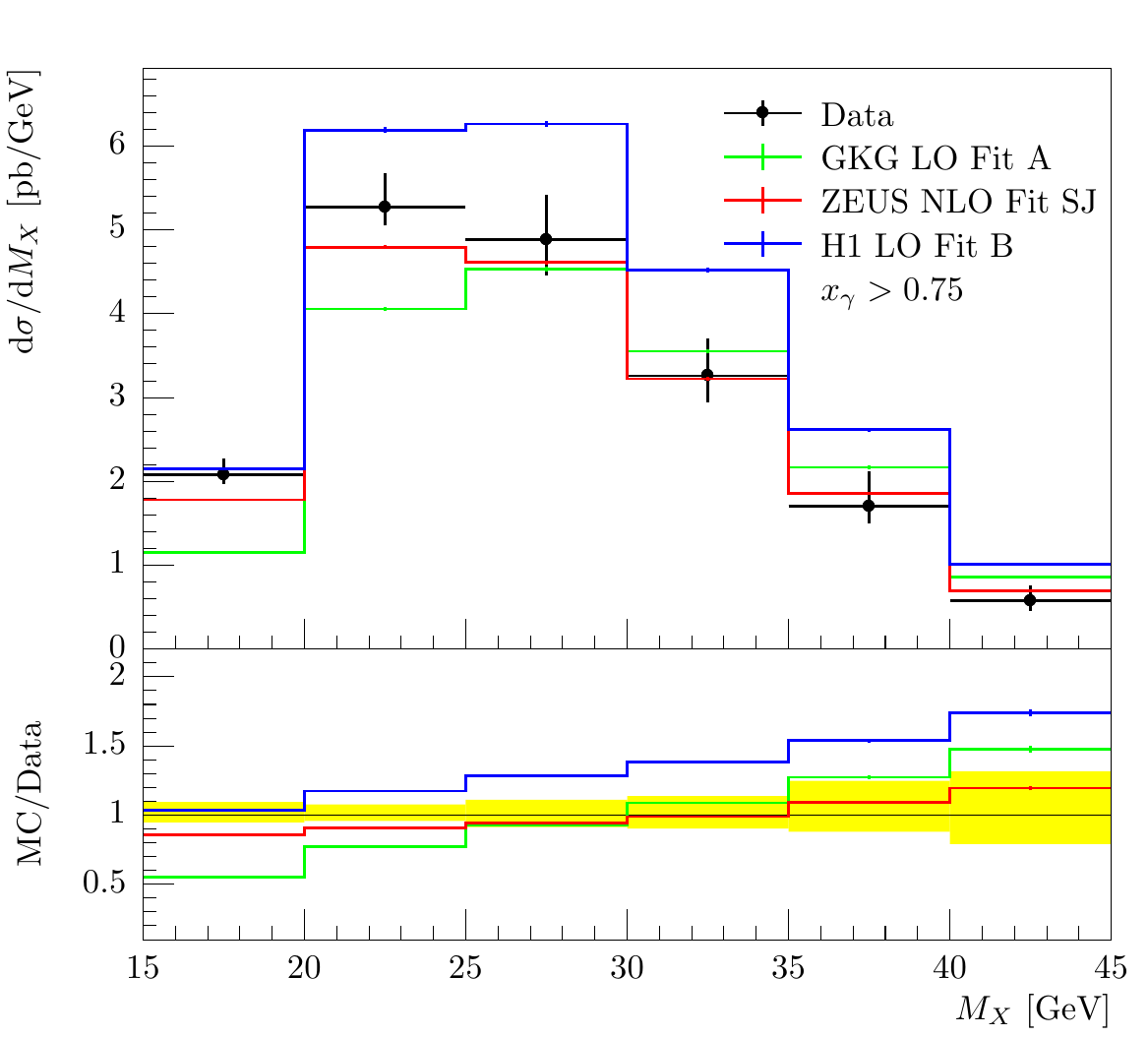}\\
(a)
\end{minipage}
\hfill
\begin{minipage}[c]{0.475\linewidth}
\centering
\includegraphics[width=0.9\linewidth]{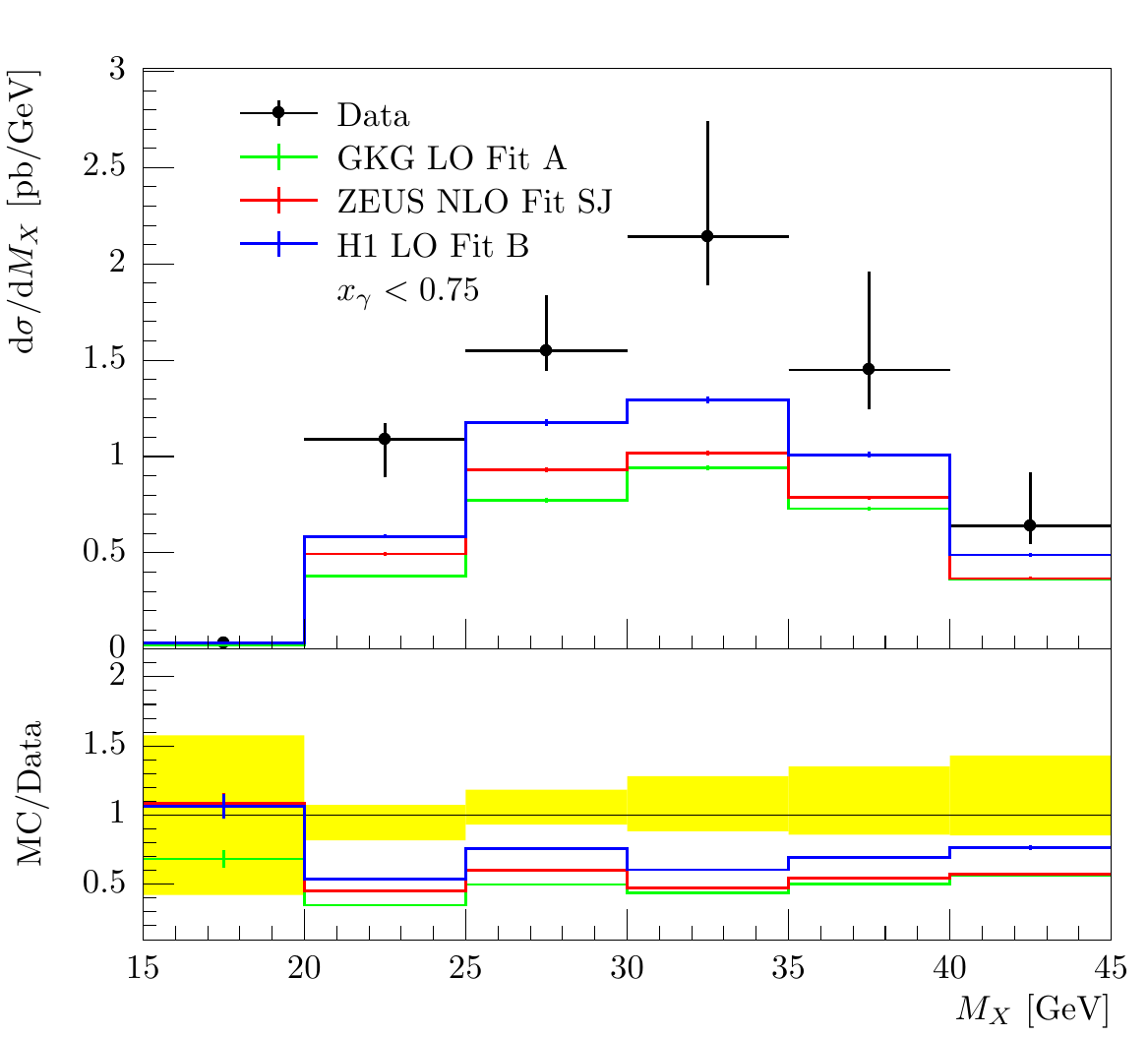}\\
(b)
\end{minipage}
\caption{\label{Fig:GKG_xgamma_MPI}
The model with gap suppression using the three dPDFs: H1 Fit B LO 
(blue lines), GKG LO Fit A (green lines) and ZEUS SJ (red lines) 
compared to ZEUS data on $M_X$ in the direct-enhanced (a) and
resolved-enhanced (b) regions. 
}
\end{figure*}

A separation of $M_X$ into the two regimes, 
Fig.~\ref{Fig:GKG_xgamma_regions}, shows that the direct-enhanced 
region is well described with the ZEUS SJ dPDFs. The GKG set improves 
the normalization but the shape of the distribution is still not 
compatible. The resolved region, however, is too suppressed with 
both of these, so the relative normalizations of the two contributions 
remain as an unresolved issue. Adding the gap suppression factor on 
top of this, Fig.~\ref{Fig:GKG_xgamma_MPI}, further suppresses the 
already suppressed resolved-enhanced region, worsening the 
agreement with the data in this regime. Little effect is seen in 
the direct-enhanced region, as expected. 

These results thus puts forth the question whether the gap suppression
is necessary if the dPDFs are refined and improved with additional
diffractive data. The improvements seen especially with the ZEUS SJ
dPDF in both the $x^{\mathrm{obs}}_{\gamma}$ and 
$z^{\mathrm{obs}}_{\Pom}$ distributions might hint
towards this. As discussed earlier, this might partly follow from 
the tight cuts applied in the ZEUS analysis which does not leave much 
room for MPIs in the $\gamma\p$ system. Also, one should keep in mind
that using NLO dPDFs with LO matrix elements might lead to different
results compared to a full NLO calculation.

\subsection{Variations of the screening parameter}

\begin{figure*}[!ht]
\begin{minipage}[c]{0.475\linewidth}
\centering
\includegraphics[width=0.9\linewidth]{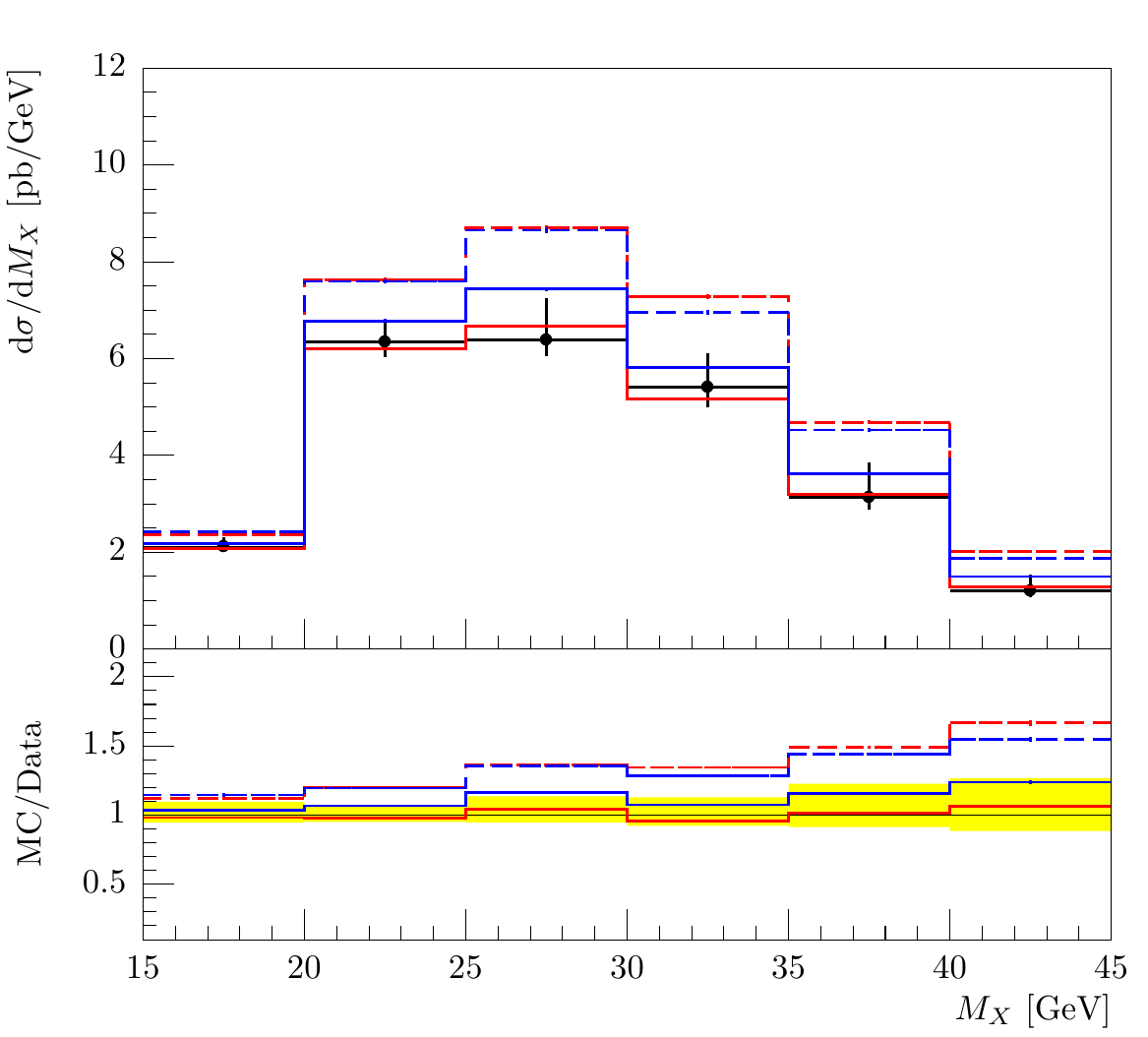}\\
(a)
\end{minipage}
\hfill
\begin{minipage}[c]{0.475\linewidth}
\centering
\includegraphics[width=0.9\linewidth]{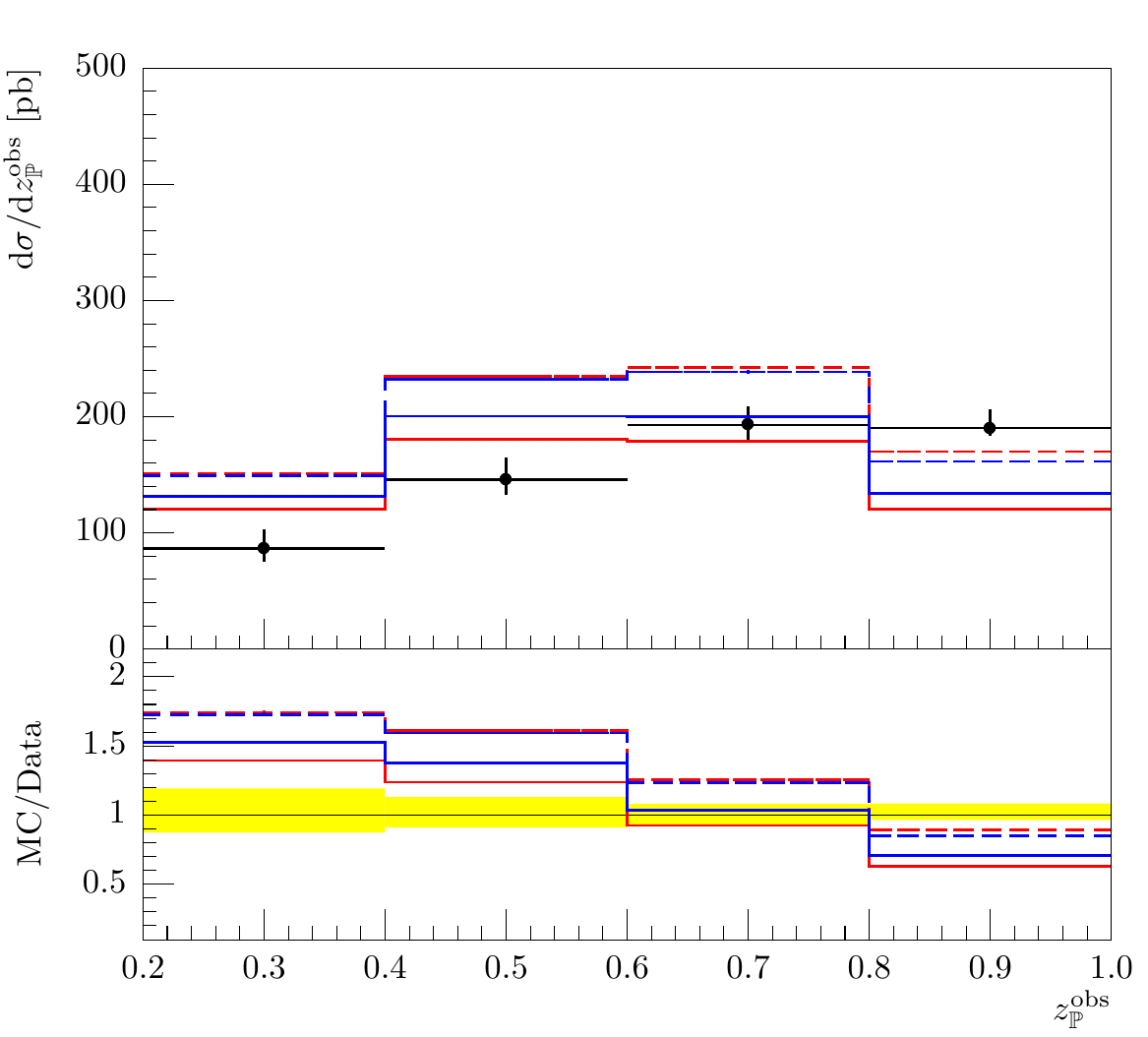}\\
(b)
\end{minipage}
\begin{minipage}[c]{0.475\linewidth}
\centering
\includegraphics[width=0.9\linewidth]{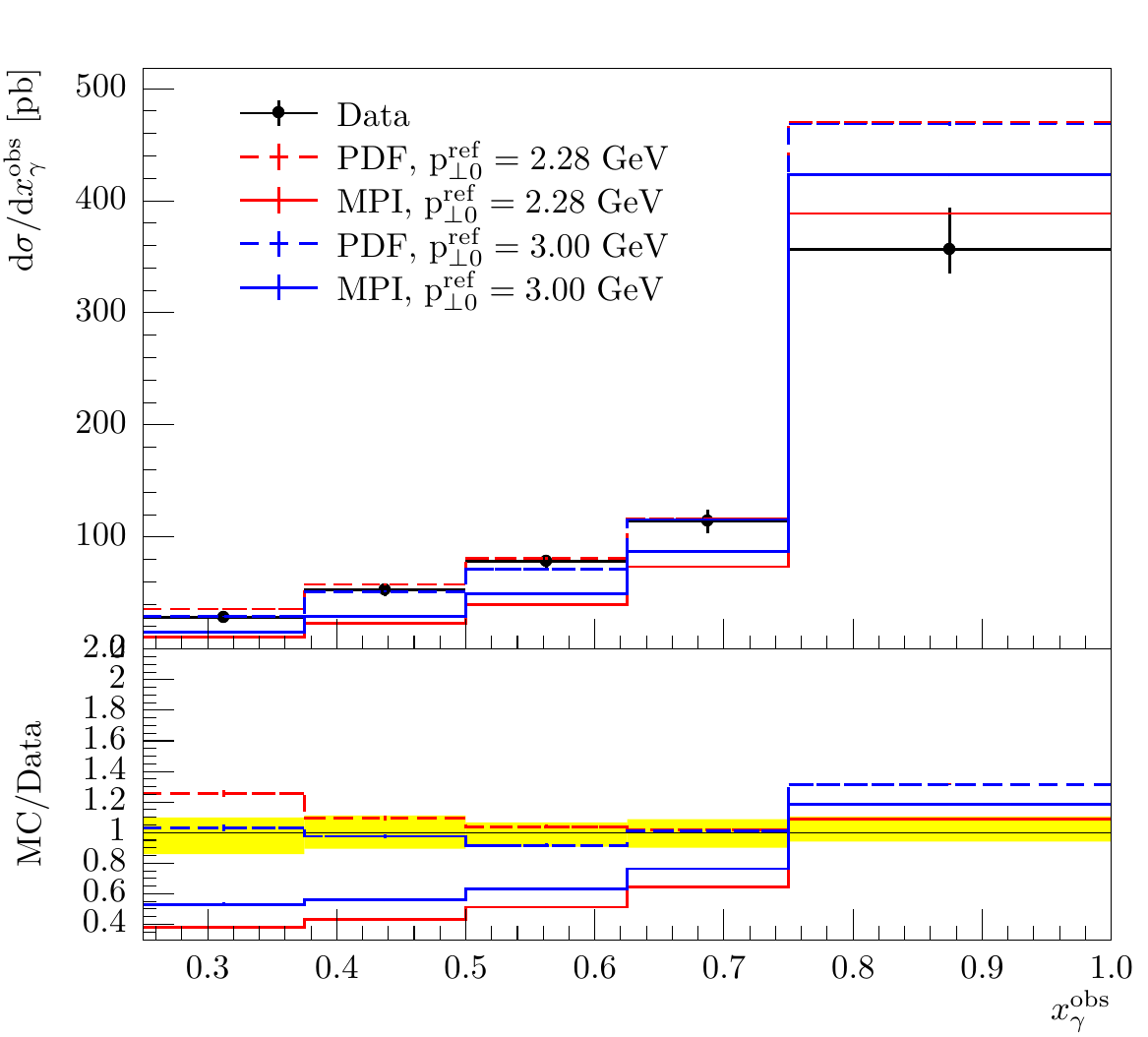}\\
(c)
\end{minipage}
\hfill
\begin{minipage}[c]{0.475\linewidth}
\centering
\includegraphics[width=0.9\linewidth]{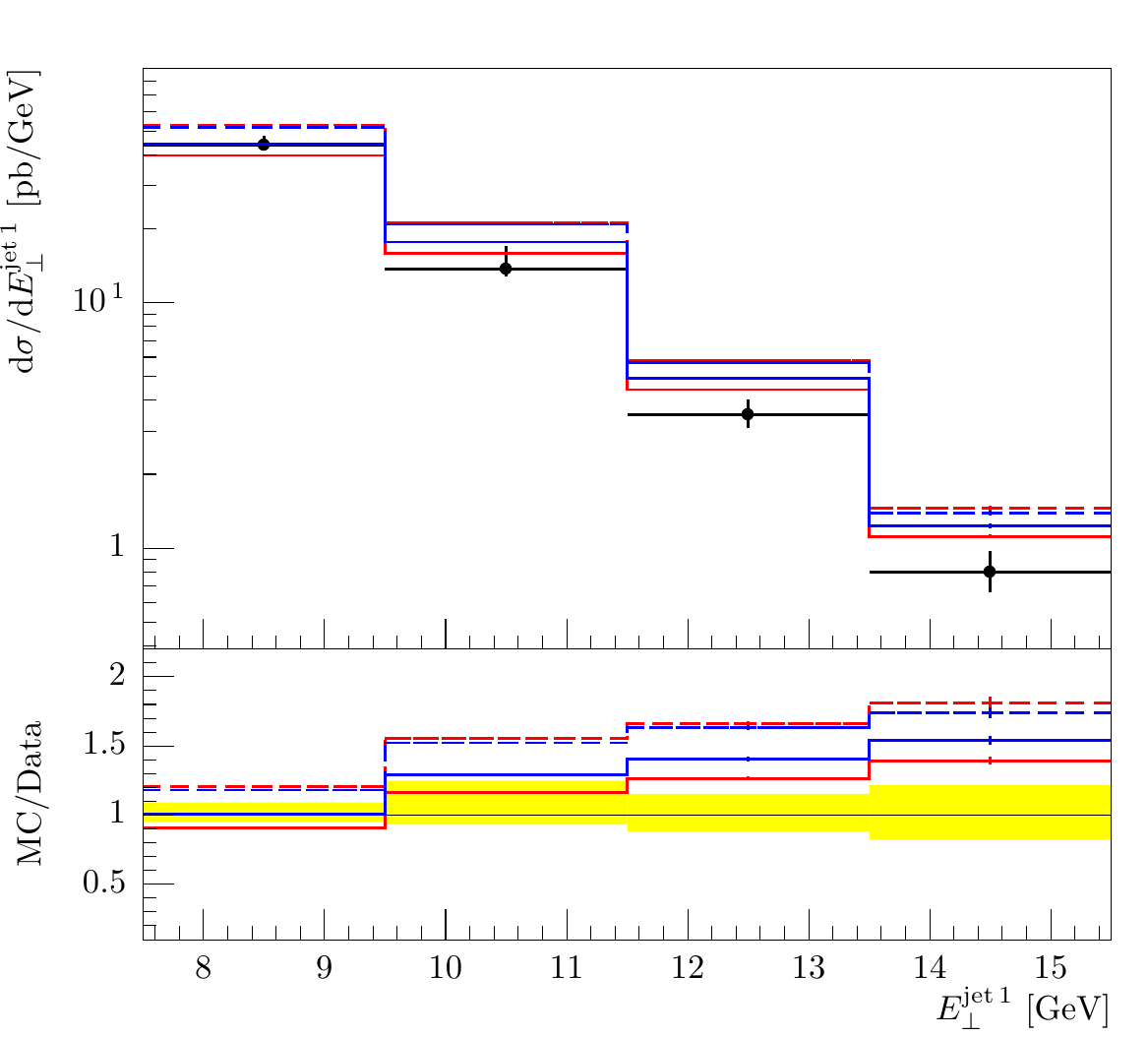}\\
(d)
\end{minipage}
\caption{\label{Fig:PDFandMPI_pTrefPP}
The model with (solid lines) and without (dashed lines) gap suppression 
using two values of $\pTo^{\mrm{ref}}$: The $\p\p$-tune,
$\pTo^{\mrm{ref}}=2.28~\GeV$ (red lines) and the $\e\p$-tune,
$\pTo^{\mrm{ref}}=3.0~\GeV$ (blue lines). Again we show the samples in
the observables $M_X$ (a), $z^{\mathrm{obs}}_{\Pom}$
(b), $x^{\mathrm{obs}}_{\gamma}$ (c) and 
$E_{\perp}^{\mrm{jet\,1}}$ (d).
}
\end{figure*}

The gap suppression method used here is highly sensitive to the model
parameters of the MPI framework. Here we especially look at the screening
parameter, $\pTo^{\mrm{ref}}$, as the value of this parameter differs between
tunes to $\e\p$ and to $\p\p$ collisions. Changing the value of
$\pTo^{\mrm{ref}}$ 
have only a small effect on
the ``PDF'' samples. The ``MPI'' samples, however, are affected by the
value of the screening parameter. A smaller value of $\pTo^{\mrm{ref}}$ 
results in more MPIs, thus we expect that the gap suppression will be 
larger if we decrease $\pTo^{\mrm{ref}}$ to its $\p\p$ value, as a 
smaller fraction of the events will survive the MPI-selection. 

This effect is exactly what is seen in 
Fig.~\ref{Fig:PDFandMPI_pTrefPP}. The ``PDF'' samples are not affected,
but the $\p\p$-tuned $\pTo^{\mrm{ref}}$ value in red causes a stronger 
suppression, best seen in the ratio plots where the solid red curves, 
the ``MPI'' sample with $\pTo^{\mrm{ref}}=2.28~\GeV$, is lower than 
the solid blue curves with $\pTo^{\mrm{ref}}=3.00~\GeV$. 
The value of $\pTo^{\mrm{ref}}$ has some effect on the shape of the 
distributions, mainly because a higher $M_X$ allows for more MPI 
activity, and thus a smaller fraction of events survive the no-MPI
requirement. This means that the gap suppression increases with
increasing energy available in the system, i.e.\ with increasing $M_X$,
seen in Fig.~\ref{Fig:PDFandMPI_pTrefPP} (a), where ratio-plot shows a
suppression factor of approximately 0.9 in the low $M_X$ bin and 0.6 
in the high $M_X$ bin.

\subsection{Gap suppression factors}

Several models have been proposed to explain the factorization breaking
in diffractive hadronic collisions. Many of these employ an overall
suppression factor, often relying primarily on the impact-parameter of
the collision, see e.g.\ \cite{Khoze:2000wk, Gotsman:2005rt, Jones:2013pga}.
Some also include a suppression w.r.t.\ a kinematical variable, such 
as the $\pT$ of the diffractive dijets. But to our knowledge, the model 
of dynamical gap survival is the first of its kind to evaluate the 
gap survival on an event-by-event basis. This means it takes into 
account the kinematics of the entire event, and is thus also able to 
provide a gap suppression factor differential in any observable. In 
the model presented here, the ratio of ``PDF'' to ``MPI'' samples 
equates the gap survival factor, as the two samples only differ by 
the no-MPI requirement that determines the models definition of a fully 
diffractive event.

\begin{figure*}[!ht]
\begin{minipage}[c]{0.475\linewidth}
\centering
\includegraphics[width=0.9\linewidth]{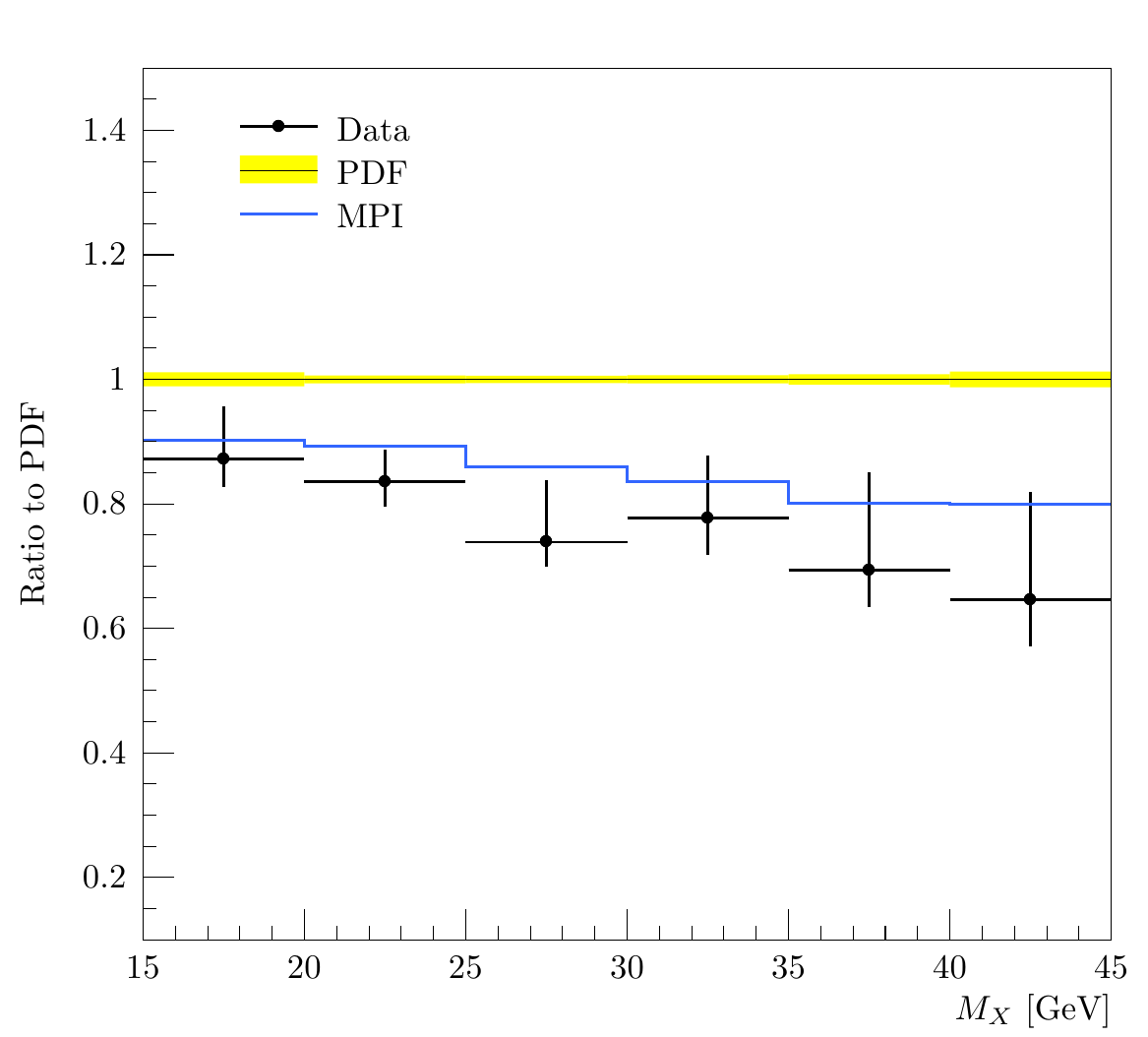}\\
(a)
\end{minipage}
\hfill
\begin{minipage}[c]{0.475\linewidth}
\centering
\includegraphics[width=0.9\linewidth]{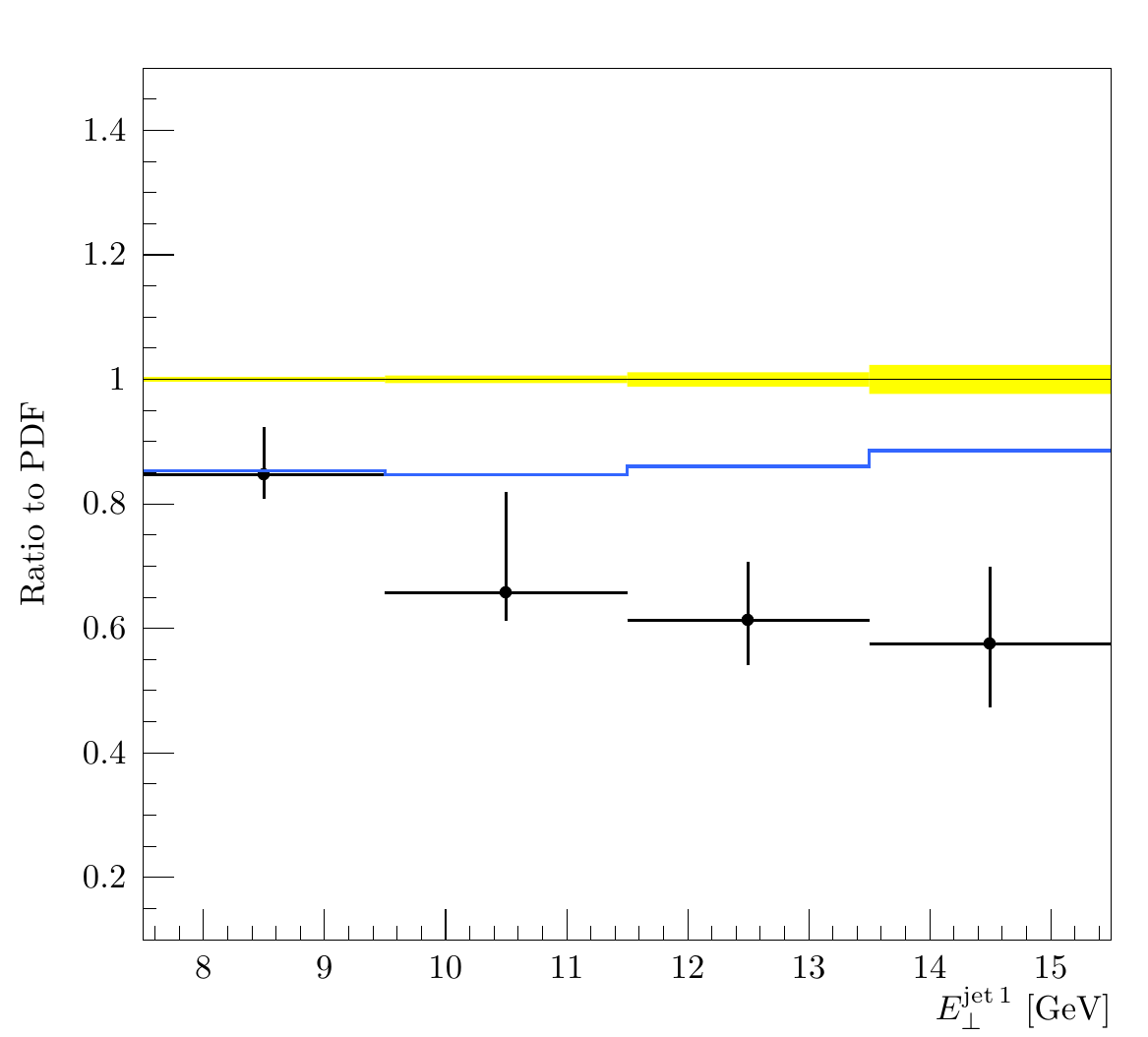}\\
(b)
\end{minipage}
\caption{\label{Fig:PDFandMPIgap_ZEUS}
The predicted gap suppression factors as a function of 
$M_X$ (a) and $E_{\perp}^{\mrm{jet\,1}}$ (b) compared to the ZEUS analysis.
}
\end{figure*}

\begin{figure*}[!ht]
\begin{minipage}[c]{0.475\linewidth}
\centering
\includegraphics[width=0.9\linewidth]{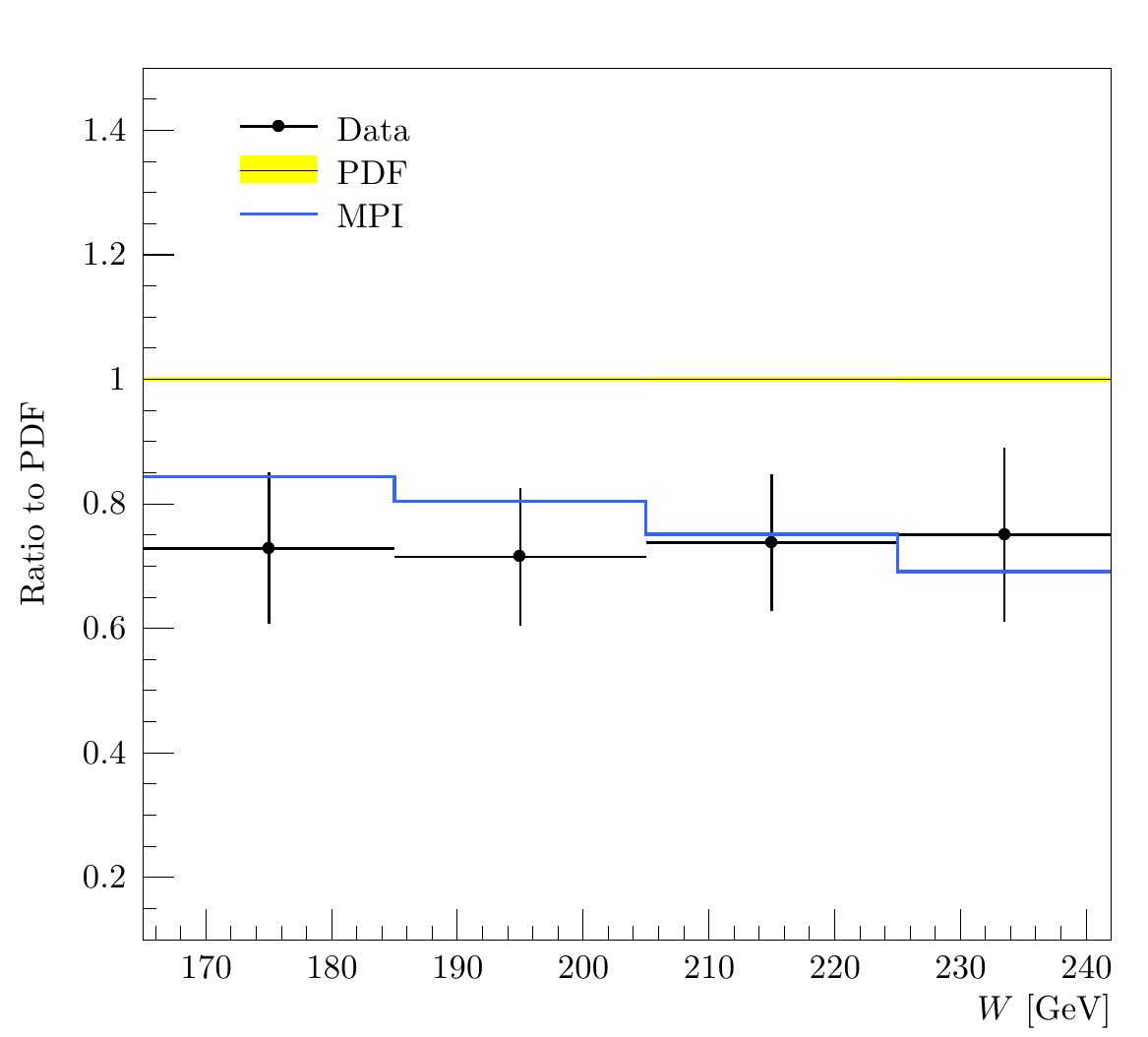}\\
(a)
\end{minipage}
\hfill
\begin{minipage}[c]{0.475\linewidth}
\centering
\includegraphics[width=0.9\linewidth]{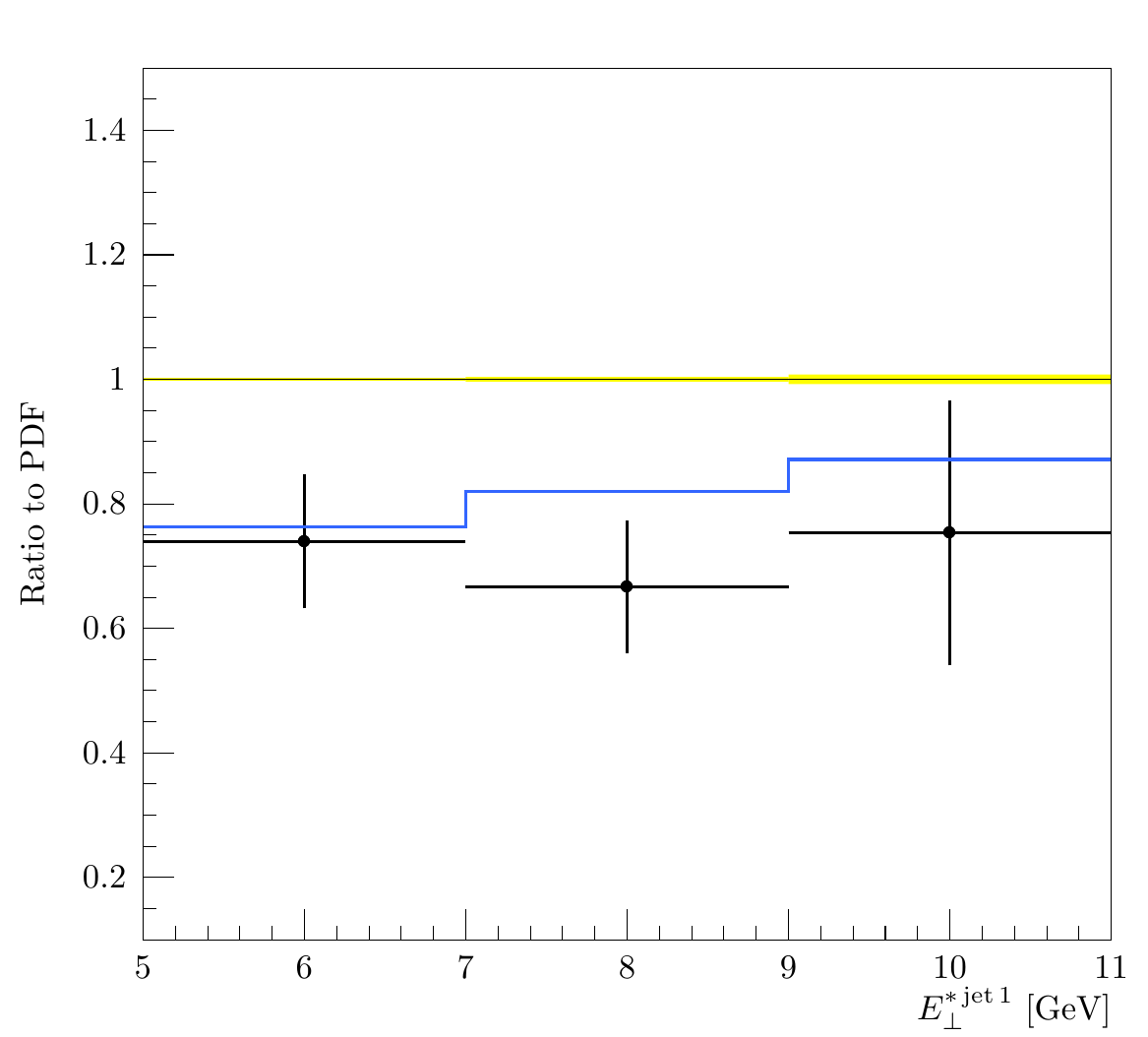}\\
(b)
\end{minipage}
\caption{\label{Fig:PDFandMPIgap_H1}
The predicted gap suppression factors as a function of 
$W$ (a) and $E_{\perp}^{\mrm{*\,jet\,1}}$ (b) compared to the H1 analysis.
}
\end{figure*}

The theoretical uncertainties not directly related to MPI probability
(e.g.\ scale variations) are expected to cancel in such a ratio. Even 
though many experimental analysis present similar ratios by using 
a NLO calculation as a baseline, such a ratio is not a measurable
quantity, as it always require a theory-based estimation for the unsuppressed result. 
These ratios, however, are useful for demonstrating the effects arising 
from different models such as our dynamical rapidity gap survival. In 
order to estimate the factorisation-breaking effect in
data w.r.t.\ our model, we show also the ratio between the data and 
the ``PDF'' sample. 

In Fig.~\ref{Fig:PDFandMPIgap_ZEUS} we show the gap suppression
differential in the observables $M_X$ and 
$E_{\perp}^{\mrm{jet\,1}}$ from the ZEUS analysis and in
Fig.~\ref{Fig:PDFandMPIgap_H1} we show the gap suppression differential 
in the observables $W$ and $E_{\perp}^{*\,\mrm{jet\,1}}$ from the H1 
analysis. These distributions demonstrate some of the main 
features of our dynamical rapidity gap survival model. We show the ratio 
of data to ``PDF'' sample (black dots) and the ratio of ``MPI'' to 
``PDF'' sample (solid blue curve). This latter ratio is exactly the gap 
suppression factor predicted by the model. The shapes of the gap 
suppression factors agree reasonably well with the suppression factors 
derived from the data (the black dots), albeit the shape of 
Fig.~\ref{Fig:PDFandMPIgap_ZEUS} (b) is off in the high-$E_{\perp}$ end, 
as already mentioned in the baseline results.

The model predicts a slowly decreasing suppression in 
$E_{\perp}^{(*)\,\mrm{jet\,1}}$, while the suppression increases 
towards larger $M_X$ and $W$. This increase follows as the 
larger diffractive masses are correlated with larger invariant masses 
of the $\gamma\p$-system, where there is more room for MPIs at fixed jet 
$E_{\perp}$. This results in a larger fraction of the events having 
additional MPIs, thus a smaller fraction of the events survive as 
diffractive. Similarly, high-$E_{\perp}$ jets takes away more 
momentum than low-$E_{\perp}$ jets, again leaving less room for MPIs 
to take place. Thus we do not predict a flat overall suppression, as 
has often been applied in the experimental analyses. 

Suppression factors in the range $0.7-0.9$ are predicted in the
shown observables. Given the uncertainty on the ``PDF'' sample, this 
is in agreement with the suppression factors of approximately 
$0.5-0.9$, as observed by H1 
\cite{Adloff:1998gg, Aktas:2007hn, Aaron:2010su, Andreev:2015cwa} and 
ZEUS \cite{Chekanov:2007rh}. A somewhat 
contradictory result was observed in ref.~\cite{Chekanov:2009aa}, in 
which the ZEUS dijet data from ref.~\cite{Chekanov:2007rh} was found 
consistent with the purely factorization-based NLO calculation when 
using the ZEUS SJ dPDFs. \\

The experimental cuts applied in the ZEUS analysis, as compared to the
analysis from H1, forces $x^{\mathrm{obs}}_{\gamma}$ to very large 
values, where the suppression from the MPIs does not have a large 
effect. Thus the ZEUS measurement requires less suppression than 
what is needed in the H1 measurement. The shown distributions, 
however, are still marred by the large theoretical uncertainties. One 
way to reduce these theoretical uncertainties would be to consider 
the ratio of photoproduced dijets to ones from DIS, as 
done e.g.\ in the recent H1 analysis 
\cite{Andreev:2015cwa}. The kinematic domain is slightly different due 
to different virtualities, but this would still greatly reduce dependency 
on dPDFs and scale variations, leaving only the mild photon PDF 
dependence in addition to the factorization breaking effects, that would 
be pronounced in this ratio. Unfortunately the current \textsc{Pythia}~8
description of DIS events at intermediate virtualities is not adequate 
to describe the inclusive DIS dijet data, so such a comparison is a 
project for the future.

\section{\label{Sec:6}Photoproduction in ultra-peripheral collisions}

Because of the more than an order of magnitude larger $\sqrt{s}$ at the 
LHC, the accessible invariant masses of the $\gamma\p$ system are much 
larger than what could be studied at HERA. This allows us to study the 
factorization-breaking effects in hard-diffractive photoproduction in a 
previously unexplored kinematical region. Such measurements would fill 
the gap between the rather mild suppression observed at HERA and the 
striking effect observed in $\p\pbar$ and $\p\p$ collisions at Tevatron 
and the LHC. This would provide important constraints for different 
models and thus valuable information about the underlying 
physics. Besides the results we present here, predictions for these 
processes have been computed in a framework based on a factorized NLO 
perturbative QCD calculation with two methods of gap survival 
probabilities, one with an overall suppression and one where the 
suppression is only present for resolved photons \cite{Guzey:2016tek}. 
The authors here expect that the two scenarios can be distinguished at 
LHC, especially in the $x_{\gamma}^{\mrm{obs}}$-distribution. The model 
presented in this work should thus be comparable to the latter 
suppression scheme from \cite{Guzey:2016tek}. Another
work considering similar processes is presented in 
Ref.~\cite{Basso:2017mue}.

In principle these measurements could be done in all kinds of
hadronic and nuclear collisions, since all fast-moving 
charged particles generate a flux of photons. There are, however, some 
differences worth covering. In $\p\p$ collisions, the photons can be 
provided by either of the beam particles with an equal probability. The 
flux of photons is a bit softer for protons than with leptons, but 
still clearly harder than with nuclei. Experimentally it might be 
difficult to distinguish the photon-induced diffraction and ``regular'' 
double diffraction in $\p\p$, since both processes would leave a similar 
signature with rapidity gaps on both sides. In $\p\Pb$ collisions the 
heavy nucleus is the dominant source of photons, as the flux is 
amplified by the squared charge of the emitting nucleus, $Z^2$. 
Thus the photon-induced diffraction should overwhelm the QCD-originating 
colorless exchanges (Pomerons and Reggeons). Similarly, in $\Pb\Pb$ 
collisions the photon fluxes are large and thus would overwhelm the 
Regge exchanges. The latter type is currently not possible to model 
with \textsc{Pythia}~8, however, as in addition to regular MPIs, one 
should also take into account the further interactions between the 
resolved photon and the other nucleons, that could destroy the rapidity gap. 
Since these are currently not implemented in the photoproduction 
framework, we leave the $\Pb\Pb$ case for a future study.
\begin{table}
\begin{center}
\caption{Kinematics for the UPC analyses.}
\label{Tab:LHCcuts}
\begin{tabular}{lll}
\hline\noalign{\smallskip}
& $\p\Pb$ & $\p\p$ \\
\noalign{\smallskip}\hline\noalign{\smallskip}
$\sqrt{s_{\mathrm{NN}}}$ & $5.0~\TeV$ & $13.0~\TeV$ \\
$E_{\perp\mathrm{,min}}^{1}$ & \multicolumn{2}{c}{$8.0~\GeV$} \\
$E_{\perp\mathrm{,min}}^{2}$ & \multicolumn{2}{c}{$6.0~\GeV$} \\
$M_{\mathrm{jets,min}}$ & \multicolumn{2}{c}{$14.0~\GeV$} \\
$x_{\Pom}^{\mathrm{max}}$ & \multicolumn{2}{c}{$0.025$} \\
$|\eta^{\mathrm{max}}|$ & \multicolumn{2}{c}{$4.4$} \\
\noalign{\smallskip}\hline
\end{tabular}
\end{center}
\end{table}

\subsection{pPb collisions}

The setup for the photoproduction in $\p\Pb$ collisions is 
the same as our default setup for $\e\p$, albeit the photon flux is 
now provided by eq.~(\ref{eq:photonFluxAint}). We here neglect the 
contribution where the proton would provide the photon flux, such that
all photons arise from the nucleus. The jets are reconstructed with 
an anti-$k_{\mathrm{T}}$ algorithm using $R=1.0$ as implemented in 
\textsc{FastJet} package \cite{Cacciari:2011ma}. The applied cuts are 
presented in table~\ref{Tab:LHCcuts} and are very similar to the ones 
used by HERA analyses. The experimentally reachable 
lower cut on $E_{\perp}$ is not set in stone, however. This depends 
on how well the jets can be reconstructed in this process. 
On one hand, the underlying event activity is greatly reduced in 
UPCs as compared to $\p\p$ collisions, thus possibly allowing for a 
decrease of the reachable jet $E_{\perp}$. On the other hand,
the increased $W$ might 
require an increase of the minimum $E_{\perp}$ w.r.t.\ the HERA 
analyses. Feasibility of such a measurement has been 
recently demonstrated in a preliminary ATLAS study \cite{ATLAS:2017kwa}
which measured inclusive dijets in ultra-peripheral $\Pb\Pb$ 
collisions at the LHC.

\begin{figure*}[!ht]
\begin{minipage}[c]{0.475\linewidth}
\centering
\includegraphics[width=0.9\linewidth]{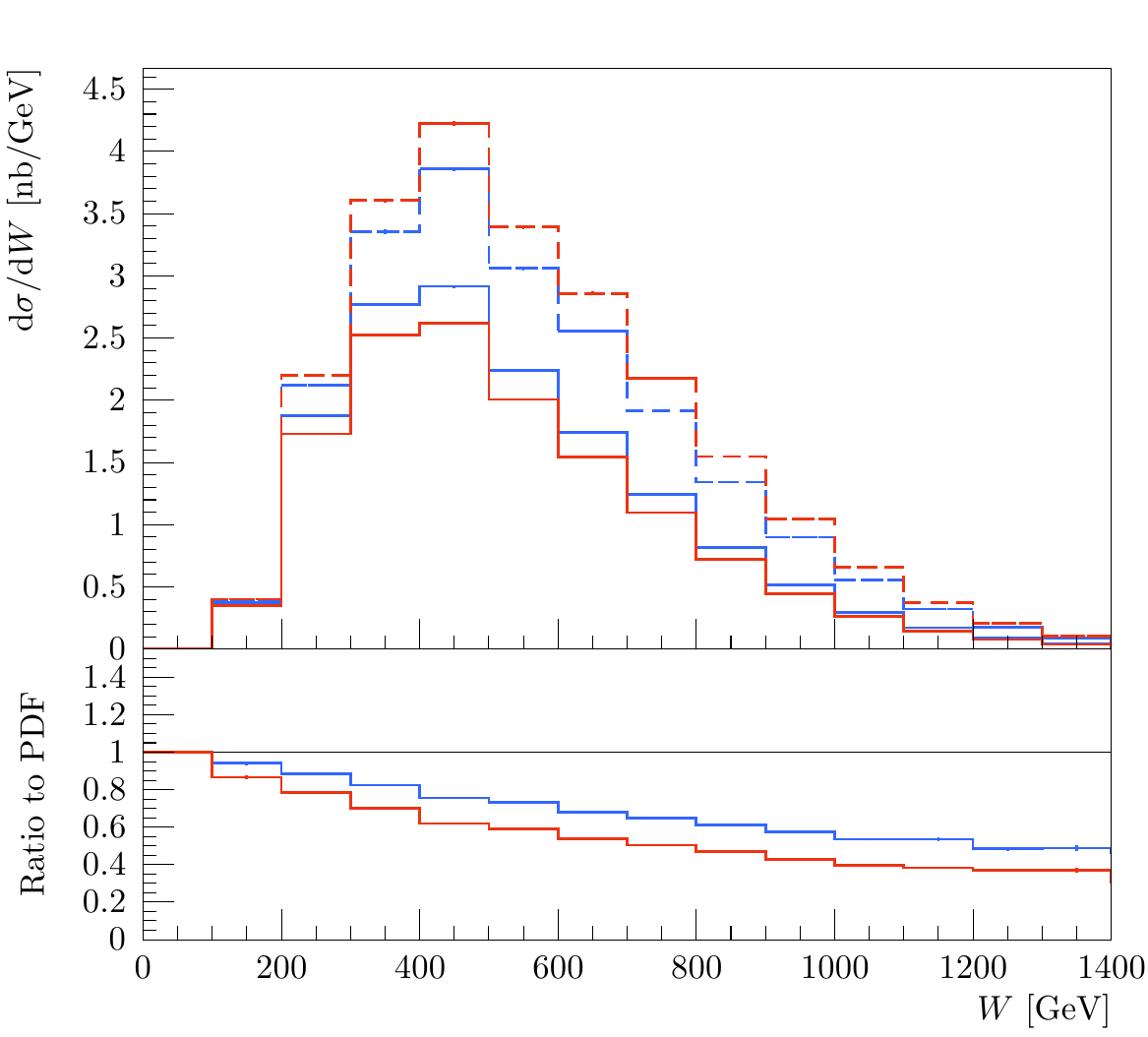}\\
(a)
\end{minipage}
\hfill
\begin{minipage}[c]{0.475\linewidth}
\centering
\includegraphics[width=0.9\linewidth]{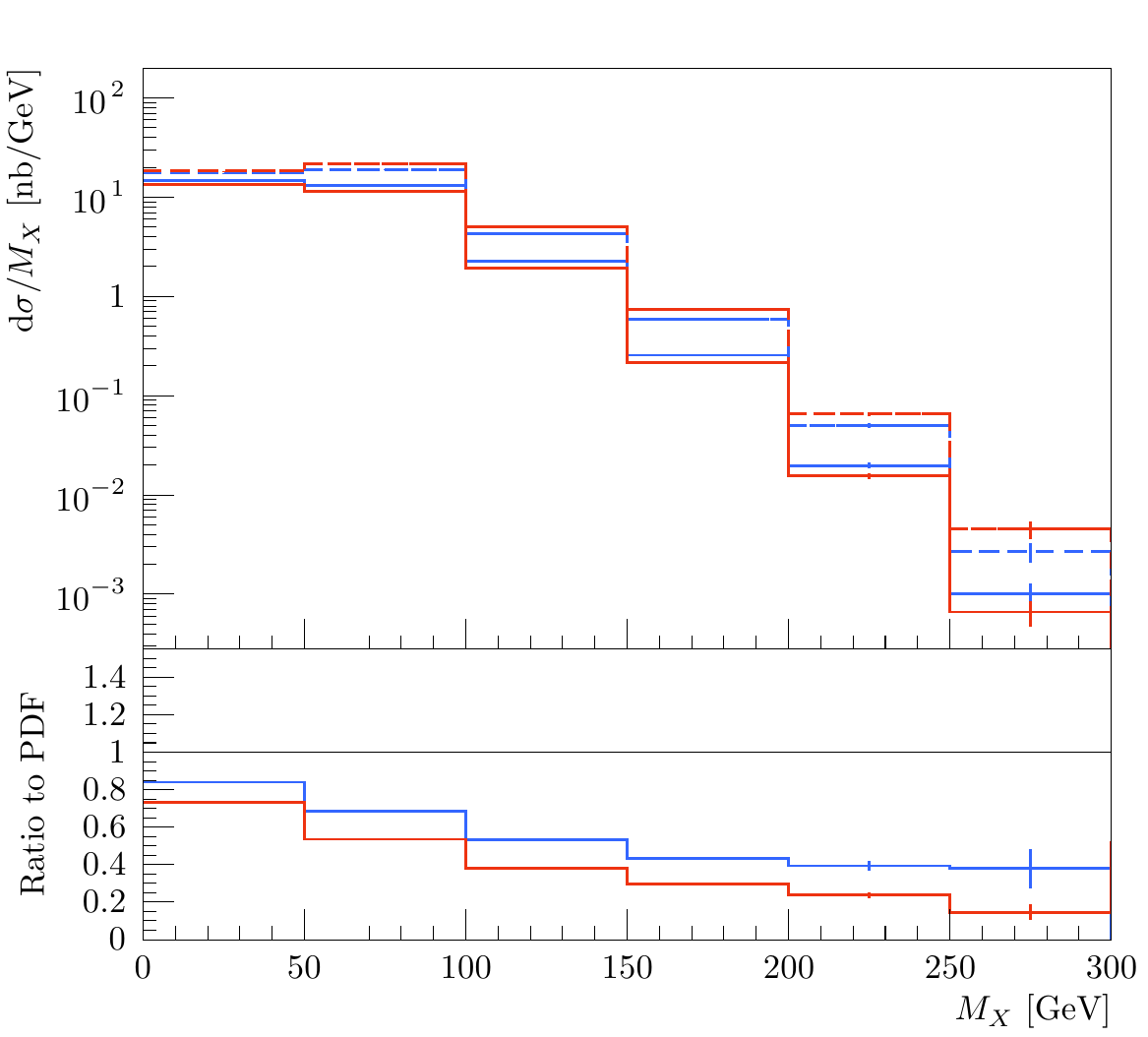}\\
(b)
\end{minipage}
\begin{minipage}[c]{0.475\linewidth}
\centering
\includegraphics[width=0.9\linewidth]{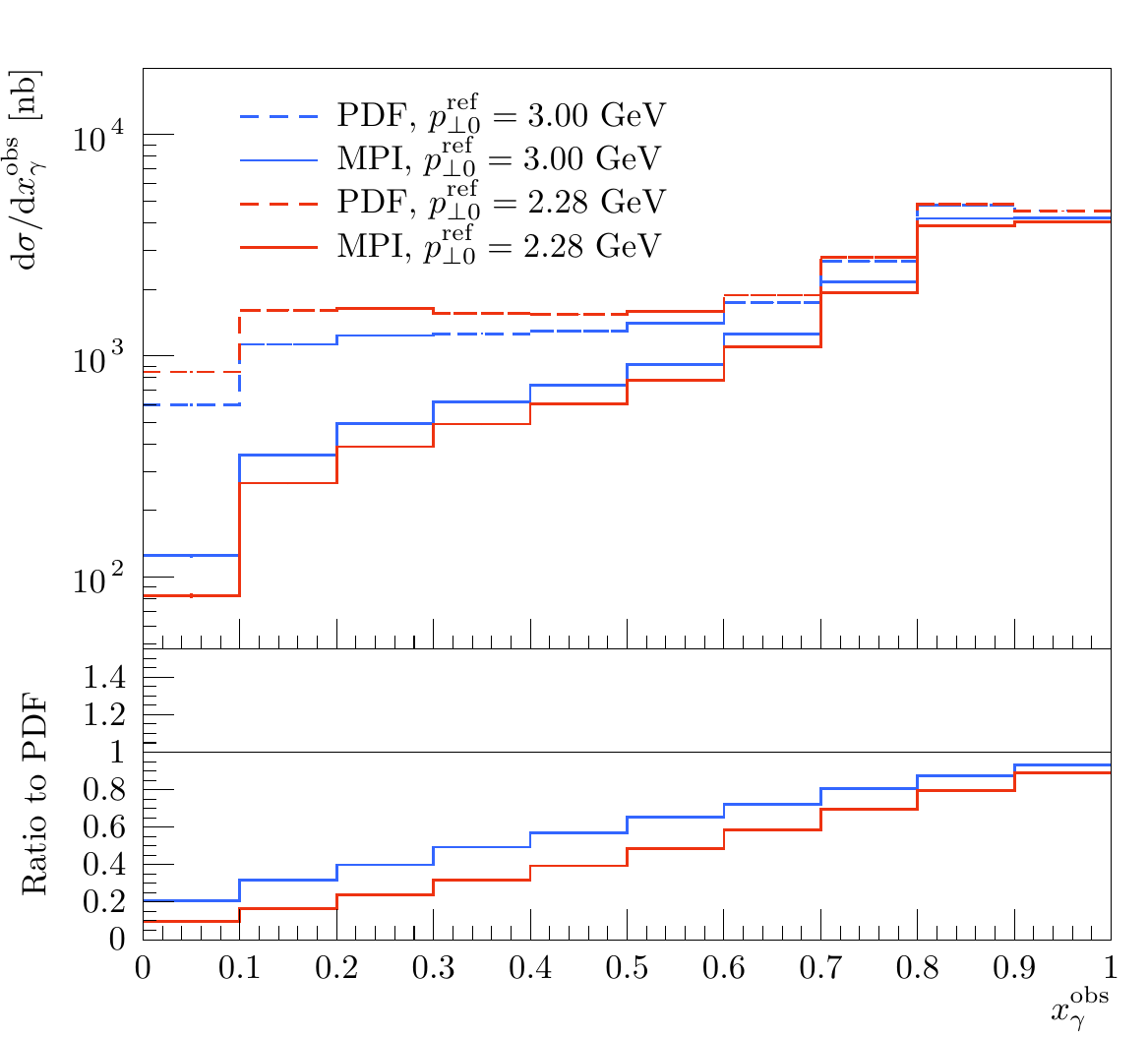}\\
(c)
\end{minipage}
\hfill
\begin{minipage}[c]{0.475\linewidth}
\centering
\includegraphics[width=0.9\linewidth]{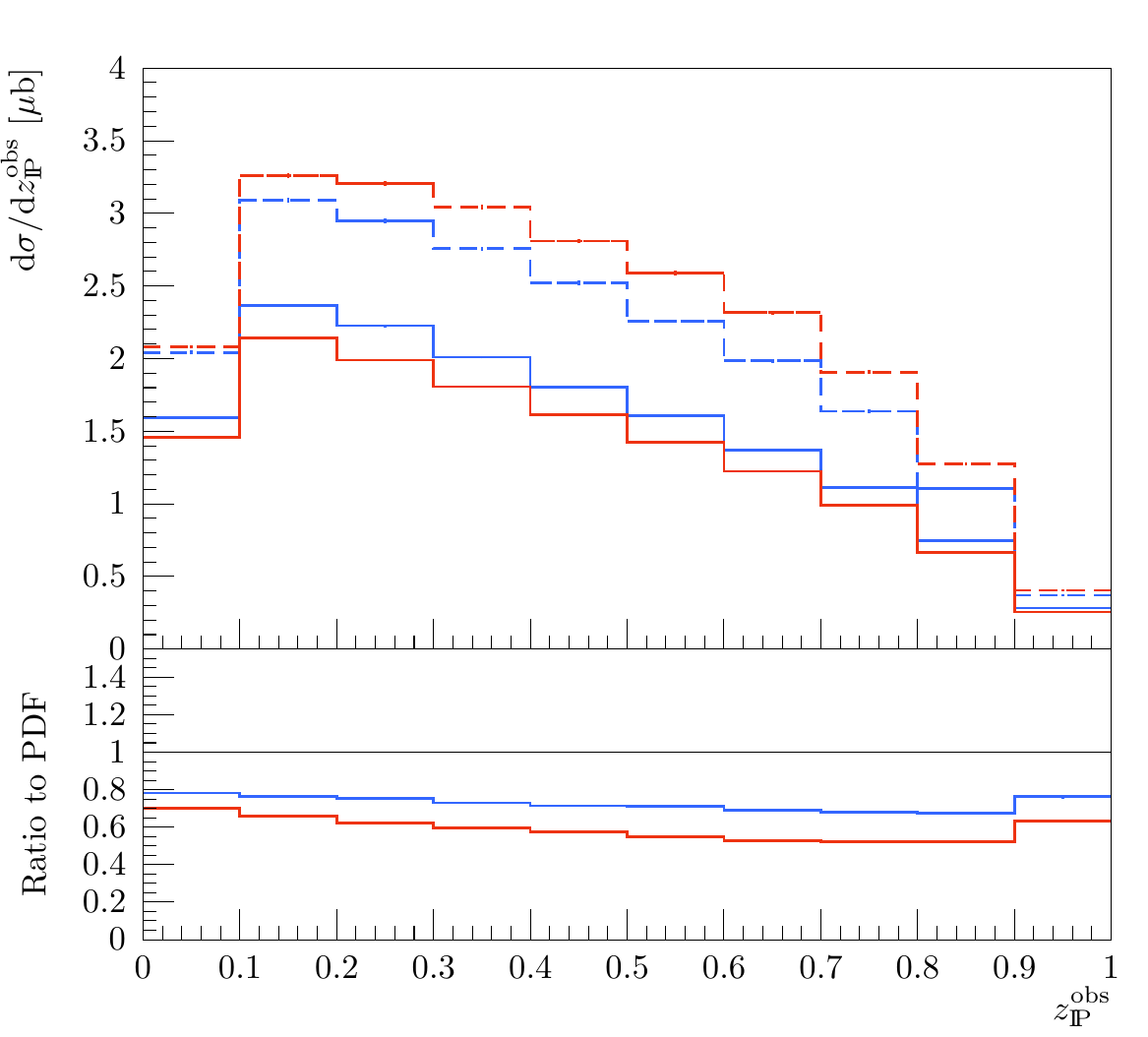}\\
(d)
\end{minipage}
\caption{\label{Fig:UPCpPb}Cross section for diffractive dijets in 
ultra-peripheral $\p\Pb$ collisions for observables $W$ (a), $M_X$ (b),
$x^{\mathrm{obs}}_{\gamma}$ (c) and 
$z^{\mathrm{obs}}_{\Pom}$ (d).
Vertical bars denote the statistical uncertainty in the MC generation.
}
\end{figure*}

The resulting differential cross sections for diffractive dijets from
UPCs in $\p\Pb$ collisions are presented in Fig.~\ref{Fig:UPCpPb}. 
Similar to sec.~\ref{Sec:4} we show the results differential in $W$, 
$M_X$, $x_{\gamma}^{\mathrm{obs}}$ and $z_{\Pom}^{\mathrm{obs}}$. 
The ``PDF'' samples (dashed lines) are without the 
gap suppression and ``MPI'' samples (solid lines) are with
the gap suppression. The lower 
panels show the ratio of the two, corresponding to the rapidity 
gap suppression factor predicted by the model. As discussed earlier, 
the energy dependence of the $\pTo$ screening parameter in 
$\gamma\p$ collisions was constrained by HERA data in a narrow $W$ 
bin around $200~\GeV$. As the UPC events at the LHC will extend to much 
higher values of $W$, the poorly-constrained energy dependence of 
$\pTo$ will generate some theoretical uncertainty for the 
predictions. To get a handle on this uncertainty we show samples with 
both the $\p\p$-tuned (red lines) and $\e\p$-tuned (blue lines) values 
for $\pTo$.

The predicted gap suppression factor is rather flat as a function of 
$z_{\Pom}^{\mathrm{obs}}$ at around $\sim 0.7$. The suppression factor
is, however, strongly dependent on $W$ and $M_X$, also observed in the 
HERA comparisons. It is more pronounced at the LHC thanks to the 
extended range in $W$, with an average suppression being roughly two 
times larger than at HERA. A similar strong dependence is also seen in 
$x_{\gamma}^{\mrm{obs}}$. As concluded earlier, the increasing 
suppression with $W$ follows from the fact that the probability for 
MPIs is increased with a higher $W$, due to the increased cross 
sections for the QCD processes. Thus a larger number of tentatively 
diffractive events are rejected due to the additional MPIs. Similarly,
decreasing $x_{\gamma}^{\mathrm{obs}}$ will leave more room
for the MPIs to take place, since the momentum extracted from
the photon to the primary jet production is decreased. 

A reduction of the $\pTo^{\mrm{ref}}$-value from $3.00~\GeV$ to 
$2.28~\GeV$ increases the MPI probability, thus having a twofold effect. 
Firstly, it increases the jet cross section in the ``PDF''-sample, as 
the additional MPIs allowed with the lower reference value increase the 
energy inside the jet cone. Secondly, the enhanced MPI probability 
rejects a larger number of tentatively diffractive events, thus giving 
a larger gap suppression effect. Collectively, these effects lead to 
$20-30~\%$ larger gap-suppression factors as compared to the $\gamma\p$ 
value for $\pTo^{\mrm{ref}}$.

\subsection{pp collisions}

\begin{figure*}[!ht]
\begin{minipage}[c]{0.475\linewidth}
\centering
\includegraphics[width=0.9\linewidth]{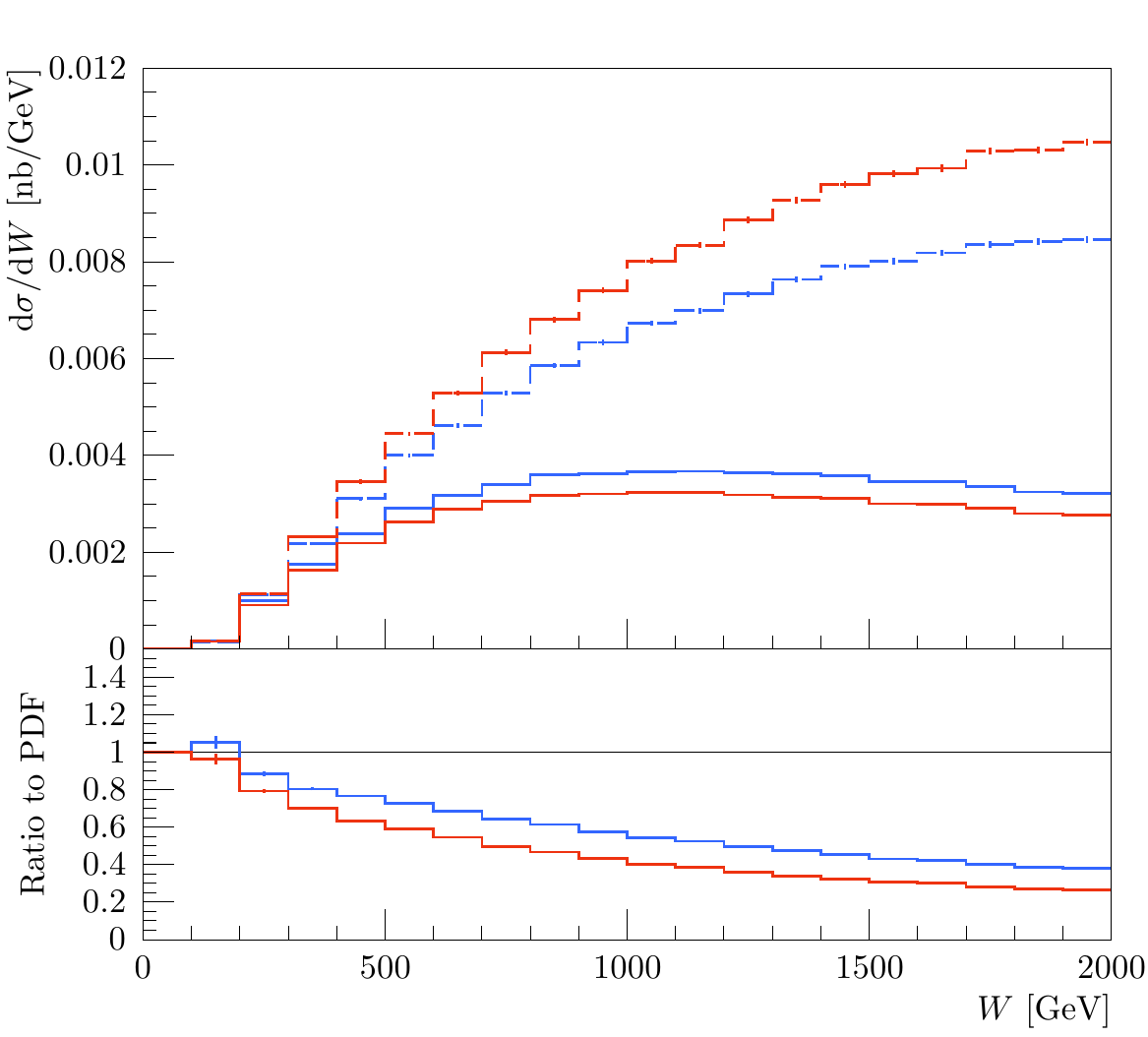}\\
(a)
\end{minipage}
\hfill
\begin{minipage}[c]{0.475\linewidth}
\centering
\includegraphics[width=0.9\linewidth]{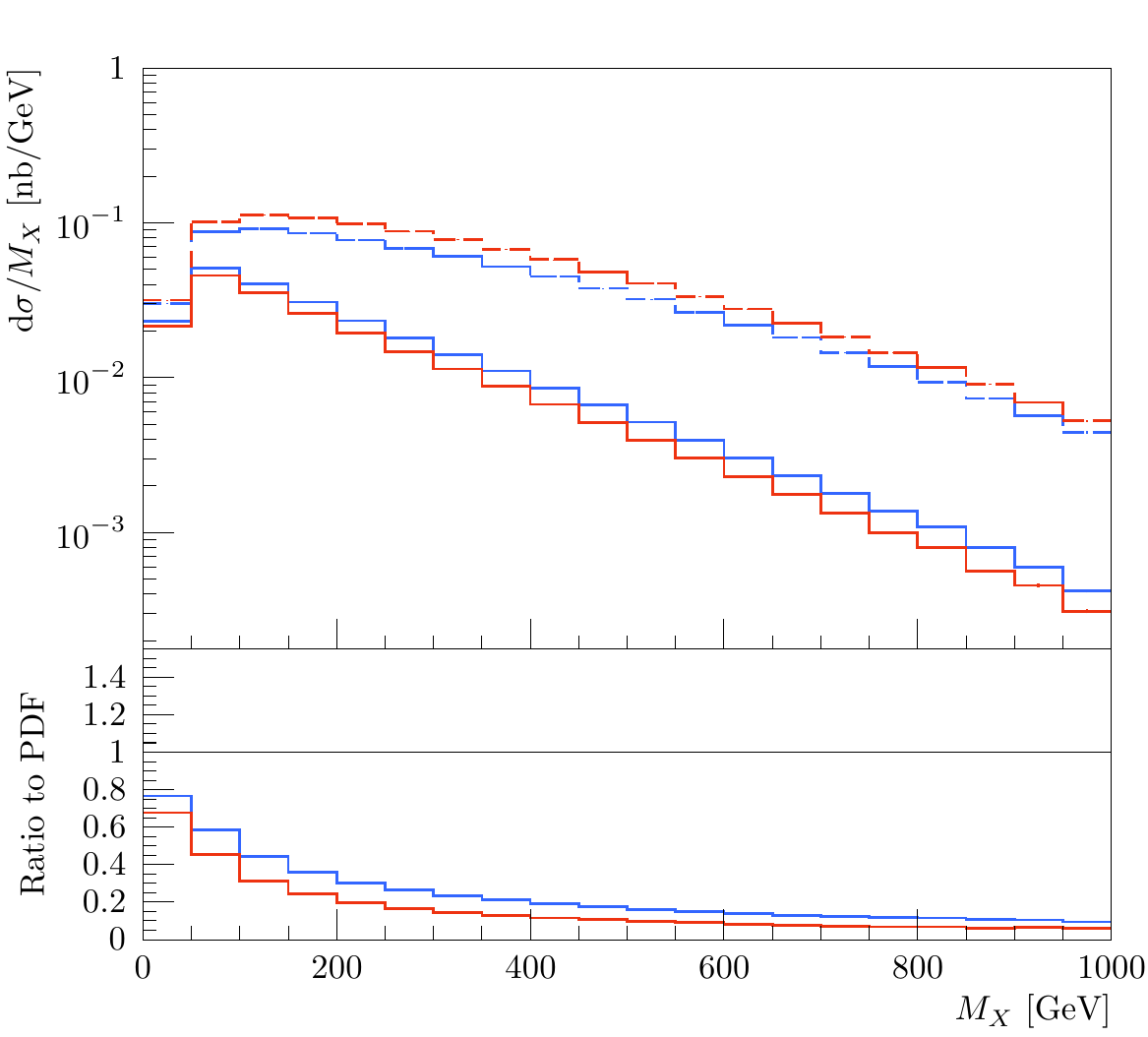}\\
(b)
\end{minipage}
\begin{minipage}[c]{0.475\linewidth}
\centering
\includegraphics[width=0.9\linewidth]{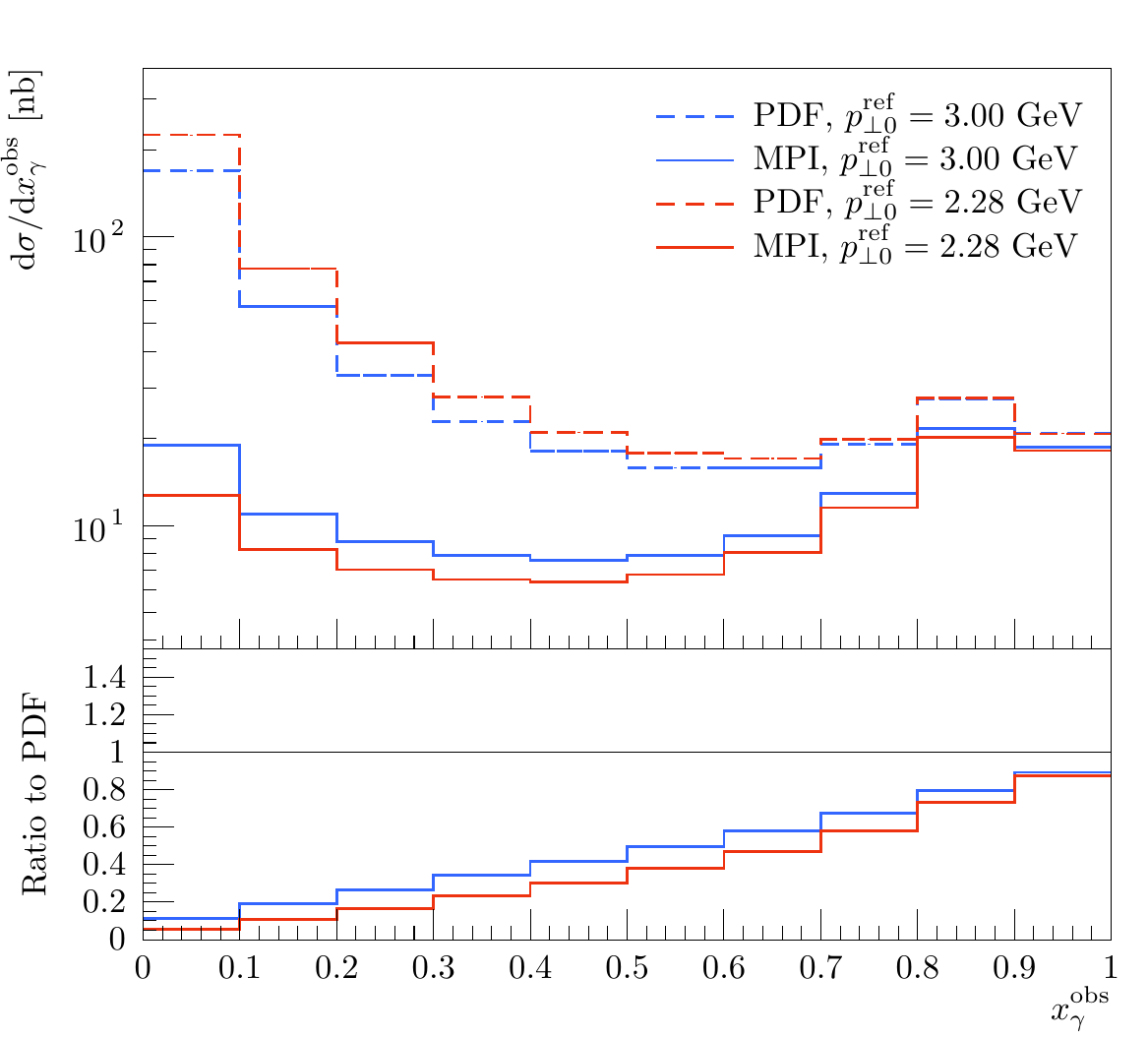}\\
(c)
\end{minipage}
\hfill
\begin{minipage}[c]{0.475\linewidth}
\centering
\includegraphics[width=0.9\linewidth]{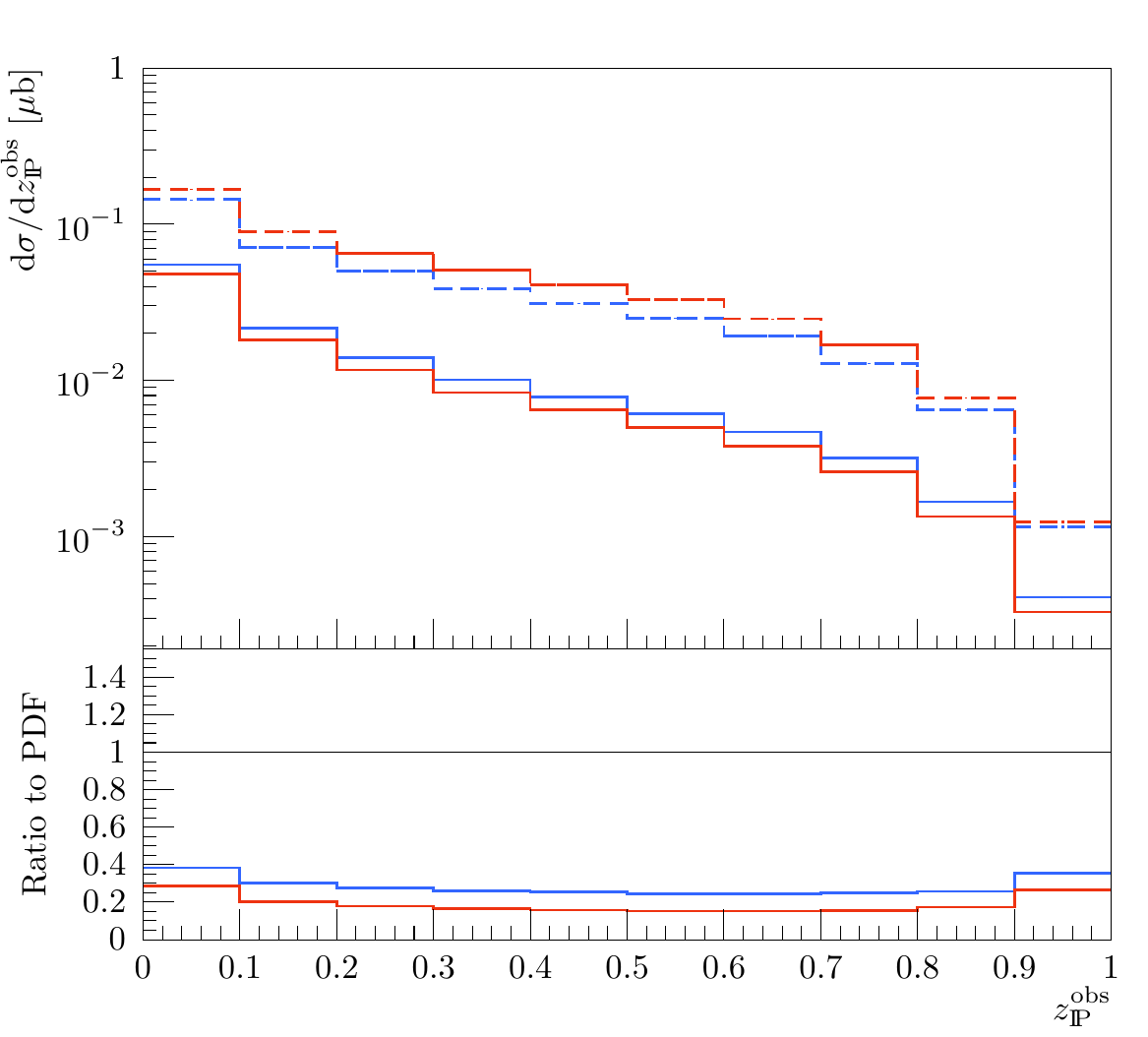}\\
(d)
\end{minipage}
\caption{\label{Fig:UPCpp}Cross section for diffractive dijets in 
ultra-peripheral $\p\p$ collisions for observables $W$ (a), 
$M_X$ (b), $x^{\mathrm{obs}}_{\gamma}$ (c) and 
$z^{\mathrm{obs}}_{\Pom}$ (d).
Vertical bars denote the statistical uncertainty in the MC generation.
}
\end{figure*}

The kinematical cuts applied in $\p\p$ equals those from $\p\Pb$.
Due to the increased $\sqrt{s}$ and the harder photon spectrum from 
protons compared to heavy ions, the $W$ range probed is extended to 
even larger values. When keeping jet kinematics fixed this leaves more 
room for MPIs in the $\gamma\p$-system, while also increasing the 
relative contribution from resolved photons. Thus the predicted 
gap-suppression factors are further increased here, as compared to 
$\p\Pb$ and $\e\p$ case, cf.\ Fig.~\ref{Fig:UPCpp}. At extreme 
kinematics -- high-$M_X$, low-$x_{\gamma}^{\mathrm{obs}}$ -- the 
gap-suppression factors are almost as large as what have been found 
in hadronic diffractive $\p\p$ events. The $\p\p$ suppression factors 
should provide an estimate of the upper limit for photoproduction, as 
the latter includes the (unsuppressed) direct contribution. The 
suppression factors show a similar sensitivity to the value of 
$\pTo^{\mrm{ref}}$ as in $\p\Pb$ collisions, such that the lower 
value gives more suppression. Notice that the cross sections are 
calculated assuming that the photon is emitted from the beam with 
positive $p_z$.

A particularly interesting observable is the $x_{\gamma}^{\mathrm{obs}}$
distribution. Due to the extended $W$ reach, the dijet production starts 
to be sensitive also to the low-$x$ part of the photon PDFs. Here, the
photon PDF analyses find that gluon distributions rise rapidly with
decreasing $x$, the same tendency as seen in proton PDFs. This 
generates the observed rise of the cross section towards low values of 
$x_{\gamma}^{\mathrm{obs}}$ when the MPI rejection 
is not applied. However, the contribution from the 
low-$x_{\gamma}^{\mathrm{obs}}$ region is significantly reduced when the 
rejection is applied, as these events have a high probability for MPIs. 
Note, however, that there are large differences in the gluon distributions 
between different photon PDF analyses in this region. Thus here a variation 
of the photon PDF in the hard scattering could have some effect on the 
predicted gap-suppression factor, even though only very mild impact was seen in the  
HERA comparisons. But as most of these events with a soft gluon in the 
hard scattering will be rejected due to the presence of additional 
MPIs, the predicted cross-sections shown in Fig.~\ref{Fig:UPCpp} 
is expected to be rather stable against such variations. 
Further uncertainty again arises from the dPDFs. But as the purpose of 
the shown UPC results is to demonstrate the gap survival effects, we do 
not discuss the sensitivity to dPDF variation here explicitly.

\section{\label{Sec:7}Conclusions}

In this paper we present a model for explaining the 
factorization-breaking effects seen in photoproduction events at HERA. 
The model, implemented in the general purpose event generator 
\textsc{Pythia}~8, is an extension of the hard diffraction model
to photoproduction. 
It is a novel combination of several existing ideas, and it is the first 
model of its kind with a dynamical gap suppression based on the kinematics 
of the entire event.

The starting point is the Ingelman-Schlein approach, where the cross 
section is factorized into a Pomeron flux and a PDF, convoluted with the 
hard scattering cross section. The Pomeron flux and PDF are 
extracted from HERA data, but if used out-of-the-box these give an 
order-of-magnitude larger cross sections in pure hadron-hadron 
collisions, while the differences in photoproduction are around a factor 
of two at most. Thus, factorization was observed to be broken in 
diffractive events with a hard scale. The dynamical model extended here, 
explain this factorization breaking with additional 
MPI activity filling the rapidity gap used for experimental 
detection of the diffractive events. Thus the MPI framework of 
\textsc{Pythia}~8 is used as an additional suppression factor on an 
event-by-event basis, giving cross sections very similar to what is 
seen in data, both in $\p\p$ and $\p\pbar$ events. As low virtuality 
photons are allowed to fluctuate into a hadronic state, MPIs are also 
possible in these systems. Thus the same mechanism is responsible for 
the factorization breaking in photoproduction events in $\e\p$ 
collisions, and also here the model predicts cross sections similar to 
what is seen in $\e\p$ data.

We present results obtained with the model compared to experimental 
data from H1 and ZEUS for diffractive dijet photoproduction. The 
agreement with the data is improved when the MPI rejection 
is applied, supporting the idea behind the factorization-breaking 
mechanism. However, the kinematical cuts applied by the experiments 
reduce the contribution from resolved photons, so the observed 
suppression is rather mild with the HERA kinematics, especially for 
the ZEUS data. The improvements in the dPDFs raises the question if such
a suppression is actually needed, as the new dPDFs seem to describe data
fairly well without, especially in the direct-enhanced region of
phase space. Furthermore, there are several theoretical uncertainties 
that hamper the interpretation of the data, and the description is far 
from perfect for all considered distributions. Many of these 
theoretical uncertainties could be reduced by considering ratios of 
diffractive dijets in DIS and photoproduction regimes, but have not 
been pursued here as the description for DIS in \textsc{Pythia}~8 is 
not yet complete.

As an additional example for the range of the model, we present 
predictions for diffractive dijets in ultra-\\peripheral $\p\p$ and 
$\p\Pb$ collisions at the LHC. In these processes a quasi-real photon 
emitted from a proton or nucleus interacts with a proton from the other 
beam. Due to the larger invariant masses of the $\gamma\p$ system in
these processes, the contribution from resolved photons is significantly 
increased. Thus UPCs is an excellent place to study the gap suppression
in photoproduction. The results demonstrate that a measurement of 
photoproduced diffractive dijet cross sections in $\p\p$ collisions 
would provide very strong constraints on our dynamical rapidity gap 
survival model, as the effects are much more pronounced than with HERA 
kinematics. The distinct features of the model are well accessible within 
the kinematical limits for UPCs at LHC. If such a measurement is not 
feasible due to the pure QCD background, a measurement in $\p\Pb$ 
collisions would be sufficient to confirm the factorization breaking in 
diffractive photoproduction and provide constraints on the underlying 
mechanism.

Future work consists of opening up for different photon PDFs in the
photoproduction framework, improving the DIS description in 
\textsc{Pythia}~8 and merging the two regimes in a consistent manner.
The first allows for probing additional theoretical uncertainties of 
the photoproduction framework, the second allows for probing the
double ratios of photoproduction to DIS cross sections for diffractive
dijets. The merging of the two regimes would allow for full event
generation of all photon virtualities needed for future collider
studies. Similarly a combination of the current model and the
\textsc{Angantyr} model for heavy ions is planned, such that $\e\A$ and
ultraperipheral UPCs in $\A\A$ collisions could be probed as well.

\section*{\label{Sec:8}Acknowledgments}

We thank Torbj\"orn Sj\"ostrand for reading through and commenting on the
manuscript and Stefan Prestel for providing plotting tools. 
We also thank Hannes Jung for support with the HERA analyses, and Radek 
Zlebcik for providing the ZEUS SJ dPDFs.

This project has received funding from the European Research
Council (ERC) under the European Union's Horizon 2020 research
and innovation program (grant agreement No 668679), and in part 
by the Swedish Research Council, contract number 2016-05996 and the 
Marie Sklodowska-Curie Innovative Training Network MCnetITN3 
(grant agreement 722104). Further support is provided by the Carl Zeiss 
Foundation and the Academy of Finland, Project 308301.


\begin{thebibliography}{99}
\bibitem{Sjostrand:2014zea}
  T.~Sj\"ostrand, S.~Ask, J.~R.~Christiansen, R.~Corke, N.~Desai, P.~Ilten, 
  S.~Mrenna and S.~Prestel {\it et al.},
  Comput.\ Phys.\ Commun.\  {\bf 191} (2015) 159
  [arXiv:1410.3012 [hep-ph]].

\bibitem{Bonino:1988ae}
  R.~Bonino {\it et al.} [UA8 Collaboration],
  Phys.\ Lett.\ B {\bf 211} (1988) 239.
  doi:10.1016/0370-2693(88)90840-4

\bibitem{Derrick:1994ze}
  M.~Derrick {\it et al.} [ZEUS Collaboration],
  Phys.\ Lett.\ B {\bf 332} (1994) 228.
  doi:10.1016/0370-2693(94)90883-4

\bibitem{Affolder:2000vb}
  T.~Affolder {\it et al.} [CDF Collaboration],
  Phys.\ Rev.\ Lett.\  {\bf 84} (2000) 5043.
  doi:10.1103/PhysRevLett.84.5043

\bibitem{Aad:2015xis}
  G.~Aad {\it et al.} [ATLAS Collaboration],
  Phys.\ Lett.\ B {\bf 754} (2016) 214
  doi:10.1016/j.physletb.2016.01.028
  [arXiv:1511.00502 [hep-ex]].

\bibitem{Aaltonen:2010qe}
  T.~Aaltonen {\it et al.} [CDF Collaboration],
  Phys.\ Rev.\ D {\bf 82} (2010) 112004
  doi:10.1103/PhysRevD.82.112004
  [arXiv:1007.5048 [hep-ex]].

\bibitem{Ingelman:1984ns}
  G.~Ingelman and P.~E.~Schlein,
  Phys.\ Lett.\  {\bf 152B} (1985) 256.
  doi:10.1016/0370-2693(85)91181-5

\bibitem{Rasmussen:2015qgr}
  C.~O.~Rasmussen and T.~Sj\"ostrand,
  JHEP {\bf 1602} (2016) 142
  doi:10.1007/JHEP02(2016)142
  [arXiv:1512.05525 [hep-ph]].

\bibitem{Bjorken:1992er}
  J.~D.~Bjorken,
  Phys.\ Rev.\ D {\bf 47} (1993) 101.

\bibitem{CMS:2018udy}
  CMS Collaboration [CMS Collaboration],
  CMS-PAS-FSQ-12-033.

\bibitem{Adloff:1998gg}
  C.~Adloff {\it et al.} [H1 Collaboration],
  Eur.\ Phys.\ J.\ C {\bf 6} (1999) 421
  doi:10.1007/s100529801046
  [hep-ex/9808013].

\bibitem{Aktas:2007hn}
  A.~Aktas {\it et al.} [H1 Collaboration],
  Eur.\ Phys.\ J.\ C {\bf 51} (2007) 549
  doi:10.1140/epjc/s10052-007-0325-4
  [hep-ex/0703022].

\bibitem{Chekanov:2007rh}
  S.~Chekanov {\it et al.} [ZEUS Collaboration],
  Eur.\ Phys.\ J.\ C {\bf 55} (2008) 177
  doi:10.1140/epjc/s10052-008-0598-2
  [arXiv:0710.1498 [hep-ex]].

\bibitem{Chekanov:2009aa}
  S.~Chekanov {\it et al.} [ZEUS Collaboration],
  Nucl.\ Phys.\ B {\bf 831} (2010) 1
  doi:10.1016/j.nuclphysb.2010.01.014
  [arXiv:0911.4119 [hep-ex]].

\bibitem{Aaron:2010su}
  F.~D.~Aaron {\it et al.} [H1 Collaboration],
  Eur.\ Phys.\ J.\ C {\bf 70} (2010) 15
  doi:10.1140/epjc/s10052-010-1448-6
  [arXiv:1006.0946 [hep-ex]].

\bibitem{Andreev:2015cwa}
  V.~Andreev {\it et al.} [H1 Collaboration],
  JHEP {\bf 1505} (2015) 056
  doi:10.1007/JHEP05(2015)056
  [arXiv:1502.01683 [hep-ex]].

\bibitem{Baltz:2007kq}
  A.~J.~Baltz {\it et al.},
  Phys.\ Rept.\  {\bf 458} (2008) 1
  doi:10.1016/j.physrep.2007.12.001
  [arXiv:0706.3356 [nucl-ex]].

\bibitem{Sjostrand:2006za}
  T.~Sj\"ostrand, S.~Mrenna and P.~Z.~Skands,
  JHEP {\bf 0605} (2006) 026
  doi:10.1088/1126-6708/2006/05/026
  [hep-ph/0603175].

\bibitem{Linssen:2012hp}
  L.~Linssen, A.~Miyamoto, M.~Stanitzki and H.~Weerts,
  doi:10.5170/CERN-2012-003
  arXiv:1202.5940 [physics.ins-det].

\bibitem{Baer:2013cma}
  H.~Baer {\it et al.},
  arXiv:1306.6352 [hep-ph].

\bibitem{Accardi:2012qut}
  A.~Accardi {\it et al.},
  Eur.\ Phys.\ J.\ A {\bf 52} (2016) no.9,  268
  doi:10.1140/epja/i2016-16268-9
  [arXiv:1212.1701 [nucl-ex]].

\bibitem{Gomez-Ceballos:2013zzn}
  M.~Bicer {\it et al.} [TLEP Design Study Working Group],
  JHEP {\bf 1401} (2014) 164
  doi:10.1007/JHEP01(2014)164
  [arXiv:1308.6176 [hep-ex]].

\bibitem{Mangano:2016jyj}
  M.~L.~Mangano {\it et al.},
  CERN Yellow Report (2017) no.3,  1
  doi:10.23731/CYRM-2017-003.1
  [arXiv:1607.01831 [hep-ph]].

\bibitem{Bierlich:2018xfw}
  C.~Bierlich, G.~Gustafson, L.~L\"onnblad and H.~Shah,
  JHEP {\bf 1810} (2018) 134
   [JHEP {\bf 2018} (2020) 134]
  doi:10.1007/JHEP10(2018)134
  [arXiv:1806.10820 [hep-ph]].

\bibitem{Friberg:2000ra}
  C.~Friberg and T.~Sj\"ostrand,
  JHEP {\bf 0009} (2000) 010
  doi:10.1088/1126-6708/2000/09/010
  [hep-ph/0007314].

\bibitem{Cabouat:2017rzi}
  B.~Cabouat and T.~Sj\"ostrand,
  Eur.\ Phys.\ J.\ C {\bf 78} (2018) no.3,  226
  doi:10.1140/epjc/s10052-018-5645-z, 10.1140/s10052-018-5645-z
  [arXiv:1710.00391 [hep-ph]].

\bibitem{Hoche:2015sya}
  S.~H\"oche and S.~Prestel,
  Eur.\ Phys.\ J.\ C {\bf 75} (2015) no.9,  461
  doi:10.1140/epjc/s10052-015-3684-2
  [arXiv:1506.05057 [hep-ph]].

\bibitem{Helenius:2017aqz}
  I.~Helenius,
  arXiv:1708.09759 [hep-ph].

\bibitem{Helenius:2018bai}
  I.~Helenius,
  arXiv:1806.07326 [hep-ph].

\bibitem{IHinProgress}
  I.~Helenius, T.~Sj\"ostrand, 
  In preparation.

\bibitem{Sjostrand:2004ef}
  T.~Sjostrand and P.~Z.~Skands,
  Eur.\ Phys.\ J.\ C {\bf 39} (2005) 129
  doi:10.1140/epjc/s2004-02084-y
  [hep-ph/0408302].

\bibitem{Sjostrand:2004pf}
  T.~Sj\"ostrand and P.~Z.~Skands,
  JHEP {\bf 0403} (2004) 053
  doi:10.1088/1126-6708/2004/03/053
  [hep-ph/0402078].

\bibitem{Andersson:1983ia}
  B.~Andersson, G.~Gustafson, G.~Ingelman and T.~Sjostrand,
  Phys.\ Rept.\  {\bf 97} (1983) 31.
  doi:10.1016/0370-1573(83)90080-7

\bibitem{Sjostrand:1987su}
  T.~Sj\"ostrand and M.~van Zijl,
  Phys.\ Rev.\ D {\bf 36} (1987) 2019.
  doi:10.1103/PhysRevD.36.2019

\bibitem{Schuler:1993wr}
  G.~A.~Schuler and T.~Sj\"ostrand,
  Phys.\ Rev.\ D {\bf 49} (1994) 2257.
  doi:10.1103/PhysRevD.49.2257

\bibitem{Stodolsky:1966am}
  L.~Stodolsky,
  Phys.\ Rev.\ Lett.\  {\bf 18} (1967) 135.
  doi:10.1103/PhysRevLett.18.135

\bibitem{Joos:1967ony}
  H.~Joos,
  Phys.\ Lett.\  {\bf 24B} (1967) 103.
  doi:10.1016/0370-2693(67)90359-0

\bibitem{Sakurai:1972wk}
  J.~J.~Sakurai and D.~Schildknecht,
  Phys.\ Lett.\  {\bf 40B} (1972) 121.
  doi:10.1016/0370-2693(72)90300-0

\bibitem{Gluck:1991jc}
  M.~Gl\"uck, E.~Reya and A.~Vogt,
  Phys.\ Rev.\ D {\bf 46} (1992) 1973.
  doi:10.1103/PhysRevD.46.1973

\bibitem{Aurenche:1994in}
  P.~Aurenche, J.~P.~Guillet and M.~Fontannaz,
  Z.\ Phys.\ C {\bf 64} (1994) 621
  doi:10.1007/BF01957771
  [hep-ph/9406382].

\bibitem{Schuler:1995fk}
  G.~A.~Schuler and T.~Sj\"ostrand,
  Z.\ Phys.\ C {\bf 68} (1995) 607
  doi:10.1007/BF01565260
  [hep-ph/9503384].

\bibitem{Cornet:2002iy}
  F.~Cornet, P.~Jankowski, M.~Krawczyk and A.~Lorca,
  Phys.\ Rev.\ D {\bf 68} (2003) 014010
  doi:10.1103/PhysRevD.68.014010
  [hep-ph/0212160].

\bibitem{Slominski:2005bw}
  W.~Slominski, H.~Abramowicz and A.~Levy,
  Eur.\ Phys.\ J.\ C {\bf 45} (2006) 633
  doi:10.1140/epjc/s2005-02458-7
  [hep-ph/0504003].

\bibitem{DeWitt:1978wn}
  R.~J.~DeWitt, L.~M.~Jones, J.~D.~Sullivan, D.~E.~Willen and H.~W.~Wyld, Jr.,
  Phys.\ Rev.\ D {\bf 19} (1979) 2046
   Erratum: [Phys.\ Rev.\ D {\bf 20} (1979) 1751].
  doi:10.1103/PhysRevD.19.2046, 10.1103/PhysRevD.20.1751

\bibitem{Skands:2014pea}
  P.~Skands, S.~Carrazza and J.~Rojo,
  Eur.\ Phys.\ J.\ C {\bf 74} (2014) no.8,  3024
  doi:10.1140/epjc/s10052-014-3024-y
  [arXiv:1404.5630 [hep-ph]].

\bibitem{vonWeizsacker:1934nji}
  C.~F.~von Weizs\"acker,
  Z.\ Phys.\  {\bf 88} (1934) 612.
  doi:10.1007/BF01333110

\bibitem{Williams:1934ad}
  E.~J.~Williams,
  Phys.\ Rev.\  {\bf 45} (1934) 729.
  doi:10.1103/PhysRev.45.729

\bibitem{Drees:1988pp}
  M.~Drees and D.~Zeppenfeld,
  Phys.\ Rev.\ D {\bf 39} (1989) 2536.
  doi:10.1103/PhysRevD.39.2536

\bibitem{Khoze:2000wk}
  V.~A.~Khoze, A.~D.~Martin and M.~G.~Ryskin,
  Eur.\ Phys.\ J.\ C {\bf 18} (2000) 167
  doi:10.1007/s100520000494
  [hep-ph/0007359].

\bibitem{Gotsman:2005rt}
  E.~Gotsman, E.~Levin, U.~Maor, E.~Naftali and A.~Prygarin,
  hep-ph/0511060.

\bibitem{Jones:2013pga}
  S.~P.~Jones, A.~D.~Martin, M.~G.~Ryskin and T.~Teubner,
  JHEP {\bf 1311} (2013) 085
  doi:10.1007/JHEP11(2013)085
  [arXiv:1307.7099 [hep-ph]].

\bibitem{Goharipour:2018yov}
  M.~Goharipour, H.~Khanpour and V.~Guzey,
  Eur.\ Phys.\ J.\ C {\bf 78} (2018) no.4,  309
  doi:10.1140/epjc/s10052-018-5787-z
  [arXiv:1802.01363 [hep-ph]].

\bibitem{Alekhin:2014irh}
  S.~Alekhin {\it et al.},
  Eur.\ Phys.\ J.\ C {\bf 75} (2015) no.7,  304
  doi:10.1140/epjc/s10052-015-3480-z
  [arXiv:1410.4412 [hep-ph]].

\bibitem{Chekanov:2007aa}
  S.~Chekanov {\it et al.} [ZEUS Collaboration],
  Eur.\ Phys.\ J.\ C {\bf 52} (2007) 813
  doi:10.1140/epjc/s10052-007-0426-0, 10.3204/proc07-01/112
  [arXiv:0708.1415 [hep-ex]].

\bibitem{Aktas:2007bv}
  A.~Aktas {\it et al.} [H1 Collaboration],
  JHEP {\bf 0710} (2007) 042
  doi:10.1088/1126-6708/2007/10/042
  [arXiv:0708.3217 [hep-ex]].

\bibitem{Lipatov:1974qm}
  L.~N.~Lipatov,
  Sov.\ J.\ Nucl.\ Phys.\  {\bf 20} (1975) 94
   [Yad.\ Fiz.\  {\bf 20} (1974) 181].

\bibitem{Gribov:1972ri}
  V.~N.~Gribov and L.~N.~Lipatov,
  Sov.\ J.\ Nucl.\ Phys.\  {\bf 15} (1972) 438
   [Yad.\ Fiz.\  {\bf 15} (1972) 781].

\bibitem{Altarelli:1977zs}
  G.~Altarelli and G.~Parisi,
  Nucl.\ Phys.\ B {\bf 126} (1977) 298.
  doi:10.1016/0550-3213(77)90384-4

\bibitem{Dokshitzer:1977sg}
  Y.~L.~Dokshitzer,
  Sov.\ Phys.\ JETP {\bf 46} (1977) 641
   [Zh.\ Eksp.\ Teor.\ Fiz.\  {\bf 73} (1977) 1216].

\bibitem{Aktas:2006hy}
  A.~Aktas {\it et al.} [H1 Collaboration],
  Eur.\ Phys.\ J.\ C {\bf 48} (2006) 715
  doi:10.1140/epjc/s10052-006-0035-3
  [hep-ex/0606004].

\bibitem{Buckley:2010ar}
  A.~Buckley, J.~Butterworth, L.~L\"onnblad, D.~Grellscheid, H.~Hoeth, J.~Monk, H.~Schulz and F.~Siegert,
  Comput.\ Phys.\ Commun.\  {\bf 184} (2013) 2803
  doi:10.1016/j.cpc.2013.05.021
  [arXiv:1003.0694 [hep-ph]].

\bibitem{Gordon:1996pm}
  L.~E.~Gordon and J.~K.~Storrow,
  Nucl.\ Phys.\ B {\bf 489} (1997) 405
  doi:10.1016/S0550-3213(97)00048-5
  [hep-ph/9607370].

\bibitem{Whalley:2005nh}
  M.~R.~Whalley, D.~Bourilkov and R.~C.~Group,
  hep-ph/0508110.

\bibitem{Guzey:2016tek}
  V.~Guzey and M.~Klasen,
  JHEP {\bf 1604} (2016) 158
  doi:10.1007/JHEP04(2016)158
  [arXiv:1603.06055 [hep-ph]].

\bibitem{Basso:2017mue}
  E.~Basso, V.~P.~Goncalves, A.~K.~Kohara and M.~S.~Rangel,
  Eur.\ Phys.\ J.\ C {\bf 77} (2017) no.9,  600
  doi:10.1140/epjc/s10052-017-5173-2
  [arXiv:1705.08834 [hep-ph]].

\bibitem{Cacciari:2011ma}
  M.~Cacciari, G.~P.~Salam and G.~Soyez,
  Eur.\ Phys.\ J.\ C {\bf 72} (2012) 1896
  doi:10.1140/epjc/s10052-012-1896-2
  [arXiv:1111.6097 [hep-ph]].

\bibitem{ATLAS:2017kwa}
  The ATLAS collaboration,
  ATLAS-CONF-2017-011.
\end{thebibliography}
\end{document}